\documentclass[final,5p,times,twocolumn]{elsarticle}
\usepackage{setspace}
\usepackage{amssymb}
\usepackage[T1]{fontenc}
\usepackage{calligra}
\usepackage{amsmath}  % for \text
\usepackage{xspace}   % optional, to manage spaces
\newcommand{\hw}[1]{{\textnormal{\calligra #1}}}

\usepackage{color}
\usepackage{epsfig}
\font\tenbf=cmbx9
\font\tenrm=cmr9
\font\ninerm=cmr9
\font\tenit=cmti9
\journal{International Journal of Solids and Structures}

\newcommand{\lle}{\mbox{$\langle$}}
\newcommand{\rle}{\mbox{$\rangle$}}

\newcommand{\bfsi}{\mbox{\boldmath$\sigma$}}
\newcommand{\bfSi}{\mbox{\boldmath$\Sigma$}}

\newcommand{\bfep}{\mbox{\boldmath$\varepsilon$}}
\newcommand{\bfka}{\mbox{\boldmath$\kappa$}}

\newcommand{\bfze}{\mbox{\boldmath$\zeta$}}

\newcommand{\bfchi}{\mbox{\boldmath$\chi$}}
\newcommand{\bfzeta}{\mbox{\boldmath$\zeta$}}

\newcommand{\bfzir}{\mbox{\boldmath$\bf 0$}}
\newcommand{\bfcK}{\mbox{\boldmath$\cal K$}}
\newcommand{\bfcL}{\mbox{\boldmath$\cal L$}}
\newcommand{\bfcG}{\mbox{\boldmath$\cal G$}}
\newcommand{\bfcM}{\mbox{\boldmath$\cal M$}}

\newcommand{\bfcU}{\mbox{\boldmath$\cal U$}}

\newcommand{\bfb}{\mbox{\boldmath$\bf b$}}

\newcommand{\bfd}{\mbox{\boldmath$\bf d$}}
\newcommand{\bfe}{\mbox{\boldmath$\bf e$}}

\newcommand{\bfm}{\mbox{\boldmath$\bf m$}}
\newcommand{\bff}{\mbox{\boldmath$\bf f$}}
\newcommand{\bfh}{\mbox{\boldmath$\bf h$}}
\newcommand{\bfg}{\mbox{\boldmath$\bf g$}}
\newcommand{\bfk}{\mbox{\boldmath$\bf k$}}

\newcommand{\bfn}{\mbox{\boldmath$\bf n$}}
\newcommand{\bfp}{\mbox{\boldmath$\bf p$}}
\newcommand{\bfq}{\mbox{\boldmath$\bf q$}}
\newcommand{\bft}{\mbox{\boldmath$\bf t$}}
\newcommand{\bfs}{\mbox{\boldmath$\bf s$}}

\newcommand{\bfu}{\mbox{\boldmath$\bf u$}}

\newcommand{\bfw}{\mbox{\boldmath$\bf w$}}
\newcommand{\bfx}{\mbox{\boldmath$\bf x$}}
\newcommand{\bfy}{\mbox{\boldmath$\bf y$}}
\newcommand{\bfz}{\mbox{\boldmath$\bf z$}}
\newcommand{\bfA}{\mbox{\boldmath$\bf A$}}
\newcommand{\bfB}{\mbox{\boldmath$\bf B$}}
\newcommand{\bfC}{\mbox{\boldmath$\bf C$}}
\newcommand{\bfD}{\mbox{\boldmath$\bf D$}}
\newcommand{\bfE}{\mbox{\boldmath$\bf E$}}
\newcommand{\bfG}{\mbox{\boldmath$\bf G$}}

\newcommand{\bfI}{\mbox{\boldmath$\bf I$}}
\newcommand{\bfJ}{\mbox{\boldmath$\bf J$}}
\newcommand{\bfL}{\mbox{\boldmath$\bf L$}}
\newcommand{\bfN}{\mbox{\boldmath$\bf N$}}

\newcommand{\bfX}{\mbox{\boldmath$\bf X$}}

\newcommand{\bfU}{\mbox{\boldmath$\bf U$}}

\newcommand{\bfR}{\mbox{\boldmath$\bf R$}}
\newcommand{\bfK}{\mbox{\boldmath$\bf K$}}
\newcommand{\bfM}{\mbox{\boldmath$\bf M$}}
\newcommand{\bfT}{\mbox{\boldmath$\bf T$}}
\newcommand{\bfS}{\mbox{\boldmath$\bf S$}}

\newcommand{\bfbM}{\mbox{$\mathbb{M}$}}

\newcommand{\bfdelta}{\mbox{\boldmath$\delta$}}
\newcommand{\bfdel}{\mbox{\boldmath$\delta$}}
\newcommand{\bfxi}{\mbox{\boldmath$\xi$}}
\newcommand{\bfta}{\mbox{\boldmath$\tau$}}
\newcommand{\bfmu}{\mbox{\boldmath$\mu$}}

\newcommand{\cH}{\mbox{$\cal H$}}
\newcommand{\cL}{\mbox{$\cal L$}}

\newcommand{\bfbA}{\mbox{$\mathbb{A}$}}
\newcommand{\bfbK}{\mbox{$\mathbb{K}$}}
\newcommand{\bfbC}{\mbox{$\mathbb{C}$}}
\newcommand{\bfbD}{\mbox{$\mathbb{D}$}}

\newcommand{\bfLa}{\mbox{\boldmath$\Lambda$}}
\newcommand{\bfcD}{\mbox{\boldmath$\cal D$}}

\newcommand{\bfcE}{\mbox{\boldmath$\cal E$}}
\newcommand{\bfcR}{\mbox{\boldmath$\cal R$}}
\newcommand{\bfcA}{\mbox{\boldmath$\cal A$}}
\newcommand{\bfcB}{\mbox{\boldmath$\cal B$}}
\newcommand{\bfcJ}{\mbox{\boldmath$\cal J$}}

\newcommand{\cV}{\mbox{$\cal V$}}

\newcommand{\bfal}{\mbox{\boldmath$\alpha$}}
\newcommand{\bfbe}{\mbox{\boldmath$\beta$}}
\newcommand{\bfga}{\mbox{\boldmath$\gamma$}}
\newcommand{\bfGa}{\mbox{\boldmath$\Gamma$}}
\newcommand{\bfpi}{\mbox{\boldmath$\pi$}}
\newcommand{\bfet}{\mbox{\boldmath$\eta$}}
\newcommand{\bftau}{\mbox{\boldmath$\tau$}}

\newcommand{\bfthe}{\mbox{\boldmath$\vartheta$}}
\newcommand{\bfeta}{\mbox{\boldmath$\eta$}}

\newcommand{\bfphi}{\mbox{\boldmath$\phi$}}

\newcommand{\bfbL}{\mbox{$\mathbb{L}$}}

\newcommand{\BB}{\begin{equation}}
\newcommand{\EE}{\end{equation}}

\newcommand{\BBEQ}{\begin{eqnarray}}
\newcommand{\EEEQ}{\end{eqnarray}}

\begin{document}

\begin{graphicalabstract}
\vspace{0.mm}
\end{graphicalabstract}

\vspace{10.mm} \noindent %\hspace{30mm} 
\parbox{8.8cm}{
\centering \epsfig{figure=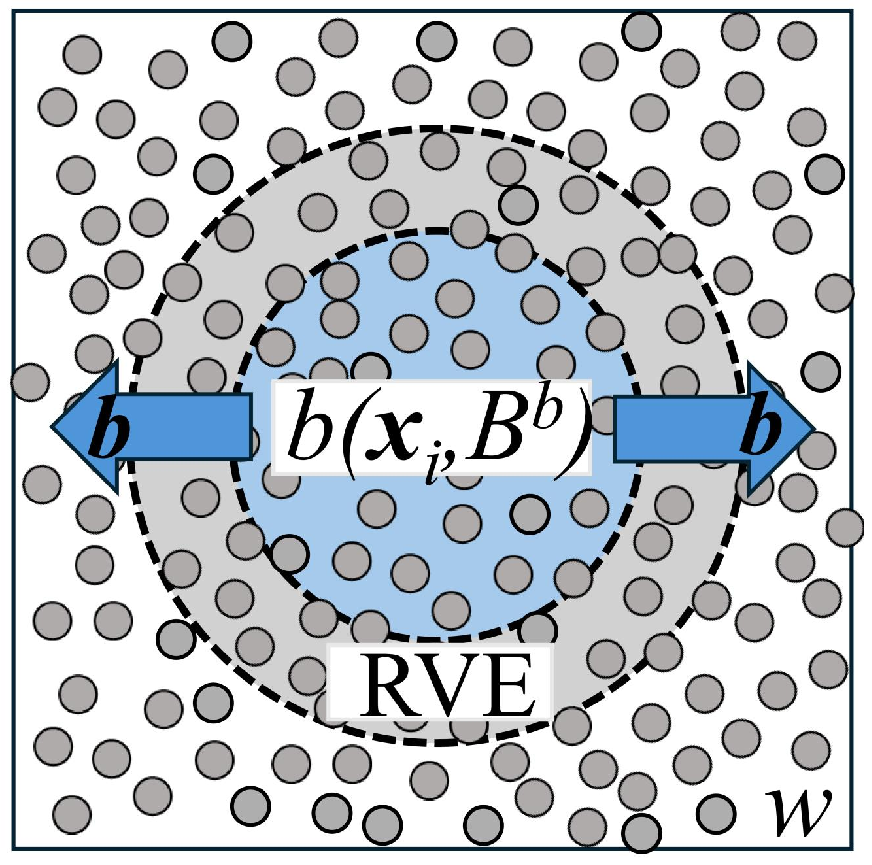, width=7.8cm}\\ \vspace{-50.mm}
\vspace{122.mm}
\vspace{-113.mm} \tenrm \baselineskip=8pt
%{ Scheme of CT image with $b({\bfx}_i,B^b)\subset{\rm RVE}\subset w$}}
{ Scheme of CT image with $b({\bfx}_i,B^b)\subset{\rm RVE}\subset w$}}
\vspace{0.mm}

\begin{frontmatter}


%% addressline={}, 
%% city={},
%% postcode={}, 
%% state={},
%% country={}}
%\def\floatpagepagefraction{1}
%\def\textpagefraction{.001}
%\shorttitle{Unified Micromechanics Theory of Composites}
%\shortauthors{V.A. Buryachenko}

\title{Unified Micromechanics Theory of Composites}  %\\
%{\color{black} Red: added and corrected parts}\\
%{\color{blue}Blue: removed parts}}  %% Article title

%% addressline={},
%% city={},
%% postcode={},
%% state={},
%% country={}}
%% addressline={},
%% city={},
%% postcode={},
%% state={},
%% country={}}

\author{Valeriy A. Buryachenko\footnote{Corresponding Author: V.A. Buryachenko, \ buryach@yahoo.com}}

\affiliation{organization={Micromechanics \& Composites LLC , Cincinnati, Ohio 45202, USA}%Department and Organization
% city={Cincinnati},
% postcode={45202}, 
% state={Ohio},
% country={USA}
}

%% Abstract
\begin{abstract}
%Abstract text.

This study introduces a modified Computational Analytical Micromechanics (CAM) framework for analyzing matrix composite materials (CMs) with arbitrary microstructures—random (statistically homogeneous or inhomogeneous), periodic, or deterministic (neither random nor periodic). The considered CMs exhibit a wide range of behaviors, including linear or nonlinear responses, coupled or uncoupled multi-physical phenomena, and local, weakly nonlocal (strain/stress gradient), or strongly nonlocal (peridynamic) phase properties.
The modified CAM approach is built on an exact Additive General Integral Equation (AGIE), applicable to CMs with any mentioned structural configuration and phase behavior. A unified iterative solution to the static AGIE is developed, using a compactly supported body force as a fundamentally new training parameter. The method introduces an extended Representative Volume Element (RVE) concept that generalizes Hill’s classical framework. This novel RVE formulation effectively filters out unrepresentative sub-datasets of effective parameters while eliminating edge, boundary layer, and sample size effects. It is universally applicable across all types of CMs, regardless of whether they exhibit local or nonlocal and linear or nonlinear phase behavior. Furthermore, the AGIE-CAM framework enables seamless integration with machine learning (ML) and neural network (NN) methods for constructing any unpredefined surrogate nonlocal operators. The methodology is designed as a modular, block-based system, supporting flexible development and refinement of computational tools. This robust framework accelerates multi-physics and multi-scale modeling in composite micromechanics. It represents a fundamental paradigm shift, redefining the analytical and philosophical foundations of micromechanics and expanding the boundaries of data-driven modeling of composite materials.
\end{abstract}

\begin{graphicalabstract}
\end{graphicalabstract}

\begin{highlights}
\item Composite materials of either random, periodic, or deterministic structures with either local or nonlocal properties of phases.
\item Exact additive general integral equation adapted to new training
parameter based on body forces with compact support.
\item New representative volume element (RVE) concept for body forces with compact support.
\item Incorporation of the new RVE concept into machine learning and neural networks
techniques for the construction of any unpredefined surrogate nonlocal operators.
\end{highlights}

%% Keywords

\end{frontmatter}



\centerline{\bf Table of Contents}
{%\tenrm
\baselineskip 10pt

\hangindent=0.4cm\hangafter=4\noindent
\noindent 1. Introduction \hfill 2\\
\hspace{5mm} 2. Preliminaries \hfill 4\\
%\hangindent=0.4cm\hangafter=1\noindent
%\hspace{8mm}
${\hspace{3.mm}}$ 2.1 Basic equations of peridynamics \hfill 4\\
%\hspace{8mm}
${\hspace{3.mm}}$ 2.2 Volumetric boundary conditions.\hfill 6\\
\hspace{5mm} 2.3 Descriptions of random and periodic structures\hfill 7\\
\hspace{8mm} 2.4 Some averages\hfill 8

\hangindent=0.4cm\hangafter=7\noindent
3. Decomposition of the material and field parameters\hfill 9\\
${\hspace{3.mm}}$ 3.1 Decomposition of the material parameters\hfill 9\\
${\hspace{3.mm}}$ 3.2 Analytica and Computational Micromechanics\\
${\hspace{8.mm}}$ in LM\hfill 9\\
${\hspace{3.mm}}$ 3.3 Modeling of one inclusion inside the infinite\\
${\hspace{8.mm}}$ matrix\hfill 10\\
%${\hspace{3.mm}}$
${\hspace{3.mm}}$ 3.4 General integral equations (GIEs)\hfill 12

\hangindent=0.4cm\hangafter=4\noindent
4. Solution of nonlinear AGIEs for random structure \\
${\hspace{3.mm}}$CMs \hfill 13\\
${\hspace{3.mm}}$ 4.1 Iterative solution of AGIE\hfill 14\\
${\hspace{3.mm}}$ 4.2 Effective constitutive law and effective dataset\hfill 14

\hangindent=0.4cm\hangafter=4\noindent
5. Periodic structure CMs\hfill 16

\noindent 6. Estimation of field fluctuations and effective\\
${\hspace{3.mm}}$energy-based criteria\hfill 18

\hangindent=0.4cm\hangafter=6\noindent
7. CMs with other constitutive laws of phases\hfill 19\\
${\hspace{3.mm}}$ 7.1 Locally elastic\hfill 19\\
%${\hspace{3.mm}}$
${\hspace{3.mm}}$ 7.2 Strongly nonlocal (strain type) model\hfill 20\\
${\hspace{3.mm}}$ 7.3 Coupled problems of composites \hfill 21\\
${\hspace{3.mm}}$ 7.4 First Strain Gradient Medium\hfill 21\\
${\hspace{3.mm}}$ 7.5 Stress-Gradient Elasticity Model\hfill 22

\hangindent=0.4cm\hangafter=12\noindent
8. Representative volume elements (EVEs)\hfill 23\\
${\hspace{3.mm}}$ 8.1 RVE for CMs subjected to remote homogeneous\\
${\hspace{8.mm}}$ loading \hfill 23\\
%${\hspace{3.mm}}$
${\hspace{3.mm}}$ 8.2 RVE for CMs subjected to inhomogeneous\\
${\hspace{8.mm}}$ loading.\hfill 24\\
${\hspace{3.mm}}$ 8.3 RVE for CMs subjected to body force with \\
${\hspace{8.mm}}$ compact support\hfill 24\\
${\hspace{3.mm}}$ 8.4 RVE for deterministic structure CMs \hfill 26\\
${\hspace{3.mm}}$ 8.5 Classification and schematic representation \\
${\hspace{8.mm}}$ of CAM\hfill 28\\
${\hspace{3.mm}}$ {\color{black}8.6 Fast Fourier transform methods in micromechanics} \\
${\hspace{8.mm}}$ {\color{black}of periodic and deterministic structure CMs\hfill 30}\

\noindent 9. Estimation of a set of surrogate operators \hfill 31

\hangindent=0.4cm\hangafter=4\noindent
\noindent 10. Conclusion\hfill 34\\
\noindent References\hfill 37}

{\color{black} \baselineskip=3pt
List of Abbreviations:\\
\noindent AGIE – Additive General Integral Equation \\
Amic - Analytical Micromechanics\\
BFCS – Body Force with Compact Support\\
CAM - Computational Analytical Micromechanics\\
CCM - Classical Continuum Mechanics \\
GIE – General Integral Equation \\
CM - Composite Materials\\
Cmic - Computational Micromechanics\\
DNS - Direct Numerical Simulations\\
EFH – Effective Field Hypothesis \\
EFM – Effective Field Method \\
FFT - Fast Fourier Transform \\
KUBC - Kinematic Uniform Boundary Conditions\\
LM - Local Micromechanics\\
ML – Machine Learning \\
MEFM – Multiparticle Effective Field Method \\
MTM - Mori-Tanaka Method\\
NN – Neural Network \\
PBC - Periodic boundary conditions\\
PD - Peridynamics\\
PINN - Physics-Informed Neural Networks\\
PM - Peridynamic Micromechanics\\
PNO - Peridynamic Neural Operator\\
RVE – Representative Volume Element \\
SUBC - Static Uniform Boundary Conditions\\
VPBC - Volumetric periodic boundary conditions}

%{\pzc Bl}
%{\brushscript This is Brush Script X style.}
%{\calligra This is Calligra handwriting.}

\section{Introduction}
\label{intro}

Predicting the behavior of composite materials based on phase properties and microstructure is a fundamental challenge in micromechanics. Stress field estimations rely on solving a basic problem: an inclusion within an infinite homogeneous matrix under a uniform effective field. When constituents exhibit local elasticity, the solution for an ellipsoidal inclusion in a homogeneous field is given analytically by Eshelby’s tensor \cite{{Eshelby1957}} (see also \cite{{Parnell2016},{Zhouet2013}}).
For inclusions of arbitrary shapes, various numerical methods have been developed. Finite element analysis (FEA) and truncation methods approximate the infinite medium with an extended sample. Direct modeling of an infinite medium is achieved using boundary integral equation (BIE) methods \cite{{Liuet2011},{Buryachenko2022a}} or volume integral equation (VIE) approaches \cite{Buryachenko2007}. Comprehensive reviews of micromechanical methods can be found in
\cite{{Buryachenko2007},{Buryachenko2022a},{Dvorak2013},{KachanovS2018},{Mura1987},{Torquato2002}}.

Conventional micromechanics assumes materials can be treated as a continuum at arbitrarily small scales, with properties and behaviors that are invariant to time and length scales, excluding size effects \cite{Huter2017}. However, real material models need to define an ``effective" set of properties, approximating microstructural details below a certain resolution \cite{Ba\v{z}antJ2002}. In cases where scales (material and field) are comparable, nonlocal elasticity theory, developed by authors like CM, incorporates lattice theory with classical elasticity to address these intermediate effects
(see \cite{{Eringen1966},{Eringen2002},{Krumhansl1965},{Kroner1967},{Kunin1966},{Kunin1967},{Kunin1983}},
as well as \cite{{Aifantis2003},{Ba\v{z}antJ2002},{Ba\v{z}antet2016},{Capriz1989},{Eringen1999},{Lazaret2020},{Maugin1998},{Voyiadjis2019}} see also comprehensive rewievs \cite{{Shaatet2020},{Tran2016},{Trovalusci2014}}).
``Generalized continuum mechanics," as defined by \cite{Maugin2017}, covers models beyond the Cauchy framework.
Generalized continua can be broadly divided into weakly and strongly nonlocal theories. {\it Weakly nonlocal theories} involve higher-order models where material points have additional degrees of freedom beyond classical translation, such as rotation \cite{CosseratC1909}, stretch \cite{Eringen1999}, or higher-order displacement derivatives \cite{{GreenR1964a},{GreenR1964b},{Mindlin1964},{MindlinE1968}}, which were essentially developed later (see, e.g., the comprehensive reviews \cite{{Aifantis2003},{BazantJ2002},{Polizzotto2015}}). The Cosserat theory \cite{CosseratC1909} treats materials as rigid particles with independent rotations, while the micromorphic theory \cite{{Eringen1964},{Eringen1968}} models a material as particles that can move, deform, and rotate. Micromorphic theory generalizes Cosserat and micropolar theories, incorporating microstretch behavior.

{\it Strongly nonlocal elasticity} theory explicitly models long-range interactions using convolutions of kinematical variables, which represent the material's degrees of freedom. This approach contrasts with weakly nonlocal theories, where nonlocality is introduced implicitly through gradients of micro-deformations and coupling variables. In strongly nonlocal theories, stress is an integral functional of strain, with a kernel that weights material properties. Key contributions to this field came from Kr\"oner \cite{{Kroner1968},{KronerD1966}} and Kunin \cite{{Kunin1967}} in the 1960s generalized in
\cite{{Eringen2002},{Kunin1983},{Rogula1982}}. These models can be divided into
{\it strain-type} and {\it displacement-type} (see \cite{{Eringen2002},{Kroner1968},{KronerD1966},{Kunin1966},{Kunin1967},{KuninV1970},{Rogula1982}}), with the displacement-type model reducible to the strain-type model \cite{Kunin1984}.

Under certain conditions, strain gradient elasticity can approximate integral non-local theory. Nonlocal microcontinuum mechanics \cite{{Eringen1999},{Eringen2002}} was developed as a more general form of nonlocal mechanics. Eringen \cite{Eringen2002} presented a unified approach to nonlocal field theories for various media: elastic solids, viscous
fluids, electromagnetic solids, and fluids, memory-dependent elastic
solids, and media with microstructure. {\it Peridynamics (PD)}, introduced by Silling \cite{Silling2000} and discussed in \cite{Maugin2017} and \cite{SillingL2010}, is a special displacement-type nonlocal theory that eliminates displacement derivatives.
Peridynamics by Silling \cite{Silling2000} (see also \cite{{Bobaruet2016},{Dorduncuet2024},{Javiliet2019},{MadenciO2014},{OterkusO2024},{Silling2010},{SillingL2008},{SillingL2010}}), revolutionized solid mechanics by replacing local partial differential equations with integral equations that avoid spatial derivatives of displacement. In this approach, the equilibrium at a material point is determined by the sum of internal forces from surrounding points within a defined "horizon." Unlike classical mechanics, which relies on local interactions, peridynamics uses a {\it state-based model} (SB) where the deformation at a point is influenced by all bonds within the horizon. This method naturally handles discontinuities like cracks, making it ideal for studying failure and damage propagation.
The {\it bond-based} (BB) peridynamic approach models interactions between pairs of points within a horizon but limits the Poisson's ratio in isotropic materials to specific values of $\nu = 1/4$ (3D and 2D plane strain) or $\nu = 1/3$ (2D plane stress) \cite{Silling2000}. The correspondence model
(see see \cite{{AguiarF2014},{Sillinget2007}})
allows peridynamics to replicate the classical elasticity tensor for fully anisotropic materials. Non-ordinary state-based models offer more flexibility but can be unstable, requiring careful numerical methods for accurate results (see \cite{Liuet2025}, \cite{Wanget2024a} for details). In contrast, the ordinary state-based model \cite{Scabbiaet2024} overcomes these issues by accurately reproducing any component of the classical continuum mechanics (CCM) elasticity tensor in both 2D and 3D. Tensor-based peridynamics \cite{Tian2024} expands on the BB model, enhancing its adaptability for both isotropic and anisotropic materials.

Foundational principles for thermoperidynamic media under uniform volumetric boundary loading were proposed using the linearized bond-based peridynamic approach (see \cite{{Buryachenko2014a},{Buryachenko2017},{Buryachenko2023b}}). A key outcome is that the effective behavior of these media can be described by a local effective constitutive equation, similar to classical thermoelasticity
\cite{Buryachenko2007}. The relationship between effective properties and influence functions, which describe mechanical interactions within the material, highlights how one region’s behavior affects others. This is achieved by decomposing local fields into load and residual fields, allowing an energetic definition of the effective elastic moduli. There is a notable similarity between locally elastic composites and peridynamic composites, as both follow principles like Hill’s condition and the self-adjoint nature of the peridynamic operator. This allows classical solutions to be extended to nonlocal peridynamic systems, offering a contrast to simplified methods like mixture theory and scale separation hypothesis (see \cite{{Askariet2006},{Askariet2008},{Askariet2015},{Chenget2024},{Diyarogluet2016},{Franket2023},{Huet2012a},{Huet2014},{LiFet2023},{MadenciO2014},{Madenciet2021},{Madenciet2023},{Mehrmashhadiet2019},{Renet2022},{WuC2023},{Wuet2021},{Xuet2008},{Yanget2024}}) {\color{black} which do not depend on the ratio of the horizon and inclusion size. It should be mentioned  the stochastic methods where
the material property is modeled by a random variable, see for references \cite{Desai2024}, \cite{Fanet2022a}, \cite{Fanet2022b},
\cite{Fanet2024b} (see also \cite{Chenet2021}, \cite{LuX2024}, \cite{Mehrmashhadiet2019}, \cite{Songet2023}, \cite{WuC2023}, \cite{Wuet2021})). The solution of this stochastic equation describes the
probability density function (PDF) of the state variable. However, the mentioned methods do not consider the standard engineering restriction, such as the nonoverlapping of randomly located inclusions. Another limiting case associated with ``randomly" inhomogeneous structure is the modeling of a critical stretch of each bond
by the Weibull distribution function, see \cite{Songet2023} and \cite{Xuet2024} (see also \cite{Zhaoet2024}). However, this direction does not correspond to the LM counterparts of random structure CMs.
}

The expansion of Local Micromechanics (LM) from classical continuum mechanics (CCM) to nonlocal phenomena (see \cite{{Buryachenko2014a},{Buryachenko2017},{Buryachenko2022a},{Buryachenko2023k}}) is driven by the General Integral Equation (GIE, see \cite{{Buryachenko2014a},{Buryachenko2017},{Buryachenko2022a}}) for microinhomogeneous media. This method, also called computational analytical micromechanics (CAM), does not rely on Green's functions or specific constitutive laws, making it more versatile. The GIE significantly improves accuracy in local field estimations and can even lead to correcting the sign of estimations inside inclusions (see \cite{{Buryachenko2022a}})

In locally elastic composite materials (CMs) under inhomogeneous loading, effective deformations are governed by nonlocal operators (differential or integral, see \cite{{Buryachenko2007},{Buryachenko2022a},{Duet2020},{Silling2014},{Wanget2020}}), which relate the statistical average of fields at a point to those in the surrounding area. These operators improve material modeling accuracy, especially for inhomogeneous media, nonlinear or nonlocal laws, and coupled processes. The CAM is particularly effective for addressing multiscale and multiphysics problems, enabling precise simulations of small-scale effects by incorporating homogenized surrogate models. This framework provides a robust approach for analyzing random and periodic structure CMs.

The study of periodic structure composites benefits from the regularity of their microstructure, allowing for specialized homogenization techniques \cite{{Fish2014},{Ghosh2011},{ZohdiW2008}}. Asymptotic homogenization, introduced by Babuska, analyzes composites where the unit cell is much smaller than the overall material size, approximating the effect of the unit cell on the global response (e.g.,\cite{{BakhvalovP1984},{Fish2014}}). In contrast, computational homogenization explicitly resolves the material’s microscopic behavior to determine macroscopic quantities like stress and deformation, making it effective for modeling complex behaviors such as nonlocal or inelastic properties (e.g.,\cite{{Kouznetsovaet2001},{Matouset2017},{TeradaK2001}}).

Madenci and colleagues \cite{Madenciet2017} (see also \cite{{Diyarogluet2019a},{Diyarogluet2019b},{Huet2022},{Liet2022b},{Madenciet2018}}, and \cite{{WangQet2024}}) advanced computational homogenization by introducing the peridynamic unit cell model, which extends traditional techniques to the peridynamic framework using classical periodic boundary conditions (PBC). The development of volumetric periodic boundary conditions (VPBC) in \cite{{Buryachenko2018},{Buryachenko2018b},{Buryachenko2022a}} further generalized these methods for peridynamics. A key innovation in peridynamic homogenization is determining effective material moduli by averaging tractions and displacements along the unit cell boundary \cite{{Buryachenko2018},{Buryachenko2018b},{Buryachenko2022a}}, offering a simpler, more efficient alternative to methods requiring volume averages of stresses and strains inside the unit cell
\cite{{Galadimaet2023},{Galadimaet2024}, {Huet2022}}. The latter methods are more complex and less general because they require differentiating displacement fields, which may lack smoothness.

The representative volume element (RVE) is crucial for accurately predicting the macroscopic properties of heterogeneous materials. The correct RVE size ensures proper representation of microstructural heterogeneity while avoiding boundary or size effects. Hill's classical definition \cite{Hill1963} requires macroscopically homogeneous boundary conditions and defines the effective behavior with a tensor of moduli. RVE size determination involves balancing the capture of microstructural heterogeneities with scale separation, achieved when $a\ll \Lambda\ll L$ (see Subsection 2.3 for details). The smallest size at which properties stabilize is the appropriate RVE. The concept of statistically equivalent representative volume element (SERVE) uses image-based modeling to generate accurate computational domains, with further details found in \cite{{Bargmannet2018},{Kanitet2003},{Matouset2017},{Moumenet2021},{Ostojaet2016}}, and others.
When the scale separation hypothesis is violated, statistically homogeneous fields no longer apply, leading to nonlocal coupling between stress and strain averages, mediated by a tensorial kernel. This requires using an effective elastic operator in integral form. Nonlocal operators include strongly nonlocal methods (strain type and displacement type, peridynamics) and weakly nonlocal models like strain- or stress-gradient methods. Micromechanics helps bridge coupled scales governed by nonlocal constitutive equations. Instead of classical effective moduli \cite{Hill1963}, effective nonlocal operators, often in integral or differential form, are required, prompting a redefinition of the representative volume element (RVE). This applies to both random (\cite{{Drugan2000},{Drugan2003},{DruganW1996}}) and periodic (\cite{{Ameenet2018},{Kouznetsovaet2004a},{Kouznetsovaet2004b},{SmyshlyaevC2000}}) composites. The RVE concept is crucial for analyzing nonlocal effects from inhomogeneous fields, material nonlocality, and inclusion interactions.

Machine learning (ML) and neural network (NN) techniques have significantly advanced nonlocal operator theory, improving generality and flexibility. Early work by Silling \cite{{Silling2021},{Youet2020}} (later expanded upon in \cite{{Youet2024}}) used Direct Numerical Simulations (DNS) for the construction of surrogate integral operators. Recent developments include nonlocal neural operators that map function spaces \cite{{Lanthaleret2024},{Liet2003}}, and various NN architectures like DeepONet, PCA-Net, and Fourier Neural Operators (FNO) \cite{{Gosmaniet2022},{HuZet2024}}. Peridynamic Neural Operator (PNO) was proposed in \cite{Jafarzadehet2024} and subsequently extended to the
Heterogeneous Peridynamic Neural Operator (HeteroPNO) in \cite{Jafarzadehet2024b}.
Additionally, Physics-Informed Neural Networks (PINNs) embed physical laws directly into the neural network as constraints
(see \cite{{Cuomoet2022},{Haghighataet2021},{Harandiet2024},{Huet2024},{Karniadakiset2021},{KimL2024},{Raissiet2019},{RenL2024}}).
Combining neural operators with PINNs enables modeling complex systems with nonlinearities, heterogeneity, and nonlocality \cite{{Faroughiet2024},{Gosmaniet2022},{WangY2024}}.

However, despite the power of ML and NN techniques, they often neglect crucial micromechanical factors like size scale, boundary layer, edge effects, and RVE, important for both linear and nonlinear materials. To overcome this, the proposed CAM generates new compressed effective datasets for complex material structures (whether random, periodic, or deterministic), using a novel RVE concept that doesn't depend on constitutive laws or surrogate operator forms. Instead, it focuses on field concentration factors within material phases. These effective datasets, incorporating the new RVE concept, must be used with any ML or NN methods predicting nonlocal surrogate operators, ensuring accurate predictions by eliminating size scale, boundary, and edge effect issues.

The structure of the paper is as follows: Section 2 offers a concise overview of peridynamic theory, tailored to support the subsequent analysis. It includes a discussion on the statistical representation of composite microstructures and describes the volumetric homogeneous displacement loading conditions, alongside certain field averages. Section 3 introduces the decomposition of both material and field parameters. General Integral Equations (GIEs) and Additive GIEs (AGIE) are considered, incorporating either statistical average fields or fields induced in an infinite matrix by a body force with compact support (BFCS), respectively.
Iterative solutions of AGIE are presented for either random or periodic structure CMs in Sections 4 and 5, respectively. Estimation of field fluctuations and effective energy-based criteria are presented in Section 6. AGIEs and effective datasets are considered in Section 7 for CMs with locally elastic, weakly nonlocal (strain gradient and stress gradient), and strongly nonlocal (strain-type) properties.
Classical RVE concepts generalized to nonlocally elastic media are considered in Section 8. A novel RVE concept is introduced specifically for cases involving a BFCS loading. By applying translation averaging to the GIE solutions for periodic structure CMs, statistical averages of fields are derived. Section 9 leverages the body force with compact support as a training parameter to develop a suite of surrogate effective operators of any
unpredefined form for CMs with either random, periodic, or deterministic structures and a wide class of phase properties.

{\color{black}
We also outline the informal structure of the paper to give the reader a clear idea of what to expect, how each section connects to the preceding and following ones, objectives, and the recommended order of reading. After the Preliminaries (Section 2)—which are tailored to support subsequent developments—the first of the two principal achievements is presented: the formulation of a new {\it AGIE} in Section 3, followed by its solutions for both random (Section 4) and periodic (Section 5) structures.
Sections 6 and 7 discuss extensions and applications of AGIE to related problems, including field fluctuation estimations and plural constitutive laws; these sections may be skipped on a first reading.
The second major achievement is the introduction of a new {\it RVE} concept, emerging from the AGIE framework (involving BFCS loading), which is developed in Subsections 8.3 and 8.4 and employed for constructing novel {\it effective datasets}.
The presentations in Subsections 8.3 and 8.4 are constructed in contrast to the traditional RVE concepts, whose adapted versions are reviewed in Subsections 8.1 and 8.2. Therefore, reading Subsections 8.1 and 8.2 beforehand is recommended.
A summary of AGIE applications in both analytical and computational micromechanics is provided in Subsection 8.5. The discussion of FFT-based solution methods for AGIE in Subsection 8.6 may be skipped on a first reading.
Finally, Section 9 transitions to a distinct but essential topic, as only {\it ML\&NN} techniques can effectively process the newly introduced effective datasets for {\it surrogate operator modeling}. The purpose of this section is to illustrate how these effective datasets can be integrated into existing ML and NN frameworks, rather than to advance the development of those techniques themselves.

In summary, we emphasize that the fundamental BFCS loading serves as the true fountainhead — the ignition spark — for all the newly proposed concepts, methods, and notions, including AGIE, RVE, effective datasets, and their implementation within ML\&NN techniques.
}

%%%%%%%%%%%%%%%%
\section{Preliminaries. }
%\vspace{-2.mm}
\subsection{Basic equations of peridynamics}
%\vspace{-2.mm}

\setcounter{equation}{0}
\renewcommand{\theequation}{2.\arabic{equation}}

One considers a linear elastic medium occupying an open simply connected bounded domain $w\subset \mathbb{R}^d$
with a smooth boundary $\Gamma_0$ and with an indicator function $W$ and space dimensionality
$d$ ($d=1,2,3$).
The domain $w$ {\color{black} with the boundary $\Gamma^0$} consists from a homogeneous matrix $v^{(0)}$ and a statistically homogeneous
{\color{black} field} $X=(v_i)$ of heterogeneity $v_i$ with indicator functions, $V_i$ and bounded by the closed
smooth surfaces $\Gamma_i$ $(i=1,2,\ldots)$.
It is presumed that the heterogeneities can be grouped into phases $v^{(q)} \quad (q=1,2,\ldots,N)$ with identical mechanical and geometrical properties.
The basic equations of local thermoelasticity of composites are considered
\BBEQ
\label{2.1}
\nabla\cdot \bfsi(\bfx)&=&-\bfb(\bfx), \\ %(2.1)
\label{2.2}
\bfsi(\bfx)&=&\bfL(\bfx)\bfep(\bfx)+\bfal(\bfx), \ \ {\rm or}\nonumber\\
\bfep(\bfx)&=&\bfM(\bfx)\bfsi(\bfx)+\bfbe(\bfx), \\ %(2.2)
\label{2.3}
\bfep(\bfx)&=&[\nabla {\otimes}{\bf u}+(\nabla{\otimes}{\bf u})^{\top}]/2, \nonumber\\
%+(\nabla {\bf u})^{\top}]/2,\ \
\nabla\times\bfep(\bfx)\times\nabla&=&{\bf 0}, %(2.3)
\EEEQ
where $\otimes$ and
and $\times$ are the tensor and vector products, respectively, and $(.)^{\top}$ denotes a matrix transposition.
It is presumed that the body force density (forcing term) $\bfb(\bfx)$ is self-equilibrated
and vanished outside some loading region $\bfcB^b$ {\color{black} (called body force with compact support, BFCS)}:
\BB
\label{2.4}
\int\bfb(\bfx)={\bf 0},\ \ \ \ \bfy\not\in b({\bf 0}, B^b):=\{\bfy|\ |\bfy|\leq B^b\},
%(2.4)
\EE
where $b(\bfx_i,B^{b})$ is the ball of radius $B^{b}$
centered at $\bfx_i={\bf 0}$.
${\bf L(x)}$ and ${\bf M(x) \equiv L(x)}^{-1}$ are the known phase
stiffness and compliance tensors.
$\bfbe(\bfx)$ and $\bfal(\bfx)=-\bfL(\bfx)\bfbe(\bfx)$ are second-order tensors of local eigenstrains and eigenstresses.
In particular, for isotropic
phases, the local stiffness tensor $\bfL(\bfx)$ is presented in
terms of the local bulk $k(\bfx)$ and shear $\mu(\bfx)$
moduli and:
\BB
\nonumber
\bfL(\bfx)=(dk,2\mu)\equiv dk(\bfx)\bfN_1+2\mu(\bfx)\bfN_2, \ \ \bfbe(\bfx)=\beta^{t}\theta\bfdel,
\EE
${\bf N}_1=\bfdelta\otimes\bfdelta/d, \ {\bf N}_2={\bf I-N}_1$ $(d=2\ {\rm or}\ 3$) whereas
$\bfdelta$ and $\bfI$ are the unit second-order and fourth-order tensors; $\theta=T-T_0$ denotes the temperature changes with respect to a reference temperature $T_0$ and $\beta^{t}$ is a thermal expansion.
For all material tensors $\bfg$ (e.g., $\bfL, \bfM,\bfbe,\bfal)$ the notation $\bfg_1(\bfx)\equiv \bfg(\bfx)-\bfg^{(0)}=\bfg^{(m)}_1(\bfx)$ $(\bfx\in v^{(m)},\ m=0,1$) is exploited.
The upper index $^{(m)}$ indicates the
components, and the lower index $i$ shows the individual
heterogeneities; $v^{(0)}=w\backslash v$, $ v\equiv \cup v_i,
\ V(\bfx)=\sum V_i(\bfx)$, and $V_i(\bfx)$ are the
indicator functions of $v_i$, equals 1 at
$\bfx\in v_i$ and 0 otherwise, $(i=1,2,\ldots)$.
Substitution of Eqs. (\ref{2.2}) and (\ref{2.4}) into Eq. (\ref{2.1}) leads to a representation of the equilibrium equation (\ref{2.1}) in the form
\BB
\label{2.5}
^L\widetilde{\bfcL}(\bfu)(\bfx)+\bfb(\bfx)={\bf 0},\ \ \ ^L\widetilde{\bfcL}(\bfu)(\bfx):=\nabla[\bfL\nabla\bfu(\bfx)+\bfal(\bfx)],
\EE
where $^L\widetilde{\bfcL}(\bfu)(\bfx)$ is an elliptic differential operator of the second order.

In this section, a summary of the linear peridynamic model introduced
by Silling \cite{Silling2000} (see also %[10, 40, 56]). %, 35, 36, 37]) is presented.
\cite{{LehoucqS2008},{MadenciO2014},{SillingA2005},{SillingL2008},{SillingL2010}}) is presented.
%Bobaru and Ha, 2011; Javili {\it et al.}, 2019; Silling and Askari, 2005; Lehoucq and Silling, 2008).
An equilibrium equation is free of any derivatives of displacement (contrary to Eq. (\ref{2.5})) and presented in the following form
\BB
\label{2.6}
%{{\cal H}_x} C(x,\hat x)
\widetilde{\bfcL}(\bfu)(\bfx)+\bfb(\bfx)={\bf 0}, \ \widetilde{\bfcL}(\bfu)(\bfx):=\!\!\!\int \!\! \hat\bff(\bfu(\hat {\bfx})-\bfu(\bfx),\hat{\bfx}-\bfx,\bfx)d\hat {\bfx} ,
\EE
where $\hat\bff$ is a {\it bond force density} whose value is the force vector
%(per unit volume squared)
that the point located at $\hat {\bfx}$ (in the reference configuration) exerts on the point located at $\bfx$ (also in the reference configuration); the third argument $\bfx$ of
$\hat\bff$ (\ref{2.6}) can be dropped for the homogeneous media.
Equations (\ref{2.5}$_1$) and (\ref{2.6}$_1$) have the same form for both local and peridynamic formulation with
the different operators (\ref{2.5}$_2$) and (\ref{2.6}$_2$). Because of this, the superscripts $^L(\cdot)$
will correspond to the local case.

The relative position of two material points in the reference configuration $\bfxi=\hat{\bfx}-\bfx$ and their relative displacement $\bfeta
=\bfu(\hat {\bfx})-\bfu(\bfx)$ are connected with the relative position vector between the two points in the deformed
(or current) configuration $\bfeta+\bfxi$.
%All deformations are
%assumed small and the reference and deformed configurations are taken to be the same.
Only points $\hat{\bfx}$ inside some neighborhood (horizon region) ${\cal H}_{\bf x}$ of $\bfx$ interact with $\bfx$:
\BB
\label{2.7}
\hat\bff(\bfeta,\bfxi,\bfx)\equiv {\bf 0}\ \ \ \forall \hat{\bfx}\not \in {\cal H}_{\bf x}.
\EE
The vector $\bfxi=\hat{\bfx}-\bfx$ ($\hat{\bfx}\in {\cal H}_{\bf x}$) is called a {\it bond} to $\bfx$, and the collection of all bonds to $\bfx$ form the horizon region ${\cal H}_{\bf x}$.
Without a loss of generality, it is assumed that a shape of ${\cal H}_{\bf x}$ is spherical: ${\cal H}_{\bf x}=\{\hat{\bfx}:\ |\hat{\bfx}-\bfx|\leq l_{\delta}\}$ and a number $l_{\delta}$, called the {\it horizon}, does not depend on $\bfx$. The properties of
$\hat\bff(\bfeta,\bfxi,\bfx)$ are concidered in \cite{Silling2000}.

Peridynamic states introduced by Silling \cite{Sillinget2007} %{\it et al.} (2007)
(for a more detailed summary, see \cite{SillingL2010})) %Silling and Lehoucq, 2010)
are the functions acting on bounds.
There are scalar, vector, and modulus states producing the scalars, vectors, and
2nd order tensors, respectively. $\underline{\bfT}[\bfx]\lle\bfxi\rle$ and $\underline{\bfT}[\hat \bfx]\lle-\bfxi\rle$ are the {\it force vector states} at $\bfx$ and
$\hat\bfx$, which return the force densities associated with $\bfxi$ and $-\bfxi$, respectively.
In the ordinary state-based peridynamics being considered, $\underline{\bfT}[\bfx]\lle\bfxi\rle$ is parallel (in contrast with the non-ordinary model) to the deformation vector state, and the equilibrium Eq. (\ref{2.6}) is expressed through the
force vector states as (see for details \cite{{Sillinget2007},{SillingL2010}})
\BB
\label{2.8}
\widetilde{\bfcL}(\bfu)(\bfx)+\bfb(\bfx)={\bf 0}, \ \widetilde{\bfcL}(\bfu)(\bfx):=\!\!\int\!\!
\{\underline{\bfT}[\bfx]\lle\bfxi\rle-\underline{\bfT}[\hat\bfx]\lle-\bfxi\rle
%\underline{\bfT}[\bfx]\lle\bfxi\rle-\underline{\bfT}[\hat\bfx]\lle-\bfxi\rle
\}
d\hat {\bfx}.
\EE

A small displacement state $\bfu$ when (see \cite{Silling2010})
\BB
\label{2.9}
\text{\hw{l}}:=\sup_{|\bf \xi|\leq {\it l}_{\delta}} |\bfeta(\bfxi)\rle|\ll l_{\delta}.
\EE
is considered.
Force vector state
\BBEQ
\label{2.10}
\underline{\bfT}&=&\underline{\bfT}^0+
\underline{\bfbK}\bullet\underline{\bfU}.
\EEEQ
is expressed through the {\it modulus state} $\bfbK$.
Here, the operation of {\it dot product} $\bullet$ of two vector states $\underline{\bfA}$, $\underline{\bfB}$ and a double state
$\underline{\bfbD}$ are introduced as
\BBEQ
\label{2.11}
\!\!\!\!\!\!\!\!\!\!\!\!\!\underline{\bfA}\bullet \underline{\bfB}\!\!&=&\!\!\big\langle\underline{\bfA}\lle\bfxi\rle\cdot
\underline{\bfB}\lle\bfxi\rle \big\rangle ^{\cal H_{\bf x}}= \int_{\cal H_{\bf x}} \underline{\bfA}\lle\bfxi\rle\cdot
\underline{\bfB}\lle\bfxi\rle ~d\bfxi,\nonumber\\
\!\!\!\!\!\!\!\!\!\!\!\!\!(\underline{\bfbD}\bullet \underline{\bfB})_i\lle\bfxi\rle\!\!&=&\!\!\int_{\cal H_{\bf x}} \underline{\bfbD}_{ij}\lle\bfxi,\bfze\rle\cdot
\underline{\bfB}_j\lle\bfze\rle ~d\bfze.
\EEEQ
Hereafter $\lle(\cdot)\rle^{\cal H_{\bf x}}(\bfx)$ (\ref{2.7}) denotes the average over the horizon region ${\cal H_{\bf x}}$ with the center $\bfx$.

A linearized model for pure mechanical loading ($\bfbe\equiv{\bf 0}$) can be written from (\ref{2.8}$_2$) as described by Silling \cite{Silling2010}
\BBEQ
\label{2.12}
\!\!\!\!\!\!\!\!\!\!\!\!\!\!\widetilde\bfcL(\bfC, \bfu)(\bfx)+\bfb(\bfx)\!\!\!\!&=&\!\!{\bf 0}, \\
\label{2.13}
\!\!\!\!\!\!\!\!\!\!\!\!\!\!\!\!\!\!\!\!\!\!\!\!\!\!\!\!\!\!\!\widetilde\bfcL(\bfC,\bfu)(\bfx):\!\!\!\!&=&\!\!\int_w\!\!\bfC^{}(\bfx,\bfq)(\bfu(\bfq)-\bfu(\bfx))~dV_q,
\EEEQ
where the integrand in Eq. (\ref{2.10}) vanishing at
$|\bfx-\bfq|\geq 2l_{\delta}$ may be non-null at $l_{\delta}<|\bfx-\bfq|< 2l_{\delta}$.
The kernel with the following symmetries holds for any $\bfx$ and $\bfq$:
\BBEQ
\label{2.14}
\bfC^{\top}(\bfx,\bfq)=\bfC^{}(\bfq,\bfx).
\EEEQ
Thermoelastic cases ($\bfbe\not \equiv{\bf 0}$) were considered in
\cite{{MadenciO2014},{Beckmannet2013},{KilicM2010},{MadenciO2016}}.
%Beckmann {\it et al.}, 2013; Kilic and Madenci, 2010; Madenci and Oterkus, 2014)
Fully coupled thermo-mechanical PD theory was proposed in \cite{{Oterkuset2014},{YangCet2024}}.

For subsequent convenience, one introduces a vector-valued function ${\bff}: \mathbb{R}^d\times \mathbb{R}^d\to \mathbb{R}^d$ by (see \cite{LehoucqS2008})
\BBEQ
\label{2.15}
\!\!\!\!\!\!\!\!\!\!{\bff}(\bfp,\bfq)=\begin{cases}
\hat\bff(\bfu(\bfp)-\bfu(\bfq),\bfp-\bfq,\bfq), & {\rm if} \ \bfp,\bfq\in w,\\
{\bf 0}, & {\rm otherwise},
\end{cases}
\EEEQ
which is presumed to be Riemann-integrable.
Then, one can
define the ``peridynamic stress" $\bfsi(\bfz)$ at the point $\bfz$
as the total force that all material points $\hat {\bfx}$ to the right of $\bfz$
exert on all material points to its left
(see e.g.,\cite{{Buryachenko2022a},{LehoucqS2008},{Sillinget2003},{WecknerA2005}}).
%[40, 60, 64]). Lehoucq and Silling, 2008; Silling {\it et al.}, 2003; Weckner and Abeyaratne, 2005).
For $dD$ ($d=1,2,3$) cases (see (\ref{2.15}))
\BBEQ
\label{2.16}
\!\!\!\!\!\!\!\!\!\!\!\!\!\!\!\bfsi(\bfx)&=&\bfcL^{\sigma}(\bfu),\\
\label{2.17}
\!\!\!\!\!\!\!\!\!\!\!\!\!\!\!\!\!\!\!\!\!\bfcL^{\sigma}(\bfu)&:=&\frac {1}{2} \int_S\int_0^{\infty}\int_0^{\infty} (y+z)^{d-1}{\bff}(\bfx+y\bfm,\bfx-z\bfm)\nonumber\\
&\otimes&\bfm dzdyd\Omega_{\bf m}.
\EEEQ
Here, $S$ stands for the unit sphere, and $d\Omega_{\bf m}$ denotes a differential solid angle on $S$ in the direction of any unit vector $\bfm$. It was proved \cite{Liet2022a} that the peridynamic stress is the same as the
first Piola-Kirchhoff static
Virial stress, which offers a simple and clear expression for numerical calculations of
peridynamic stress.

The equilibrium Eqs. (\ref{2.8}), (\ref{2.21}) and (\ref{2.16}) of ordinary state-based PD are considered.
When the interactions between material points are only pairwise, the equilibrium equations are reduced to the bond-based PD equations. One of the simplest nonlinear is
the proportional microelastic material model \cite{SillingA2005}
%Beckmann {\it et al.}, 2013; Kilic and Madenci, 2010; Madenci and Oterkus, 2014)
\BBEQ
\label{2.18}
\!\!\!\!\!\!\!\!\!\!\!\hat\bff^{\rm bond}(\bfeta,\bfxi,\bfx)\!\!&=&\!\!f(|\bfeta+\bfxi|,\bfxi)\bfe,\nonumber\\
\!\!\!\!\!\!\!\!\!\!\! f(|\bfeta+\bfxi|,\bfxi)\!\!&=&\!\!c(\bfxi)s,
\\
\label{2.19}
\!\!\!\!\!\!\!\!\!\!\!\!\!\!\bfe\!\!&:=&\!\!
\frac {\bfeta+\bfxi}{|\bfeta+\bfxi|},\ \ \ s:=\frac {|\bfeta+\bfxi|-|\bfxi|}{|\bfxi|},
\EEEQ
where $s$ denotes the bond stretch (also called bond strain) which is the relative change of the length of a bond, and $c$ is referenced as the ``bond constant". % (see Silling and Askari (2005)).
Although the model (\ref{2.18}) and (\ref{2.19}) is linear in terms of the bond stretches, it is nonlinear in terms of displacements;
thermoelastic cases ($\bfbe\not \equiv{\bf 0}$) were considered in
\cite{{Beckmannet2013},{KilicM2010},{MadenciO2014},{MadenciO2016}}. A nonlinear model in terms of
bond stretches %Jafarzadeh {\it et al.} (2022)
(at $\bfbe\equiv {\bf 0}$)
\BB
\label{2.20}
f(|\bfeta+\bfxi|,\bfxi)=c(\bfxi)[s+3s^2/2+s^3/2]
\EE
is considered in \cite{Jafarzadehet2022}. The potential role of employing a nonlinear peridynamic kernel
in predicting the onset of fractures has been explored in \cite{Dimolaet2022} (see also \cite{{Cocliteet2022a},{Cocliteet2022b}}).

A linearized version of the theory (for small displacement (\ref{2.9})) for a microelastic homogeneous
material (\ref{2.21}) takes the form ($\forall \bfeta,\bfxi)$
\BB
\label{2.21}
\hat\bff^{\rm bond}(\bfeta,\bfxi,\bfx)=\hat\bff_{\rm lin}^{\rm bond}(\bfeta,\bfxi,\bfx)=\bfC^{\rm bond}(\bfxi,\bfx)\bfet,
%\ \ \ \bfet^t(\hat{\bfx},\bfx):=\bfet(\hat{\bfx},\bfx)-\widetilde{\bfbe}^x(\hat{\bfx},\bfx).
\EE
Here, the material's {\it micromodulus} function $\bfC$
contains all constitutive information and its
value is a second-order tensor given by
\BB
\label{2.22}
\bfC^{\rm bond}(\bfxi,\bfx)=\frac {\partial \hat\bff({\bf 0},\bfxi,\bfx)}{\partial \bfeta }\ \ \ \forall \bfxi.
\EE
Substitution of Eq. (\ref{2.21}) into Eq. (\ref{2.6}) leads to the equilibrium equation
\BBEQ
\label{2.23}
\!\!\!\!\!\!\!\!\!\!\!\!\!\widetilde{\bfcL}(\bfC^{\rm bond},\bfu)(\bfx)\!\!\!\!&+&\!\!\!\!\bfb(\bfx)={\bf 0}, \\
\label{2.24}
\!\!\!\!\!\!\!\!\!\!\!\!\!\!\widetilde{\bfcL}(\bfC^{\rm bond},\bfu)(\bfx)\!\!\!\!&:=&\!\!\!\!\!\int\!\!\bfC^{\rm bond}(\bfx,\bfq)(\bfu(\bfq)-
\bfu(\bfx))dV_q.
\EEEQ

For consistency with Newton's third law,
the micromodulus function $\bfC^{\rm bond}$ (\ref{2.21}) and (\ref{2.22}) for the homogeneous materials must be symmetric
to its tensor structure as well as to arguments
\BB
\label{2.25}
\bfC^{\rm bond}(\bfx,\bfq)=\bfC^{\rm bond}(\bfq,\bfx)=(\bfC^{\rm bond})^{\top}(\bfx,\bfq) \ \ \ \forall \bfx,\bfq,
\EE
where the properties of $\bfC^{\rm bond}$ are discussed in detail in Silling \cite{Silling2000}
(see also\cite{{Huet2012a},{EmmrichW2007b}}). %Bobaru and Ha, 2011; Hu {\it et al.}, 2012a; Emmrich and Weckner, 2007a, 2007b).
For example, for the micromodulus functions with the step-function and triangular profiles,
\BB
\label{2.26}
\bfC^{\rm bond}(\bfxi)= \bfC^{\rm bond} V({\cal H}_{\bf x}), \ \
\bfC^{\rm bond}(\bfxi)=\bfC^{\rm bond}(1-|\bfxi|/l_{\delta})V({\cal H}_{\bf x}),
\EE
respectively, where $V({\cal H}_{\bf x})$ is the indicator function of ${\cal H}_{\bf x}$.
The peridynamic solution
of Eq. (\ref{2.23}) with $\bfC^{\rm bond}$ described by Eq. (\ref{2.26}) is investigated in detail by both
numerical and analytical methods in 1D (see \cite{{Bobaruet2009},{Mikata2012},{Sillinget2003},{WecknerA2005}}),
2D (see \cite{{Huet2012a},{Huet2012b}}), and 3D cases \cite{{Mikata2023},{Weckneret2009}}.

For bond-based peridynamics, the stress representation (\ref{2.17}) can be recast in terms of displacements
\BBEQ
\label{2.27}
\!\!\!\!\!\!\!\!\!\!\!\!\!\!\bfcL^{\sigma}\!\!\!\!\!\!&&\!\!\!\!\!\!(\bfC,\bfu):=\frac {1}{2} \int_S\int_0^{\infty}\!\!\int_0^{\infty}\!\! (y+z)^{d-1}
\bfC^{\rm bond}((y+z)\bfm,
\nonumber\\
\!\!\!\!\!\!\!\!\!\!\!\!\!\!\!\!\!\!\!\!\!\!\!\!\!\!\!\!\!\!&&\!\!\!\!\!\!\!\!\!\!\bfx-z\bfm)[\bfu(\bfx+y\bfm)-\bfu(\bfx-z\bfm)%\nonumber\\
]\otimes\bfm dzdyd\Omega_{\bf m}.
\EEEQ

It is interesting that equilibrium Eqs. (\ref{2.12}) and (\ref{2.23}) formally coincide, although the kernels
$\bfC(\bfx,\bfq)$ (\ref{2.13}) and $\bfC^{\rm bond}(\bfx,\bfq)$ (\ref{2.24}) are conceptually different with different symmetry properties (\ref{2.14}) and (\ref{2.25}), respectively. Moreover, in the state-based version, the maximum interaction distance between the points is $2l_{\delta}$, whereas this distance in bond-based peridynamics coincides with the horizon $l_{\delta}$.
However, this similarity provides a possibility to reformulate the results obtained before for the linear bond-based peridynamic micromechanics (see for details \cite{{Buryachenko2022a},{Buryachenko2023k}}) to their counterparts for the linear ordinary state-based ones.

\subsection{ Volumetric boundary conditions}
Owing to nonlocality, the equilibrium equation (\ref{2.6})
is combined with a ``boundary" condition, used as a volumetric constraint in the so-called interaction domain $w_{\Gamma}$ (in opposite to the local case where the boundary conditions are imposed directly at the bounding surface $\Gamma^{(0)}$, see for details \cite{{Silling2000},{Kilic2008}}; %Silling, 2000; Kilic, 2008);
i.e., the nonlocal boundary $w_{\Gamma}$ is a $d$-dimensional region
unlike its $(d-1)$-dimensional counterpart $\Gamma^0$ in local problems.
The interaction domain $w_{\Gamma}$
contains points $\bfy$ not in $w$ interacting with points $\bfx\in w$.
The most popular shape for $w_{\Gamma}$ with prescribed either the forces or displacements is a boundary layer
of thickness given by $2l_c$
(see \cite{MacekS2007}):
$w_{\Gamma}=\{w\oplus{\cal H}_{\bf 0}\}\backslash w$, where $w\oplus{\cal H}_{\bf 0}$ is the Minkowski sum $w$ (${\cal A}\oplus{\cal B}:=\cup_{\bfx\in{\cal A},\bfy\in {\cal B}}\{\bfx+\bfy\}$); then $w$ is the internal region of
$\overline{\overline{w}}$ (see \cite{Silling2000}).

It is presumed that $\overline{\overline{w}}$ is considered as a cutting out of a macrodomain (containing a statistically large number of inclusions' realization) from the random
heterogeneous medium covering the entire space $\mathbb{R}^d$.
Then, the Dirichlet, or Neumann volumetric boundary conditions (VBC, see \cite{{Duet2013},{MengeshaD2014}})
%Du {\it et al.}, 2013)
are called homogeneous volumetric loading conditions if there exist some symmetric constant tensors, either
$\bfep^{w_{\Gamma}}$ or $\bfsi^{w_{\Gamma}}$ such that
\BBEQ
\label{2.28}
\bfu(\bfx)&=&\bfh(\bfy)= \bfep^{w_{\Gamma}}\bfy, \ \forall\bfy\in w_{\Gamma u}=w_{\Gamma},\\
\label{2.29}
\widetilde{\bfcL}(\bfx)&=&-\bfg(\bfy)=-\bfsi^{w_{\Gamma}}\bfn(\bfy), \ \forall\bfy\in w_{\Gamma \sigma}=w_{\Gamma},
\EEEQ
respectively.
There are no specific restrictions on the smoothness and shape of $\Gamma_0$, which is defined only by the convenience of representation.
It should be mentioned that in the LM, the analogs of the VBC (\ref{2.28}) and (\ref{2.29}) (at the nonlocality vanishing $l_{\delta}\to 0$) are the homogeneous boundary conditions (also called the
kinematic uniform boundary conditions (KUBC) and static uniform boundary
conditions (SUBC), respectively)
\BBEQ
\label{2.30}
\bfu(\bfy)&=& \bfep^{w_{\Gamma}}\bfy, \ \forall\bfy\in \Gamma_{0u}={\Gamma}_0,\\
\label{2.31}
\bft(\bfy)&=&\bfsi^{w_{\Gamma}}\bfn(\bfy), \ \forall\bfy\in \Gamma_{0\sigma}={\Gamma}_0,
\EEEQ
correspond
to the analyses of the equations for either strain or stresses, respectively, which are formally similar to each other. However,
in peridynamic micromechanics, a primary unknown variable is displacement (rather than stresses), and, because of this, the VBC
(\ref{2.28}) is assumed.

Seemingly to the volumetric boundary domain $w_{\Gamma i}$, we introduce a volumetric interface boundary (called also interaction interface, see \cite{{Selesonet2013},{AllaliG2015}})
%Seleson, {\it et al.}, 2013; Alali and Gunzburger, 2015)
$v_{\Gamma i}=v_{\Gamma i}^{+}\cup v_{\Gamma i}^{-}$ where
$v_{\Gamma i}^{+}$ and $v_{\Gamma i}^{-}$ are the boundary layers (internal and external, respectively) divided by the geometric boundary $\Gamma_i$ and have a thickness expressed through the horizon as $2l_{\delta}$
{\color{black}The geometrical boundaries of the boundary layers
$v_{\Gamma i}^{+}$ and $v_{\Gamma i}^{-}$ are
$\Gamma_i^{+}$ and $\Gamma_i^{-}$, respectively.}
For a general form of the inclusion $v_i$ the external volumetric interface $\Gamma^-_i$ can be expressed through the Minkowski sum
$\Gamma^-_i=\{v_i\oplus2{\cal H}_{\bf 0}\}\backslash v_i$, where
$2{\cal H}_{\bf 0}:=\{\bfx|\ \bfx/2\in {\cal H}_0\}$.
A nonlocal closure of the inclusion $v^l_i:=v_i\oplus2{\cal H}_{\bf 0}$ (with an indicator function $V_i^l(\bfx)$) is called an {\it extended inclusion} while $v^l:=\cup v_i^l$ ($i=1,2,\ldots$) (with the indicator function $V^l(\bfx)=\sum_i V_i^l(\bfx)$) stands for the extended inclusion phase. In so doing $v^{l(0)}:=w\backslash v^l\subset v^{(0)}$ is called a {\it truncated matrix}.

{\color{black} In the simplest case, a micromodulus of interaction interface of the inclusion $v_i$ (see, e.g. \cite{{AllaliG2015},{AllaliL2012},{SillingA2005}}) is presented as an average value of the material properties at two material points connecting dissimilar materials
($V^{(0)}(\bfx)+V_i(\hat\bfx)=V^{(0)}(\hat\bfx)+V_i(\bfx)=1$)}
\BBEQ
\label{2.32}
\bfC^i(\bfx,\hat\bfx)=[\bfC^{(0)}(\bfx,\hat\bfx)+\bfC_i(\bfx,\hat\bfx)]/2
\EEEQ
{\color{black} although more sophisticated models of interaction interface properties (see, e.g., \cite{{Madenciet2017},
{Laurienet2023}}, see for references \cite{Bieet2024}), can be incorporated into
the general subsequent representations. In particular, a variation of $\bfC^i(\bfx,\hat\bfx)$ was presented in \cite{Madenciet2017}
and \cite{Qiet2024}
in a spirit of the functionally graded materials theory described in, e.g. \cite{Buryachenko2007}.}

\subsection {Descriptions of random and periodic structures of the composite microstructures}
We consider a popular group of composites, called matrix composites, which consists of a continuous matrix phase reinforced by isolated inhomogeneities of various shapes.
Three material length scales (see, e.g., \cite{{Torquato2002},{Zaoui2002}}) are considered:
the macroscopic scale $L$, characterizing the extent of $w$, the microscopic scale $a$, related with the
heterogeneities $v_i$, and the horizon $l_{\delta}$. Moreover, one supposes that the applied field varies on a characteristic length scale $\Lambda$.
The limit of our interests for both the material
scales and field one are either
\BBEQ
\label{2.33}
\!\!\!\!\!\!\!\!\!\!\!\!\!\!\!L\geq\Lambda\geq a^{\rm int}\geq a\geq l_{\delta}\ \ \ \!\!\!\!\!\!\!\! &{\rm or}&\!\!\!\!\!\!\!\!\ \ \ L\gg \Lambda\gg a^{\rm int}\geq a\gg l_{\delta},\\
\label{2.34}
\!\!\!\!\!\!\!\!\!\!\!\!\!\!\!\!\!\!\!\!L\geq \Lambda\geq|\Omega_{00}| \geq l_{\delta}\ \ \ \!\!\!\!\!\!\!\!&{\rm or}&\!\!\!\!\!\!\!\! \ \ L\gg \Lambda\gg|\Omega_{00}|\gg l_{\delta},
\EEEQ
where the inequalities (\ref{2.33}$_2$) and (\ref{2.34}$_2$) are called a scale separation hypotheses.
The inequalities (\ref{2.33}) correspond to the case of random structure CMs, where $a^{\rm int}$ stands for the scale of long-range interactions of inclusions (e.g. $a^{\rm int}=6a$). The inequalities (\ref{2.34}) describe the scale representations for periodic structure CMs, where $\Omega_{00}$ is a unite cell (see for details Subsection 2.4), and, for shortening, we use $|\Omega_{00}|$ instead of linear $|\Omega_{00}|^{1/d}$.

The random quantities of the statistically homogeneous random fields (see, e.g. \cite{{Buryachenko2022a},{MalyarenkoO2019}})
are described by a conditional
probability density $\varphi (v_i,{\bf x}_i \vert v_1,{\bf x}_1$
for finding a heterogeneity of type $i$ with the center $\bfx_i$ in the domain $v_i$, with the fixed heterogeneities $v_1$ centered at ${\bf x}_1$.
The notation $\varphi (v_i , {\bf x}_i\vert ;v_1,{\bf x}_1)$ denotes the case ${\bf x}_i\neq
{\bf x}_1$.
We have $\varphi(v_i, {\bf x}_i\vert ;v_1,{\bf x}_1)=0$ {\color{black} (since heterogeneities cannot overlap) for values of ${\bf x}_i$ placed inside the
some domain $ v^0_{1}$ called ``excluded volumes'',} where $v^0_{1}\supset v_1$
with indicator function $V^0_{1}$ is the ``excluded volumes'' of $\bfx_i$ with respect to $v_1$,
and $\varphi (v_i, {\bf x}_i\vert ;v_1,{\bf x}_1)\to \varphi(v_i, {\bf x}_i)$
as $\vert {\bf x}_i-{\bf x}_m\vert\to \infty$, $m=1,\ldots,n$ (since no long-range order is assumed).
$\varphi (v_i,{\bf x})$ is a number density, $n^{(k)}=n^{(k)}({\bf x})$ of component $v^{(k)}\ni v_i$
at the point ${\bf x}$ and $c^{(k)}=c^{(k)}({\bf x})$ is the concentration, i.e. volume fraction,
of the component $v_i\in v^{(k)}$ at the point ${\bf x}$:
$ c^{(k)}({\bf x})=\langle V^{(k)}\rangle ({\bf x})=\overline v_in^{(k)}({\bf x}),
\ \overline v_i={\rm mes} v_i\ \ (k=1,2,\ldots,N;\ i=1,2,\ldots),\quad
c^{(0)}({\bf x})=1-\langle V\rangle ({\bf x}).$
{\color{black} Additionally to the average $\langle (.)\rangle ({\bf x})$,} the notation
$\langle (.)\vert; v_1,{\bf x}_1\rangle ({\bf x})$
will be used for the conditional average taken
for the ensemble of a statistically homogeneous
{\color{black} set} $X=(v_i)$ at the point ${\bf x}$,
on the condition that there is heterogeneity at
the point ${\bf x}_1$ and
${\bf x}_i\neq\bfx_1$.
The notation $\langle (.)\vert; v_1,{\bf x}_1\rangle ({\bf x})$
is exploited for the additional condition ${\bf x}\notin v_1$. In a general case
of {\it statistically inhomogeneous} media with the homogeneous matrix (e.g., for so
called {\it Functionally Graded Materials}, FGM), the conditional probability density is not invariant with respect to
translation, that is, the microstructure functions depend on their absolute positions
\cite{QuintanillaT1997}:
\BBEQ
\!\!\!\!\!\!\!\!\!\!\!\!\!\!\varphi (v_i , {\bf x}_i\vert ;v_1,{\bf x}_1)\!\!&=&\!\!
\varphi (v_i , {\bf x}_i +\bfx)\vert ;v_1,{\bf x}_1+\bfx), \ \ \ \nonumber\\
\label{2.35}
\!\!\!\!\!\!\!\!\!\!\!\!\!\!\!\!\!\!\!\!\!\varphi (v_i , {\bf x}_i\vert ;v_1,{\bf x}_1)&\not=&
\varphi (v_i , {\bf x}_i +\bfx)\vert ;v_1,{\bf x}_1+\bfx),\ \ \
\EEEQ
for statistically homogeneous $(\forall\bfx\in R^d)$ and inhomogeneous $(\exists \bfx\in R^d)$ media, respectively.

Of course, any deterministic (e.g., periodic) field of inclusions $v$ with the centers $\bfx_{\alpha}\in\bfLa$ can be presented by
the probability density $\varphi (v_i,{\bf x}_i )$ and conditional probability density
$\varphi (v_i,{\bf x}_i \vert; v_j,{\bf x}_j)$ expressed through the $\delta$ functions ($\bfx_{\bf \alpha}\in \bfLa$)
\BBEQ
\varphi (v_i,{\bf x}_i )&=&\sum_{\bf \alpha}\delta (\bfx_i-\bfx_{\bf \alpha}), \nonumber\\
\label{2.36}
\varphi (v_i,{\bf x}_i\vert; v_j,\bfx_j)&=&\sum_{\bf \alpha}\delta (\bfx_i-\bfx_{\bf \alpha})-\delta(\bfx_i-\bfx_j).
\EEEQ

In more detail, a periodic structure of CM is considered.
For simplicity of notations for periodic media, we consider 2D cases
$w=\cup \Omega_{ij}$ ($i,j=0,\pm 1, \pm 2,\ldots$)
with the square
unit cells $\Omega_{ij}$ and the centers of the unite cells $\bfLa=\{\bfx_{ij}\}$. Let a representative unit cell $\Omega_{00}$
with the corner points $\bfx^c_{kl}$ ($k,l=\pm 1$) has the boundary
$\Gamma^0=\cup\Gamma^0_{ij}$ where the boundary partition $\bfx_{ij}^{0}\in \Gamma^0_{ij}$ separates the UCs $\Omega_{00}$ and $\Omega_{ij}$ ($i=0,\pm 1,\ j=\pm(1-|i|)$, see Fig. 19.2 in \cite{Buryachenko2022a}).
A representative volume element (or the unit cell, UC) $\Omega_{00}$ is deformed in a repetitive way, as its neighbors.
The position vectors $\bfx^0_{ij}\in\Gamma^0_{ij}$ or
$\bfx^0_{kl}\in\Gamma^0_{kl}$ are presented by the corner
points $\bfx^c_{mn}$ ($i=0,\pm 1,\ j=\pm(1-|i|),\ k=-i,\ l=-j,\ m,n=\pm 1)$; e.g.,
$\bfx^0_{ij}=\bfx^0_{kl}+\bfx^c_{1,-1}-\bfx^c_{-1,-1}$
for the $i=1$, $j=0$.
$N$ field points $\bfx_i$ ($i = 1,\ldots,N$) in the central
cube $\Omega_{00}=\{[- l^{\Omega}, l^{\Omega}]^d\}$ are periodically reflected as
$\bfx_i^{\bf \alpha}$ into the neighboring cubes $\Omega_{\bf \alpha}$, where
$\bfal = (\alpha_1, \ldots, \alpha_d)\in Z^+$, and $\alpha_i = 0,\pm 1$
$(i = 1,\ldots,d)$.

A volumetric unit cell boundary ${\Omega}_{\Gamma}=\Gamma^{+}\cup\Gamma^{-}$ contains
the boundary layers $\Gamma^{+}$ and $\Gamma^{-}$ (internal and external, respectively) separating by the geometric boundary $\Gamma^0$ and having a thickness given by the horizon $l_{\delta}$ (the UC $\Omega_{00}$ with some parts of the internal and external volumetric boundaries is presented in \cite{Buryachenko2022a}).
For a general form of the UC $\Omega_{00}$ the external volumetric boundary $\Gamma^-$ can be expressed through the Minkovski addition
$\Gamma^-=\{\Omega_{00}\oplus{\cal H}_{\bf 0}\}\backslash \Omega_{00}$
with the partitions $\Gamma_{\bf \alpha}^-=\Gamma^-\cap\Omega_{\alpha}$
(if $|\alpha_1|+\ldots+|\alpha_d|=d-1$)
and $\Gamma_{\bf \alpha}^{c-}=\Gamma^-\cap\Omega_{\alpha}$ (if $|\alpha_1|+\ldots+|\alpha_d|=d$); for $d=3$ and $|\alpha_1|+|\alpha_2|+|\alpha_3|=1$,
the case $\Gamma^-\cap\Omega_{\alpha}$ corresponds to the edges.
For $\bfx\in \Omega_{00}$ and ${\cal H}_{\bf x}\subset\Omega_{00}$ (i.e. $\bfx\not\in \Gamma^{+}$), we don't need to take the interaction of the opposite volumetric boundaries into account.
However, for $\bfx\in\Gamma^+$, some points of the family $\bfy\in{\cal H}_{\bf x}$ don't fall into the UC $\bfy\not\in\Omega_{00}$ that required establishment of the tying constraints linking both the position points and the fields of the opposite volumetric boundaries $\Gamma^{+}$ and $\Gamma^{-}$.
If the source point
$\bfx_p+\bfxi\in\Omega_{\bf \alpha}$ $(\bfxi\in{\cH}_p$) then
the peridynamic counterpart of the local
periodic boundary conditions (PBC) called
{\it volumetric periodic boundary conditions} (VPBC, see \cite{{Buryachenko2018a},{Buryachenko2022a}}) represent periodic displacements and antiperiodic tractions. So, the homogeneous VBC at the remote boundary (\ref{2.28}) is equivalent to
the VPBC
\BBEQ
\label{2.37}
\bfu(\bfx_p+\bfxi)&=&\bfu(\bfx_p+\bfxi-2\bfal^l)
+2\bfep^{w_\Gamma}\bfal^l,\\
\label{2.38}
\bft(\bfx^0_{\bf\alpha})&=&-\bft(\bfx^0_{\bf\gamma}),
\EEEQ
where $\bfga=-\bfal$ and $\bfx^0_{\bf\alpha}$ are defined analogously
to $\bfx^0_{ij}$. The VPBCs (\ref{2.37}) and (\ref{2.38}) coincide with the classical PBC
\BBEQ
\label{2.39}
\bfu(\bfx_p)&=&\bfu(\bfx_p-2\bfal^l)
+2\bfep^{w_\Gamma}\bfal^l,\\
\label{2.40}
\bft(\bfx^0_{\bf\alpha})&=&-\bft(\bfx^0_{\bf\gamma})
\EEEQ
only for $l_{\delta}=0$.
The VPBC (\ref{2.37}) and (\ref{2.38}) were proposed in \cite{{Buryachenko2018a},{Buryachenko2022a}} for any peridynamic constitutive lows of phases; it was applied to CMs
with both the bond-based peridynamic properties of constituents \cite{{Buryachenko2018a},{Buryachenko2022a}}
and non-ordinary state-based \cite{{Galadimaet2023}} peridynamic properties of phases. For the general forcing term
(\ref{2.4}) (instead of VBC (\ref{2.28})), analogs of VPBC (\ref{2.37}) and (\ref{2.38}) are not defined.

%\sffamily
%{\noindent \bf Comment 2.1.}
%; for shortening, we use $|\Omega|$ instead of $|\Omega|^{1/d}$.
The VPBC (\ref{2.37}) was proposed for the case that the inclusion $v_i$ does not intersect the boundary $\Gamma^0$ of the UC $\Omega^{00}$.
So, for ${\rm dist}(v_i(\bfx), \Gamma^0)\geq l_{\delta}^{(0)}$ (for $\forall \bfx\in v_i$ and $\forall v_i\subset \Omega_{00}$) and $l_{\delta}^{(1)}\leq l_{\delta}^{(0)}$, the VPBC (\ref{2.37}) holds.
However, if the inclusion $v_i$ does not intersect the boundary $\Gamma^0$ of the UC $\Omega^{00}$
then the VPBC (\ref{2.37}) should be corercted by replacement of $\bfxi\in{\cH}_p$ by $\bfxi\in{\cH}^{\rm 0+1}_p$, where
${\cH}^{\rm 0+1 }_p$ is a horizon region with the radius $l_{\delta}^{(0)}+l_{\delta}^{(1)}$.
Micromodulus $\bfC(\bfx,\hat{\bfx})$ corresponding to the points $\bfx=\bfx_p$ and $\hat{\bfx}=\bfx_p+\bfxi$
is given by the formula
\BBEQ
\label{2.41}\!\!\!\!\!\!\!\!\!\bfC(\bfx,\hat{\bfx})=
\begin{cases}
\bfC^{(1)}(\bfx,\hat{\bfx}), &\!\! {\rm for} \ \bfx,\hat{\bfx}\in v,\\%\cr
\bfC^{(0)}(\bfx,\hat{\bfx}), & \!\!{\rm for} \ \bfx,\hat{\bfx}\in v^{(0)},\\%\cr
\bfC^{i}(\bfx,\hat{\bfx}), & \!\!{\rm for} \ \bfx\in v,\hat{\bfx}\in v^{(0)} \\
{ }& \!\! {\rm or}\ %$\bigwedge
\ \bfx\in v^{(0)},\hat{\bfx}\in v,
\end{cases}
\EEEQ
where the bonds connecting points in the different materials are characterized
by micromodulus $\bfC^i$, which can be chosen such that, e.g. (\ref{2.37}).
Although there are no technical issues with the correction of
discretized peridynamic equation in the extended interface $\bfx\in v_{\Gamma}$, it is helpful to assume
for simplicity, only that the peridynamic operator in the extended interface (fuzzy interface, i.e.
not as sharp as it would be in a local formulation) is
described by Eq. (\ref{2.6}) with the constant horizon and micromodulus (\ref{2.41}) determined as an average
value of the micromoduli in the matrix and inclusion (\ref{2.32}).
The adaptive grid refinement technique was proposed in \cite{BobaruH2011}
for the analysis of peridynamic problems in the vicinity of the interface that involves a variable horizon
size.

It should be mentioned that the most attractive tool of ML and NN techniques in PM of periodic structure CMs is using a general case of forcing term (\ref{2.4}) as a training parameter (see Subsection 4.7). Then solution periodicity is lost and PBC (\ref{2.39}) (and VPBC (\ref{2.37})) cannot be fulfilled. This problem of correction of PBC (\ref{2.39}), even in LM is not solved for the general case of forcing term (\ref{2.4}). However, this challenge is effortlessly surmounted in Subsection 8.3.

%\rmfamily

\subsection {Some averages}
%\smallskip
In the case of statistically homogeneous random functions $\bfg (\bfx)$, the ergodicity condition is assumed when the spatial average is estimated over
one sufficiently large sample and statistical mean coincide for both the whole volume $w$ and the individual constituent $v^{(k)}$ ($k=0,1,\ldots,N$):
\BBEQ
\label{2.42}
\!\!\!\!\!\!\!\!\!\! \lle\bfg\rle \!& =&\!\!
\{\bfg\}\equiv \lim _{w\uparrow \mathbb{R}^d} |w|^{-1}\int _w
\bfg(\bfx)(\bfx)d\bfx, \\
\label{2.43}
\!\!\!\!\!\!\!\!\!\!\!\!\!\!\lle \bfg\rle^{(k)} \!\!& =&\!\! \{\bfg\}^{(k)}\equiv\lim _{w\uparrow \mathbb{R}^d} |w|^{-1}\int _{w}
\bfg(\bfx)(\bfx)V^{(k)}d\bfx,
\EEEQ
were $|w|={\rm mes}\, w$.
Under the Gauss theorem, the volume averages by integrals over the corresponding boundary are expressed.
In particular, for the homogeneous boundary conditions either
(\ref{2.28}) or (\ref{2.29})
the mean value $\{\bfep\}$ or $\{\bfsi\}$ of $\bfep$ or $\bfsi$
coincide with
$\bfep^{w_{\Gamma}}$ and $\bfsi ^{w_{\Gamma}}$ (see, e.g. \cite{Buryachenko2022a}).

The volume averages of the strains and stresses inside the extended inclusion $v^l_i$
can be presented by the averages over the external inclusion boundary
$\Gamma_i^-$
by the use of Gauss's theorem
\BBEQ
\label{2.44}
\!\!\!\!\!\!\!\!\!\!\!\lle\bfep V^l\rle=\bfep^{l(1)} \!\!\!\!& =&\!\!\!\! \bfep^{l\omega(1)}:= \frac {1}{\bar w}\sum_i\!\!\int_{\Gamma_i^-}\!\!
\bfu(\bfs){\,}{\,}^{\,^S}\!\!\!\!\!\!\otimes\bfn(\bfs)\,d\bfs,\\
\label{2.45}
\!\!\!\!\!\!\!\!\!\!\!\lle\bfsi V^l\rle=\bfsi^{l(1)} \!\!\!\!& =&\!\!\!\! \bfsi^{l\omega(1)}:= \frac {1}{\bar w}\sum_i\!\!\int_{\Gamma^-_i}\!\!
\bft(\bfs){\,}{\,}^{\,^S}\!\!\!\!\!\!\otimes\bfs\,d\bfs.
\EEEQ
The absence of numerical differentiation defines an advantage of the surface average.
(\ref{2.44}) with respect to the volume average.
The surface average of stresses
(\ref{2.45}) has the well-known advantage of the
reduction of dimension by one that can be crucial for the analyst. In Eqs. (\ref{2.44}) and (\ref{2.45}) the averages over the modified phases (e.g., the extended inclusions or truncated matrix) are used
\BBEQ
\label{2.46}
\!\!\!\!\!\!\!\!\!\!\!\!\!\lle \bfg\rle^{l(k)} = \{\bfg\}^{l(k)}\equiv\lim _{w\uparrow \mathbb{R}^d} |w|^{-1}\int _{w}
\bfg(\bfx)(\bfx)V^{l(k)}d\bfx,
\EEEQ
instead of the averages (\ref{2.42}$_2$) exploited in local micromechanics.

It should be mentioned that all averages(\ref{2.42})-(\ref{2.46}) 
are fulfilled only for statistically homogeneous media subjected to the homogeneous boundary conditions.

For periodic structure CMs, the CMs are constructed using the
building blocks or cells:
$w=\cup\Omega_{\bf m}$ containing the inclusions $ v_{\bf m}\subset \Omega_{\bf m}$.
Hereafter the notation
%$\langle (.)\rangle ({\bf x})$
${\bf f}^{\small \Omega}(\bfx)$
will be used for the average of the function ${\bf f}$
over the cell $\bfx\in\Omega_i$ with the center
$\bfx^{\Omega}_i\in \Omega_i$:
\BB
{\bf f}^{\Omega}(\bfx)={\bf f}^{\Omega}(\bfx_i^{\Omega})\equiv
n(\bfx)\int_{\Omega_i}{\bf f}(\bfy)~d\bfy,\quad \bfx\in \Omega_i,
\label{2.47}
\EE
$n(\bfx)\equiv 1/\overline \Omega_i$
is the number density of inclusions in the cell $\Omega_i$.

Let $\cV_{\bf x}$ be a ``moving averaging" cell (or moving-window \cite{{Buryachenko2022a},{GrahamBradyet2003}}) with the center $\bfx$
and characteristic size $a_{{\cal V}}=({\overline{\cV}})^{1/d}$, and let
for the sake of definiteness
$\bfchi$ be a random vector uniformly distributed on $\cV_{\bf x}$
whose value at $\bfz\in \cV_{\bf x}$ is
$\varphi_{\small \bfchi}(\bfz)=1/\overline {\cV}_{\bf x}$ and
$\varphi_{\small \bfchi }(\bfz)\equiv 0$ otherwise. Then
we can define the average of the function ${\bf f}$ with respect to
translations of the vector $\bfchi:$
\BB
\langle {\bf f} \rangle _{\bf x}(\bfx-\bfy)={1\over\overline
{\cV}_{\bf x}}\int_{\cV_{\rm \bf X}}{\bf f}(\bfz -\bfy)~d\bfz,
\quad \bfx\in {\Omega} _i.
\label{2.48}
\EE
Among other things, ``moving averaging" cell $\cV_{\bf x}$
can be obtained by translation of a cell $\Omega_i$ (\ref{2.47})
and can vary in size and shape during motion from point
to point.
To make the exposition clear, we will
assume that $\cV_{\bf x}$ results from $\Omega_i$ by translation of the vector
$\bfx-\bfx^{\Omega}_i$; it can be seen, however, that this assumption is not
mandatory.

\section{Decomposition of the material and field parameters}

{\color{black} IIn Section 3, two branches of Micromechanics are identified — analytical micromechanics and computational micromechanics — each encompassing several specific methods. This section is not “a chest of ancient relics,” but rather a final nostalgic reflection on the achievements we are now leaving behind in favor of the new approach (see Conclusion 9).
The second direction concerns the introduction of both GIE and AGIE. We have presented a vivid, colorful history of the development of GIE from Rayleigh (1892) to the present day. In contrast, AGIE emerges suddenly, like a “jack-in-the-box,” governed by a body force with compact support and without any prior foundation in micromechanics. Nevertheless, it is precisely AGIE that serves as the key element — the ignition spark — for the entire newly proposed approach in Sections 4-9. Solution of AGIE at the BFCS loading generate conceptually new {\it effective dataset}.
}

\setcounter{equation}{0}
\renewcommand{\theequation}{3.\arabic{equation}}
\subsection{Decomposition of the material parameters}
At first, the operators $\widetilde{\bfcL}(\bfC,\bfu)$ (\ref{2.12}) (and ${\bfcL}^{\sigma}(\bfC,\bfu,\bfbe)$ (\ref{2.16}),
are linear ones with respect to the arguments, e.g.,
\BBEQ
\label{3.1}
\widetilde{\bfcL}(\bfC_1+\bfC_2,\bfu)&=&\widetilde{\bfcL}(\bfC_1,\bfu)+\widetilde{\bfcL}(\bfC_2,\bfu),
\nonumber\\
\widetilde{\bfcL}(\bfC,\bfu_1+\bfu_2)&=&\widetilde{\bfcL}(\bfC,\bfu_1)+\widetilde{\bfcL}(\bfC,\bfu_2).
\EEEQ
For the matrix CMs with a homogeneous matrix, we decompose the material parameters
%: the elastic modulus $\bfL$, micromodulus, $\bfC$, peristatic operator $\widetilde{\bfcL}$ (\ref{2.6}), and peristatic stress %operator $\hat{\bfcL}^{\sigma}$ (\ref{2.43})
as
\BBEQ
\label{3.2}
\bfL(\bfx)&=& \bfL^{(0)}+\bfL_1(\bfx),
\nonumber \\
\bfC(\bfx,\hat {\bfx})&=& \bfC^{(0)} (\bfx,\hat {\bfx})+\bfC_1(\bfx,\hat {\bfx}),
\EEEQ
where $\bfL_1(\bfx):=\bfL(\bfx)-\bfL^{(0)}$ and
$\bfC_1(\bfx,\hat {\bfx}):=\bfC(\bfx,\hat {\bfx})-\bfC^{(0)}(\bfx,\hat {\bfx})$ are the jumps of material properties with respect to the matrix.
Obviously, $\bfL_1(\bfx)\equiv {\bf 0}$ vanishes outside the inclusion phase $\bfx\not\in v$ whereas $\bfC_1^{\rm}(\bfx,\hat {\bfx})\equiv{\bf 0}$ outside the extended inclusion phase
$v^{l}:=v\oplus {\cH}_0$ (obtained by a Minkovski sum of
$v$ when either $\bfx\not\in v^{{\rm}l}$ or
$\hat {\bfx}\not \in v^{{\rm}l}$).

In a similar manner, the peridynamicc operators $\widetilde{\bfcL}(\bfC,\bfu)$ (\ref{2.12}),
and the stress operator
${\bfcL}^{\sigma}(\bfC,\bfu)$ (\ref{2.27}) can be decomposed as
\BBEQ
\label{3.3}
\widetilde{\bfcL}(\bfC,\bfu)&=&\widetilde{\bfcL}^{(0)}(\bfC,\bfu)+\widetilde{\bfcL}_1(\bfC,\bfu),\nonumber \\
%\label{3.3}
{\bfcL}^{\sigma}(\bfC,\bfu)&=& {\bfcL}^{\sigma(0)}(\bfC,\bfu)+
{\bfcL}_1^{\sigma}(\bfC,\bfu),
\EEEQ
where $\widetilde{\bfcL}^{(0)}(\bfC,\bfu)$ (\ref{2.12}) and
${\bfcL}^{\sigma(0)}(\bfC,\bfu)$ (\ref{2.27}) are estimated at $\bfC(\bfx,\hat {\bfx})=
\bfC^{(0)}(\bfx,\hat {\bfx})$.
In so doing, $\widetilde{\bfcL}_1(\bfC,\bfu)$ (\ref{2.24}) and ${\bfcL}^{\sigma}_1(\bfC,\bfu)$ (\ref{2.27})
vanish inside the truncated matrix $\bfx\in v^{(0)l}:=w\backslash v^l$.

Peridynamic counterpart of the tensorial decomposition for local elasticity
\BB
\label{3.4}
\bfsi(\bfx)=^L\!\bfL^{(0)}\bfep(\bfx)+^L\!\bftau(\bfx), \ 
^L\!\bftau(\bfx):=\bfsi(\bfx) - ^L\!\bfL^{(0)}(\bfx)\bfep(\bfx)
\EE
can be presented in the next form for
the operator (\ref{2.17})
\BB
\label{3.5}
\bfcL^{\sigma}(\bfu)(\bfx)=\bfcL^{\sigma (0)}(\bfu)(\bfx)+\bfcL^{\sigma}_1(\bfu) (\bfx),
\EE
were $\bfcL^{\sigma (0)}$ denotes an action of the operator
$\bfcL^{\sigma}$ on the medium with the material properties of the matrix defined by the bond force
$\hat\bff^{(0)}$ [for example, for the bond force (\ref{2.8}), we have $\bfC^{\rm }(\bfxi,{\bfx})\equiv \bfC^{(0)}(\bfxi)$]
and the displacement fields $\bfu({\bfy})$ of the real CM.
The jump operator $\bfcL^{\sigma}_1(\bfu) (\bfx)$ defined by Eq. (\ref{3.4})
is called the {\it local stress polarization tensor} [compare with Eq. (\ref{3.4})] and represented
as
\BBEQ
\label{3.6}
\bftau(\bfx)\!\!&=&\!\!\bfcL^{\sigma}_1(\bfu) (\bfx)=
{1\over 2} \int_S\int_0^{\infty}\int_0^{\infty} (y+z)^{d-1}\nonumber\\
\!\!&\times&\!\!{\bff}_1(\bfx+y\bfm,\bfx-z\bfm)
\otimes\bfm dzdyd\Omega_{\bf m},
\EEEQ
where
\BB
%\label{3.7}
{\bff}_1(\bfp,\bfq):={\bff}(\bfp,\bfq)-{\bff}^{(0)}(\bfp,\bfq).\nonumber
\EE
In particular, for the linear bond force (\ref{2.8}), 
we define the {\it micropolarization} tensors
\BBEQ
\label{3.7}
\widetilde{\bftau}(\bfx,\hat{\bfx})&:=&
%\!\!\!&=&\!\!\!
\bfC_1(\bfx,\hat{\bfx})\bfeta(\hat{\bfx},\bfx),\nonumber\\
\widetilde{\bftau}^{(0)}(\bfx,\hat{\bfx})&:=&
%\hat{\bff}^{(0)}(\bfx,\hat{\bfx})=
%\!\!\!&=&\!\!\!
\bfC^{(0)}(\bfx,\hat{\bfx})\bfeta(\hat{\bfx},\bfx)
\EEEQ
are expressed through the material parameters (\ref{3.2}).

\subsection{
Analytica and Computational Micromechanics in LM}

The effective field hypothesis (EFH) dates back to Poisson, Faraday, Mossotti, Clausius, Lorenz, and Maxwell (1824-1879, see \cite{Buryachenko2022a})
who proposed EFH {\bf H1a} as a local homogeneous
field acting on the inclusions ($\bfx\in v_i$)
\BBEQ
\label{3.8}
\overline{\bfep}_i(\bfx)={\rm const}
\EEEQ
and differing from the applied macroscopic one (at infinity).
This concept of EFH (forming the first background of analytical micromechanics, see for details \cite{Buryachenko2007}) has directed the development of micromechanics (even if the term EFH was not used) over the last 150 years (daily and globally) and contributed to their progress incomparable with any other concept of analytical micromechanics.

Buryachenko \cite{Buryachenko2024a} and \cite{Buryachenko2024b} presented general classifications of Analytical Micromechanics (AMic) and Computational Micromechanics (CMic) methods.
So, the concept of the EFH (even if this term is not mentioned) in combination with subsequent assumptions
totally predominates (and creates the fundamental limitations) in all four groups of {\it Analytical Micromechanics} (AMic, classification by
Willis \cite{Willis1981}) of {\it random} random structure matrix CMs in physics and mechanics of heterogeneous media:
\vspace{-1mm}
\BBEQ
\label{3.9}
&&{\rm \underline{Gr1)}\ model\ methods}, \nonumber \\
&&{\rm \underline{Gr2)}\ perturbation\ methods,} \nonumber \\
&& { \rm \underline{Gr3)}\ variational\ methods}, \nonumber \\
&& {\rm \underline{Gr4)}\ self-consistent\ methods},
\EEEQ
see for references and details
\cite{{Buryachenko2007},{Buryachenko2022a},{Dvorak2013},{KachanovS2018},{Torquato2002}}.
The ultimate goal of AMic is to develop more cheap, fast, robust, and more flexible methods (for making $\bfL^*$ estimations)
than direct numerical simulation (DNS), although it takes additional intellectual complexity to
the implementations.

In contrast, {\it Computational Micrmechanics} (CMic) for CM of {\it deterministic} structures is based on DNS, which can be found by different numerical methods. Computational micromechanics can be classified into three broad categories (blocks):
\BBEQ
\label{3.10}
\!\!\!\!\!\!&& \!\!\! \!\!\! \!\!\! \!\!\! \! {\rm \underline{\rm Block\ 1)}\ Asymptotic \ homogenization,}\nonumber\\
\!\!\!\!\!&& \!\!\! \!\!\! \!\!\! \!\!\! \! \underline{\rm B1ock \ 2)}\ {\rm Computationa\ homogenization}\nonumber\\ 
\!\!\!\!\!&& \!\!\!\!\!\!\!\!\!\!\!\underline{\rm Block \ 3)}\ \ \ {\rm Finite\ set \ of}\ \ {\rm inclusions.}
\EEEQ
Blocks 1 and 2 are applied to periodic structure composites (see Introduction). Block 3 corresponds to one, and the finite set of inclusions for either the finite-size sample
or the infinite matrix with a finite set of inclusions (in such a case, the problem can be solved by either the volume integral equation methods or the boundary integral equation one, see for references, e.g. \cite{Buryachenko2022a}).
The key distinction between CMic and AMic lies in the reliance on DNS, which sets computational methods apart from analytical ones. This differentiation is not based on the conventional definitions of ``analytical" and ``computational" but instead arises from the specific techniques employed to model and estimate material properties.
It should be mentioned that the classification of AMic (\ref{3.9}) and CMic (\ref{3.10}) holds also for PM (see for details \cite{Buryachenko2024b}).

A key concept in analytical micromechanics is the GIE, which precisely relates random fields at a given point to those in its surroundings.
Derivation of the GIE begins with the static governing equation for displacement, which follows a Navier-like form
\BBEQ
\label{3.11}
\nabla{ ^L\!}\bfL^{(0)}(\bfx)\nabla\bfu(\bfx)=- \nabla{^L\!}\bfL_1(\bfx)\nabla\bfu(\bfx).
\EEEQ
Various forms of GIEs, ranked by increasing generality, are detailed in \cite{Buryachenko2015a} and reproduced in \cite{Buryachenko2022a} ($\bfx\in w$):
\vspace{-2mm}
\BBEQ
\label{3.12}
\!\!\!\!\!\!\!{\bfep}({\bf x}) \!\!\!&=&\!\! {\bfep}^{w \Gamma}
+\int{\bfU}(\bfx-\bfy)^L\!\bftau(\bfy)~d{\bf y},\\
\label{3.13}
\!\!\!\!\!\!\!{\bfep}({\bf x}) \!\!&=&\!\! \langle {\bfep}\rangle ({\bf x})
+\int{\bfU}(\bfx-\bfy)[^L\!\bftau(\bfy)\nonumber\\
\!\!&-&\!\!
\overline{\lle}{^L\!\bftau\rle(\bfy)}]~d{\bf y},\\
\label{3.14}
\!\!\!\!\!\!{\bfep}({\bf x})
\!\!&=&\!\!
\langle {\bfep}\rangle ({\bf x})
%\nonumber\\
+\int[{\bfU}(\bfx-\bfy)^L\!\bftau(\bfy)
\nonumber\\
\!\!&-&\!\!\underline{\lle}{\bfU}(\bfx-\bfy)^L\!\bftau\rle(\bfy)]
~d{\bf y},
\EEEQ
%\vspace{-1mm}
%\noindent
$\lle(\cdot)\rle$ and $\lle(\cdot)\rle(\bfx)$ are the statistical averages introduced in Subsection 2.4. Hereafter, one introduces the infinite body Green's function ${\bf G}^{(0)}$ of the Navier equation with homogeneous elastic modului $^L\!{\bfL}^{(0)}$:
%defined by
$
\nabla \left\lbrace{^L\!{\bfL}^{(0)}[{\nabla \otimes
{\bf G}^{(0)}(\bfx)}
]}\right\rbrace $ $=-\bfdel \delta
({\bf x}),
$
of order $O\big(\int |\bfx|^{1-d}d|\bfx|\big)$ as $|\bfx|\to\infty$ and vanishing at infinity ($|\bfx|\to\infty$). %%%%%%%%%%%%%%%%%%%
The {\color{black} Green's} tensors for the strains is used
$ \bfU(\bfx)=\nabla\nabla\bfG^{(0)}(\bfx).$
\sffamily
\noindent{\bf Comment 3.1.} The historical development of the GIE from Eq. (\ref{3.12}) to Eq. (\ref{3.14}), with key milestones ( \cite{{Rayleigh1892},{Shermergor1977},{Khoroshun1978},{OBrian1979}}), is outlined in \cite{Buryachenko2022a}. So, the integral in Eq.
(\ref{3.12}) has no absolute convergence, whereas in Eq. (\ref{3.13}) (proposed in \cite{Rayleigh1892} at $\lle\bfep\rle(\bfx)\equiv$const.), there are no difficulties connected with the asymptotic behavior of the generalized
functions $\bfU$ at infinity (as $|\bfx-\bfy|^{1-d}$) and there is no need
to postulate either the shape or the size of the integration domain $w$ \cite{{HoriK1998},{FassiFehriet1989},{JuT1992},{JuT1995}} (see also \cite{Sobczyket2007}) or to
resort to either regularization \cite{{Kanaun1977},{KanaunL2008},{Kroner1974},{Kroner1986},{Kunin1983}} or renormalization \cite{{ChenA1978b},{McCoy1979}}, or to consider an auxiliary problem with mixed boundary conditions \cite{{Brisard2017},{Brisardet2013}} of integrals (\ref{3.12})
that are divergent at infinity \cite{Willis1981}.
The shift of the statistical average bracket from $\overline{\lle}$ (\ref{3.13}) to $\underline{\lle}$ (\ref{3.14}) marks the emergence
the second background of micromechanics (called also Computational Analytical Micromechanics , CAM) \cite{Buryachenko2022a}, a pivotal advancement in micromechanics since Rayleigh's first intuitive formulation of GIE (\ref{3.13}) \cite{Rayleigh1892} (at $\lle\bfep\rle(\bfx)\equiv$const.).
\rmfamily

\subsection{ Modeling of one inclusion inside the infinite matrix}
We consider the equilibrium equation for an infinite $R^d$ ($d=1,2,3$) homogeneous peridynamic medium
(\ref{2.16}) subjected to the body force with compact support (BFCS) $\bfb(\bfx)$ (\ref{2.4})
\BB
\label{3.15}
\widetilde{\bfcL}^{(0)}(\bfC^{(0)},\overline{\bfu})(\bfx)+\bfb(\bfx)={\bf 0}.
\EE
A prescribed displacement $\overline{\bfu}(\bfx)$ (called the {\it effective displacement field}) corresponds to the self-equilibrated body-force density $\bfb(\bfx)$. In particular, a case $\overline{\bfu}(\bfx)$ corresponding to the homogeneous effective strain is
$
\overline{\bfu}(\bfx)=\overline{\bfep}\cdot\bfx, \ \ \ \overline{\bfep}={\rm const.}
$
Let us consider a wacrodomain $w$ with one inclusion $v_i$ subjected to the prescribed effective field loading $\overline\bfep(\bfx)$ corresponding to the BFCS $\bfb(\bfx)$ (\ref{2.4}). Then, Eq.
(\ref{2.6}) for a general peridynamic operator $\widetilde\bfcL$ can be presented as
\BBEQ
\label{3.16}
\widetilde\bfcL(\bfu)(\bfx)=\widetilde\bfcL^{(0)}(\overline\bfep)(\bfx)=-\bfb(\bfx).
\EEEQ
The main advantage of the representation (\ref{3.16}) (see also p. 774 in \cite{Buryachenko2022a}) is that it avoids the challenges associated with the fuzzy boundaries that are characteristic of nonlocal theories (see, e.g.,\cite{{Duet2013},{Kilic2008},{MacekS2007},{Silling2000}}). Equation (\ref{3.16}) simplifies these challenges by eliminating the need for volumetric boundary conditions and sidestepping the difficulties of properly imposing surface effects (see for references and details, e.g. \cite{{Scabbiaet2023},{WecknerE2005},{YuZ2024},{Bobaruet2016}}, Chapter 14) in nonlocal models. This approach contributes to a more tractable and efficient formulation for analyzing materials under nonlocal influences.

Hereafter, for the construction of a solution of Eq. (\ref{3.16}) for the nonlinear elastic case (see, e.g. 
(\ref{2.18})-(\ref{2.20})),
the {\it perturbators} are defined in the reduced form
($\bfx\in R^d$)
\BBEQ
\label{3.17}
\bfthe(\bfz)-\overline{\bfthe}(\bfz)=\bfcL^{\theta\zeta}_i(\bfz,\overline{\bfze}),
%\equiv \mathbb{L}^{u u}\overline{\bfu},
\EEEQ
where we introduce the substitutions
\BB
\label{3.18}
(\bfu,\bfeta)\leftrightarrow \bfthe,\ \ \ (\bfu,\bfep)\leftrightarrow\bfze, \ \ \ [\bfx,( \hat\bfx, \bfx)]\leftrightarrow \bfz.
\EE
The elements of the doublet $\bfthe$ correspond to the variables on the left-hand side of Eq. (\ref{3.18}),
the elements of the doublet $\bfze$ correspond to the effective fields on the right-hand sides of the definitions, whereas
the arguments $\bfx$ and $( \hat\bfx, \bfx)$ of the doublet $\bfz$ correspond to the parameters $\bfu$ and $\bfeta$, respectively.
The superindeces $^{\theta\zeta}$ of the perturbators $\bfcL^{\theta\zeta}$ correspond to the variables in the left-hand side $\bfthe$ and right-hand side $\bfze$, respectively.
In particular,
the operators $\bfcL_k^{u u}(\bfx-\bfx_k,\overline\bfu)$ and $\bfcL_k^{u\varepsilon}(\bfx-\bfx_k,\overline\bfep)$
(called the {\it perturbators} of the displacement)
have the physical interpretation of perturbations
introduced by a single heterogeneity $v_k$ in the infinite matrix subjected to the
effective fields $\overline {\bfu}_i(\bfx)$ and
$\overline {\bfep}_i(\bfx)$, respectively, where at first no restrictions are imposed on the inhomogeneities of the effective fields $\overline {\bfu}_i(\bfx)$ and $\overline {\bfep}_i(\bfx)$.

Let us consider two inclusions $v_i$ and $v_j$ placed in an infinite homogeneous matrix and subjected to the inhomogeneous field $\widetilde{\bfze}_{i,j}(\bfx)$ ($\bfu,\bfeta =\bfthe;\ \ \bfu,\bfep=\bfze;\ \ [\bfx, ( \hat\bfx, \bfx)]=\bfz;\ \ \bfx\in \mathbb{R}^d$).
We can transform Eq. (\ref{3.4}) into the following ones ($\bfz\in v_i^l$)
\BB
\label{3.19}
\bfthe(\bfz)-\widetilde {\bfthe}_{i,j}(\bfz) -\bfcL^{\theta\zeta}_i(\bfz-\bfx_i,{\widetilde {\bfze}}_{i,j}):=
%\widetilde {\bfthe}_{i,j}(\bfx)+
\bfcL^{\theta\zeta}_{i,j}(\bfz ,{\widetilde {\bfze}}_{i,j})
\EE
defining the perturbator $\bfcL^{\theta\zeta}_{i,j}(\bfz ,{\widetilde {\bfze}}_{i,j})$ which can be found
by any numerical method analogously to the operator $\bfcL^{\theta\zeta}_i(\bfz-\bfx_i,{\overline {\bfze}}_{i})$ (\ref{3.17}).
It should be mentioned that the operators $\bfcL^{\theta\zeta}_i(\bfz-\bfx_i,{\widetilde {\bfze}}_{i,j})$ and
$\bfcL^{\theta\zeta}_{i,j}(\bfz ,{\widetilde {\bfze}}_{i,j})$ (\ref{3.19}) act on the effective fields $\widetilde{\bfze}_{i,j}(\bfx)$ at $\bfx\in v_i$ and $\bfx\in v_i,v_j$, respectively, and the kernel of the operator $\bfcL^{\theta\zeta}_{i,j}$ can be decomposed
($K=I,J)$:
\BBEQ
\label{3.20}
\bfcL^{\theta\zeta}_{i,j}(\bfz,\bfy)\!\!&=&\!\!\bfcL^{I\theta\zeta }_{i,j}(\bfz,\bfy)+\bfcL^{J\theta\zeta }_{i,j}(\bfz,\bfy),\nonumber\\
\bfcL^{K\theta \zeta }_{i,j}(\bfz,\bfy)\!\!&=&\!\!\bfcL^{\theta\zeta}_{i,j}(\bfz,\bfy)V_k(\bfy),
\EEEQ
where one follows Mura's tensor indicial notation (see for details \cite{Buryachenko2022a}).
The double superindices $^{\theta\zeta}$
%$^{\varepsilon \varepsilon}$, and $^{\sigma\varepsilon}$ show the variables
is used analogously to the double superindices $^{uu}$ and
$^{u \varepsilon}$
%, and $^{\sigma\varepsilon}$
in Eqs. (\ref{3.17}).

Similarly, the effective field perturbators $\bfcJ^{\theta \zeta}_{i,j}$
and $\bfcJ^{\theta\zeta\infty }_{i,j}$
%$\bfT^{\theta\infty}_{i,j}(\bfz)$
can be defined;
they describe the perturbation of the effective field
$\overline{\bfthe}_i(\bfz)-\widetilde {\bfthe}_{i,j}(\bfz)$ introduced by both the heterogeneity $v_j$ (interacting with $v_i$) and the fictitious inclusion
with the response operator $\bfcL^{(0)}$ and eigenfield $\bfbe_1^{\rm fict}(\bfy)$ corresponding to the field in the remote inclusion $v_j$ (without interaction with $v_i$) ($\bfy\in v_j,\ \bfx\in v_i,\
\bfz \in \mathbb{R}^d$)
\BBEQ
\label{3.21}
\overline{\bfthe}_i(\bfz)-\widetilde {\bfthe}_{i,j}(\bfz) &=&
\bfcJ_{i,j}^{\theta \zeta}(\widetilde{\bfze}_{i,j})(\bfz)\\
\label{3.22}
\overline{\bfthe}_i(\bfz)-\widetilde {\bfthe}_{i,j}(\bfz) &=&
\bfcJ_{i,j}^{\theta\zeta\infty }(\widetilde{\bfze}_{i,j})(\bfz),
\EEEQ
(see for details \cite{Buryachenko2022a}).

\sffamily
\noindent{\bf Comment 3.2.}
Estimation of the perturabator $\bfcL^{\theta\zeta}_i(\bfz,\overline{\bfze})$ (\ref{3.17})
is, in fact, a basic problem of micromechanics (see Introduction) for one inclusion inside the infinite homogeneous
matrix.
In the PM, estimation of the perturabator $\bfcL^{\theta\zeta}_i(\bfz,\overline{\bfze})$ (\ref{3.17}) was considered by four different methods (see
\cite{{Buryachenko2022a},{Buryachenko2019b}}) for the linear bond-based medium with the same horizon $l_{\delta}$ in both the inclusion and matrix.
The generalization to the linear state-based model as well as to the multiphysics coupled problem
(see the LM applications in \cite{Buryachenko2015a}) are straightforward.
The popular discretization methods for the solution of PD equations (see for references \cite{{DEliaet2020},{DEliaet2017},{Littlewoodet2024}}) are the meshfree method with one-point Gaussian quadrature referring
to it as “meshfree PD” \cite{SillingA2005} (see also the solutions for 1D case
\cite{{Bobaruet2009},{EmmrichW2007a},{EmmrichW2007b},{ErikssonS2021},{Sillinget2007},{WecknerA2005},{WecknerE2005}}
and 2D case
\cite{{BobaruH2011},{Huet2012b},{Leet2014},{Saregoet2016}}), finite element methods (FEM,\cite{{ChenG2011},{MacekS2007},{Renet2017},{SunF2021},{Sunet2020},{TianD2015},{Wildmanet2017}}), quadrature and collocation approaches \cite{{Selesonet2016},{Zhanget2016a},{Zhanget2016b}}, and the boundary element method \cite{Lianget2021}.
{\color{black} Adaptive algorithm (see, e.g., \cite{{BobaruH2011},{Bobaruet2009},{Buryachenko2020}}) using a multi-grid approach with fine grid spacing only in critical regions
is designed in a multi-adaptive approach (\cite{Ongaroet2023})
to dynamically switch both the discretization scheme
and the grid spacing of the regions.}

\noindent{\bf Comment 3.3.}
The significantly less trivial phenomenon is another limiting case of
CCM-PD coupling, where usually small areas of a domain $\mathbb{R}^d$, which might be affected by
the presence of discontinuities is described with a PD model, whereas the remaining parts of the system are described through a more efficient CCM model. The goal of Local-to-Nonlocal (LtN) coupling (see \cite{{Birneret2023},{DEliaet2022},{Ongaroet2021},{Yuet2018},{Zaccariottoet2018},{ZhangJet2023},{Zhonget2024}}) is to alleviate the computational burden
by combining the computational efficiency of PDEs with the accuracy of nonlocal models
under the assumption that the location of nonlocal effects can be preliminarily prescribed (it looks like a localized plasticity model in the vicinity of inclusions in the LM of CMs, see pp. 556-565 \cite{Buryachenko2022a}).
It would be interesting to estimate $\bfcL^{\theta\zeta}_i(\bfz,\overline{\bfze})$ for the different nonlocal models in the phases (up to the vanishing length scale $l_{\delta}/a\to 0$ in some areas).
\rmfamily

\subsection{General integral equations (GIEs)}

In this subsection, at first, we consider peridynamic statistically inhomogeneous CMs
occupying the space $R^d$ ($d=1,2,3)$. We consider the so-called {\it general integral equation} (GIE) connecting the random fields at the point
being considered and all surrounding points. At first, the loading by the BFCS
$\bfb(\bfx)$ (\ref{2.4}), the direct summations of all surrounding perturbators $\bfcL^{\theta\zeta}_j(\bfz,\overline{\bfze})$
(\ref{3.17}) exerting on the fixed inclusion $v_i$ is described by the GIE (called the Additive GIE, AGIE, see for details \cite{Buryachenko2023k})
($\bfz\in v_i$)
\BBEQ
\label{3.23}
!\!\!\!\!\!\langle {\bfthe} \rle_i (\bfz)\!\!&=&\!\! \bfthe^{b(0)} ({\bf z})+\!\!
\int \!\!\bfcL^{\theta\zeta}_j(\bfz-\bfx_j,\overline{\bfze})\nonumber\\
&\times&\!\!\varphi (v_j,{\bf x}_j\vert v_1,{\bf x}_1)d{\bf x}_j
\EEEQ
with the deterministic fields $\bfthe^{b(0)} ({\bf z})$ produced by the body force $\bfb(\bfx)$ in the infinite homogeneous matrix.
The term {\it Additive} GIE is used similarly to {\it Additive} Manufacturing because the perturbations in Eq. (\ref{3.23}) are directly added without any renormalization terms, such as those in Eq. (\ref{3.24}).
A centering of Eqs. (\ref{3.23}) is considered, which performs a subtraction from both sides of Eq. (\ref{3.23}) of their statistical averages. It leads to the GIE
\BBEQ
\label{3.24}
\!\langle {\bfthe} \rle_i (\bfz)\!\!&=&\!\! \lle \bfthe\rle ({\bf z})+
\int \big[\bfcL^{\theta\zeta}_j(\bfz-\bfx_j,\overline{\bfze})
\varphi (v_j,{\bf x}_j\vert v_i,{\bf x}_i)\nonumber\\
\!\!\!\!\!\!\!\!\!\!\!\!\!\!\!\!&-&\!\! \lle\bfcL^{\theta\zeta}_j(\bfz-\bfx_j,\overline{\bfze})\rle(\bfx_j)\big]d{\bf x}_j,
\EEEQ
which is more general and valid for any inhomogeneous $\langle \bfthe\rangle ({\bf z})$ while Eqs. (\ref{3.23}) are only correct for the BFCS loading (\ref{2.4}). Owing to the centering of Eq. (\ref{3.16}), Eq. (\ref{3.24})
contains the renormalizing term $\lle \bfcL^{\theta\zeta}_j(\bfz-\bfx_j,\overline{\bfze})\rle(\bfx_j)$ providing an absolute convergence
of the integrals involved in Eqs. (\ref{3.24}).

\sffamily
\noindent{\bf Comment 3.4.}
The derivation of the exact form of the GIE (\ref{3.24}) has a long and intricate history, particularly in the context of LM of CMs (see \cite{{Buryachenko2014b},{Buryachenko2015a}} for references and details). The first correct (approximte) GIE (\ref{3.13}) was heuristically proposed by Lord Rayleigh \cite{Rayleigh1892} (for $\lle\bfep\rle(\bfx)\equiv \bfep^{w\Gamma}$) and was later rigorously established in 1977 (see \cite{Buryachenko2022a}).
A second exact GIE (\ref{3.14}) in the framework of LM, along with its generalizations to strongly and weakly nonlocal linear heterogeneous materials, was introduced between 2010 and 2022 (see for references and details \cite{Buryachenko2022a}). However, in all prior developments, specific instances of GIE (\ref{3.24}) were obtained through intricate multistep procedures, constrained by particular linear constitutive relations governing the phases of the material. In contrast, a more general form of GIE (\ref{3.24}), applicable to CMs with strongly and weakly nonlocal as well as nonlinear phase properties (see, e.g. 
(\ref{2.18})-(\ref{2.20})), was derived directly from the newly formulated AGIE (\ref{3.23}) via a single-step centering operation. This advancement eliminates the need for case-specific formulations and significantly broadens the applicability of the GIE and AGIE framework.

It is remarkable that, in earlier developments, the incorrect formulation of Eq. (\ref{3.12}) was addressed by none other than Lord Rayleigh \cite{Rayleigh1892}, who pioneered the idea of modifying the integral term itself. This bold step laid the foundation for a dominant trajectory in micromechanics research, one that would shape the field for the next 130 years (from Eq. (\ref{3.13}) to Eq. (\ref{3.14})). Traditional corrections to Eq. (\ref{3.12}) followed this path, involving sophisticated transformations of the integral operator through the introduction of renormalized expressions such as $\bfU(\bfx-\bfy)\lle\bftau\rle(\bfy)$ [Eq. (\ref{3.13})] and $\lle\bfU(\bfx-\bfy)\bftau\rle(\bfy)$ [Eq. (\ref{3.14})]. These efforts, while technically brilliant, were built on the premise that the operator itself must be altered to reconcile theory with physical reality. 
In striking methodological contrast, the present work introduces a radically new approach corresponding to a simple yet profound substitution: $\bfep^{w\Gamma} \to \bfep^{b(0)}(\bfx)$ in Eq. (\ref{3.12})
\BBEQ
\label{3.25}
{\bfep}({\bf x}) = \bfep^{b(0)}(\bfx)
+\int{\bfU}(\bfx-\bfy)^L\!\bftau(\bfy)~d{\bf y}.
\EEEQ
To the best of the author's knowledge, the BFCS loading (\ref{2.4}) has not been previously applied in micromechanics--whether for random or periodic structure--likely because its practical significance was not initially recognized. However, the primary motivation for introducing BFCS loading is the fundamentally new opportunity it offers: to explore its role as a training parameter for estimating any unpredefined surrogate nonlocal operators (see Section 9). What might appear at first glance as a minor technical adjustment ($\bfep^{w\Gamma} \to \bfep^{b(0)}(\bfx)$ in Eq. (\ref{3.12})) is, in fact, a fundamental philosophical breakthrough. It inaugurates a second philosophy of micromechanics (see for details Conclusion). Whereas the traditional philosophy, grounded in the effective field hypothesis (EFH) [Eq. (\ref{3.8})], relies implicitly on Hill’s classical RVE concept \cite{Hill1963} (discussed further in Section 8), the new philosophy opens a fundamentally different and far more transparent pathway for the field. By preserving the original form of the integral operator, this alternative strategy paves the way for the natural emergence of the nonlinear AGIE [Eq. (\ref{3.23})]. 
\noindent {\bf Comment 3.5.} Thus, two distinct lines of development have emerged in the quest to correctly modify the flawed Eq. (\ref{3.12}), each giving rise to fundamentally new directions in the evolution of micromechanics. The first pathway, initiated by Lord Rayleigh \cite{Rayleigh1892}, involved a pivotal modification to the integral term of Eq. (\ref{3.12}). This conceptual innovation laid the foundation for over a century of subsequent progress—from Eq. (\ref{3.13}) through its more advanced forms, Eqs. (\ref{3.14}) and (\ref{3.24}). This trajectory ultimately culminated in Eq. (\ref{3.24}), which may be regarded as the second background of micromechanics and represents the most significant development in the field since Rayleigh’s original contribution.
The second approach preserves the original integral structure of Eq. (\ref{3.12}) but introduces a fundamental shift by replacing the classical free term $\bfep^{w\Gamma}$ with the more general $\bfep^{b(0)}(\bfx)$, as defined in Eq. (\ref{3.25}). This modification is equivalent to replacing the conventional homogeneous boundary conditions (Eq. (\ref{2.30})) or volume boundary conditions (Eq. (\ref{2.28})) with the BFCS described by Eq. (\ref{2.4}). The resulting framework, further developed in Eq. (\ref{3.23}), underpins what is herein referred to as the new philosophy of micromechanics (see Conclusion).
This conceptual transformation transcends a mere methodological refinement. It represents a paradigm shift in the theoretical foundations of micromechanics—arguably the most profound advance in the field since the seminal contributions of Poisson, Faraday, Mossotti, Clausius, Lorenz, and Maxwell (1824–1879; see \cite{Buryachenko2022a}), and Lord Rayleigh \cite{Rayleigh1892}.

\rmfamily

After a statistical average of Eqs. (\ref{3.23}) and (\ref{3.24}), the conditional perturbator $\lle\bfcL^{\theta\zeta}_j(\bfz ,\overline{\bfze}) \vert ; v_i,{\bf x}_i\rle_j$ can be expressed through the
explicit perturbator for two interacting heterogeneities subjected to the field $\widetilde{\bfze}_{i,j}$
\BBEQ
\label{3.26}
\!\!\!\!\!\!\!\!\!\!\!\!\!\! \langle \overline{\bfthe}_i\rangle(\bfz) \!\!\!& =&\!\!\! \bfthe^{b(0)} ({\bf z})
+\!\!\int\!\! \bfcJ^{\theta\zeta}_{i,j}(\lle\widetilde{\bfze}_{i,j}\rle)(\bfz)\nonumber\\
&\times&\!\!\varphi (v_j,{\bf x}_j\vert; v_i,{\bf x}_i)%\nonumber\\ & -&
\,d{\bf x}_j,\\
\label{3.27}
\!\!\!\!\!\!\!\!\!\!\langle \overline{\bfthe}_i\rangle(\bfz) \!\!\!& =\!&\!\! \langle \bfthe\rangle ({\bf z})
+\int \bigl\{\bfcJ^{\theta\zeta}_{i,j}(\lle\widetilde{\bfze}_{i,j}\rle)(\bfz)\varphi (v_j,{\bf x}_j\vert; v_i,{\bf x}_i)\nonumber\\
&-&\bfcJ^{\theta\zeta\infty}_{i,j}(\lle\widetilde{\bfze}_{i,j}\rle)(\bfz)
\bigl\}d{\bf x}_j,
\EEEQ
where the conditional probability density $\varphi (v_j,{\bf x}_j\vert; v_1,{\bf x}_1)$ for one fixed inclusion was used
(see for details \cite{Buryachenko2022a}); $\bfcL^{\theta\zeta}_j(\bfz-\bfx_j,\overline{\bfze}
\vert ; v_1,{\bf x}_1)$ stands
a perturbation $\bfcL^{\theta\zeta}_j(\bfz-\bfx_j,\overline{\bfze})$ at the fixad inclusion $v_1\not=v_j$.
The deterministic field $\bfthe^{b(0)}(\bfz)$ (\ref{3.26}) produced by the forcing term $\bfb(\bfx)$ in the infinite homogeneous matrix.
Equations (\ref{3.26}) and (\ref{3.27}) are central to the overall formulation, capturing the material’s response under the applied effective field loading. These equations are constructed based on numerical solutions for one or two inclusions, and they incorporate statistically averaged fields $\lle\overline{\bfze}_{i}\rle(\bfx)$ and $\lle\widetilde{\bfze}_{i,j}\rle(\bfx)$
to represent the overall macroscopic behavior of the composite or heterogeneous material.
There are a couple of fundamental advantages of Eqs. (\ref{3.26}) and (\ref{3.27}).
So, these equations have no Green functions and, moreover, the most intriguing feature of Eqs. (\ref{3.26}) and (\ref{3.27})
is the absence of a constitutive law in these equations. As we can see, Eqs. (\ref{3.26}) and (\ref{3.27}) for the peridynamic micromechanics coincides with the corresponding equation for the LM CMs (see \cite{Buryachenko2015a}). It means that the same equation can be used for both the peridynamic CMs and locally elastic CMs.
The GIEs (\ref{3.26}) and (\ref{3.27}) proposed are adapted to the straightforward generalizations of corresponding methods of local thermoelastic micromechanics (see, e.g., \cite{{Buryachenko2007},{Buryachenko2014b},{Buryachenko2015a}}) to their peridynamic counterparts.
\sffamily

\noindent{\bf Comment 3.6.}
The analytical micromechanics (Amic) formulation (\ref{3.9}) is generally derived from the solutions to Eqs. (\ref{3.26}) and (\ref{3.27}), along with their specific cases. In particular, within the LM framework, the perturbations $\bfcL^{\theta\zeta}_j(\bfz-\bfx_j,\overline{\bfze})$ are determined using various methods like FEA ( \cite{{BuryachenkoB2011},{BuryachenkoB2012a},{BuryachenkoB2012b},{BuryachenkoB2013}}), VIE method ( \cite{{Buryachenko2010a},{Buryachenko2010b}}), BIE approach ( \cite{{Buryachenko2013},{Buryachenko2016},{Buryachenko2017c},{Buryachenko2018a}}).
These solutions are then substituted into Eq. (\ref{3.14}) to estimate the effective moduli $^L\!\bfL^*$ and stress concentration factors.
The classical hypotheses for the LM, such as the Effective Field Hypothesis (EFH) (\ref{3.1}) and its extension {\bf H1}, closing assumptions 
(see hypotheses {\bf H2a} and {\bf H2b} in Section4) along with the assumption of ellipsoidal symmetry {\bf H3}, have been analyzed in detail (\cite{Buryachenko2022a}). These hypotheses are fundamental to the classical LM methods, including the Effective Field Method (EFM), the Mori-Tanaka Method (MTM), and the Multiparticle Effective Field Method (MEFM). It is worth emphasizing that the solutions of Eq. (\ref{3.14}), along with those of Eqs. (\ref{3.26}) and (\ref{3.27}), are entirely independent of assumptions {\bf H1} and {\bf H3}, as they do not rely on these hypotheses at any stage of the derivation.
The GIE (\ref{3.14}) enhances the accuracy of local field estimations, even correcting the sign of predictions within inclusions ( \cite{{Buryachenko2022a}}).
Equation (\ref{3.14}) has been extended to address a variety of other problems within the LM framework: thermoelasticity ( \cite{{Buryachenko2017b},{Buryachenko2017c},{BuryachenkoB2012a},{BuryachenkoB2012b},{BuryachenkoB2013}}), coupled problems ( \cite{{Buryachenko2014b},{Buryachenko2015a}}), and wave propagation (\cite{{Buryachenko2014b},{Buryachenko2015a}}), and infiltration in porous media \cite{Buryachenko2015c}. It has also been adapted for CMs with both strongly nonlocal (e.g., strain-type, displacement-type, and peridynamics) and weakly nonlocal (e.g., strain-gradient, stress-gradient, and higher-order models) phase properties, as discussed in Chapter 14 of \cite{Buryachenko2022a}.
However, a key limitation of all these GIEs and their extensions (considered in the papers referred to in Subsection 3.4) is that they remain linear in relation to the primary unknown variable. To the author’s knowledge, there are no existing nonlinear versions (such as, e.g., (\ref{2.18})-(\ref{2.20}))
of the GIE (\ref{3.24}) and the AGIE (\ref{3.23})—whether for LM or PM—that directly or implicitly account for nonlinear constitutive phase properties.
In the following section, the author proposes a solution to the nonlinear AGIEs (\ref{3.23}), offering a potential way to address this gap.

{\color{black} \noindent{\bf Comment 3.7.} 
The second pathway, driven by the AGIE under the BFCS formulation (\ref{2.4}), has no precedent in the micromechanics literature. In fact, it introduces an entirely new paradigm—one that cannot be realized without simultaneously invoking the GIE solution strategy (\ref{3.24}) and modern ML\&NN methodologies. At first sight, incorporating all existing micromechanical models (see Eq. (\ref{3.9}) and Blocks 1 and 2 in (\ref{3.10})) may seem redundant. However, the rationale for this structure becomes fully transparent by the end of the paper (see Figs. 7 and 10). Specifically, we demonstrate that traditional micromechanics—its notions, its limited framework, and its techniques (see Eq. (\ref{3.9}) and Blocks 1 and 2 in (\ref{3.10}))—are neither necessary nor applicable within the BFCS setting (\ref{2.4}). We therefore opted for a single, unified exposition rather than a sequence of separate publications, as fragmenting the material would deprive readers of the broader conceptual context essential to understanding the objectives of the earlier components.}

\rmfamily

\section {Solution of nonlinear AGIEs for random structure CMs}

{\color{black} In Section 4, the iterative solution of AGIE (\ref{3.23}) for random-structure composite materials under BFCS loading generates a conceptually new effective dataset. This solution is simpler than that of GIE (3.24) (see Buryachenko, 2022), since AGIE does not include a renormalizing term (as in Eqs. (\ref{3.13}), (\ref{3.14}), and (\ref{3.24})). Another reason for its relative simplicity lies in the finite size of the domain, which naturally leads to the introduction of a fundamentally new RVE concept (see details in Section 8)}.

\setcounter{equation}{0}
\renewcommand{\theequation}{4.\arabic{equation}}

\subsection{Iterative solution of AGIE}

The effective field hypothesis, which serves as the fundamental approximation in numerous micromechanical methods, is mathematically formulated by Eq. (\ref{3.8}) (for details, see \cite{Buryachenko2007}).
To achieve closure of the AGIE (\ref{3.26}), the following additional hypothesis is introduced:

\noindent {{\bf Hypothesis H2a)}. {\it Each pair of inclusions $v_j$ and $v_j$
is subjected to the inhomogeneous field
$\widetilde{\bfze}_{i,j}(\bfx)$, and statistical average
$\lle\widetilde{\bfze}_{i,j}\rle(\bfx)$ is defined by the formula $(\bfze=\bfu,\bfep$)
\BBEQ
\label{4.1}
\lle\widetilde{\bfze}_{i,j}\rle(\bfx)&=&\lle\overline{\bfze}_{k}\rle(\bfx)
\EEEQ
at $\bfx\in v_k,\ k=i,j$}.

The hypothesis {\bf H2a}, when reformulated in terms of the strain fields $\bfep(\bfx)$ for $\bfx\in v_i$, represents a conventional closure assumption (for further details, see \cite{{Buryachenko2007},{KachanovS2018},{Willis1981}}).
This assumption reduces to the "quasicrystalline" approximation introduced by Lax \cite{Lax1952}, which neglects the pairwise interactions between heterogeneities and assumes spatial uniformity of the effective fields.

\noindent {\bf Hypothesis H2b, ``quasi-crystalline" approximation}.
{\it It is supposed that the mean value of the effective field at a point
$\bfx\in v_i$ does not depend on the field inside other heterogeneities
$v_j\not = v_i$, $\bfx\in v_k, \ (k=i,j)$}:
\BBEQ
\label{4.2}
\!\!\lle\overline{\bfze}|v_i,\bfx_i;v_j, \rle(\bfx)\!\!&=&\!\!\lle\overline{\bfze}_k\rle(\bfx)
\\%\nonumber\\
\label{4.3}
\lle\overline{\bfze}_{k}\rle(\bfx)&\equiv&{\rm const.}
\EEEQ

Acceptance of hypothesis {\bf H2a} closes the system (\ref{3.26}) in the following forms,
\BBEQ
\label{4.4}
\!\!\!\!\!\!\!\!\!\!\!\!\!\langle \overline{\bfthe}_i\rangle(\bfz)\!\!&=&\!\!\bfthe^{b(0)} ({\bf z})
+\!\!\int\! [\bfcJ^{I\theta\zeta}_{i,j}(\bfz ,\lle\overline{\bfze}_{i}\rle)
\nonumber\\
\!\!\!&+&\!\!\!\bfcJ^{J\theta\zeta}_{i,j}(\bfz ,\lle\overline{\bfze}_{j}\rle)]
%\nonumber\\
%&\times&\!\!\!
\varphi (v_j,{\bf x}_j\vert; v_i,{\bf x}_i)\,d{\bf x}_j,
\EEEQ
were a decomposition $\bfcJ^{\theta\zeta}_{i,j}=\bfcJ^{I\theta\zeta}_{i,j}+\bfcJ^{J\theta\zeta}_{i,j}$
was introduced analogously to Eq. (\ref{3.20}).
The integral equation (\ref{4.4}) can be solved using an iterative approach based on a recursive formula
\BBEQ
\label{4.5}
\!\!\!\!\!\!\!\!\!\!\!\!\!\!\!\!\!\!\!\!\!\!\!\langle \overline{\bfthe}_i^{[n+1]}\rangle(\bfz)\!\!\!&=&\!\!\!\bfthe^{b(0)} ({\bf z})
+\!\!\int
[\bfcJ^{I\theta\zeta}_{i,j}(\lle\overline{\bfze}^{[n]}_{i}\rle)(\bfz) \nonumber\\%\nonumber\\
\!\!\!\!&+&\!\!\!\!
\bfcJ^{J\theta\zeta}_{i,j}(\lle\overline{\bfze}^{[n]}_{j}\rle)(\bfz) ]
%\!\!&\times&\!\!
\varphi (v_j,{\bf x}_j\vert; v_i,{\bf x}_i)d{\bf x}_j,\\
\label{4.6}
\!\!\!\!\!\!\!\!\!\!\!\!\!\!\!\lle\bfthe^{[n+1]}\rle_i(\bfz)\!&=&\!\bfcA^{\theta\zeta}_i(\lle\overline{\bfze}^{[n+1]}_{i}\rle)(\bfz), \\
\label{4.7}
\!\!\!\!\!\!\!\!\!\!\!\!\!\!\!\bar v_i\lle\bftau^{[n+1]}\rle_i({\bf x})
\!\!\!&=&\!\!\!{\bfcR}_i(\lle\overline {\bfze}^{[n+1]}_i\rle) ({\bfx}).
\EEEQ
The iterative solution is constructed with an initial approximation derived as an explicit solution from Equations (\ref{4.3}), within the framework of the effective field hypothesis {\bf H1a} (\ref{3.8})
(see for details \cite{Buryachenko2022a}).
Equation (\ref{4.5}) admits a simplified form under Hypothesis Hb2, as shown in Equations (\ref{4.2}), which are also utilized in the LM method (see \cite{Buryachenko2007}).
However, a direct generalization of Lax's ``quasicrystalline" approximation is possible (see Equation (\ref{4.2})) when the assumptions in (\ref{4.2}) are relaxed, allowing the statistical average of the perturbed strain field, $\lle\overline{\bfze}_{k}\rle(\bfz)$,
to vary spatially rather than remain constant, i.e.,
$\lle\overline{\bfze}_{k}\rle(\bfz)\not\equiv{\rm const}$ for $\bfz\in v_k^l,$ \ $(k=i,j)$
\BBEQ
\label{4.8}
\lle\bfcL^{\theta\zeta}_j(\bfz ,\overline{\bfze}_{j})\vert ; v_i,{\bf x}_i\rle
= \lle\bfcL^{\theta\zeta}_j(\bfz ,\overline{\bfze}_{j})\rle.
\EEEQ
This significantly simplifies the problem formulated in Eq. (\ref{4.3}), where, under these conditions, the equation takes a more tractable form
$\bfcJ^{\theta\zeta}_{i,j}=\bfcJ_{i,j}^{\theta\zeta\infty}$
($\bfx\in v_i$) that reduces Eq. (\ref{4.3}) to
\BBEQ
\label{4.9}
\!\!\!\!\!\!\!\!\!\!\!\!\!\!\!\!\langle \overline{\bfthe}_i^{[n+1]}\rangle(\bfz)\!\!&=&\!\!\bfthe^{b(0)} ({\bf z})
+\!\!\int \!\!\bfcJ^{\theta\zeta\infty}_{i,j}(\lle\overline{\bfze}_{j}^{[n]}\rle)(\bfz)
\nonumber\\
&\times&\!\! 
\varphi (v_j,{\bf x}_j\vert; v_i,{\bf x}_i)\, d{\bf x}_j.
\EEEQ
The second background of LM, introduced in \cite{{Buryachenko2010a},{Buryachenko2010b}} and mathematically formulated in Eq. (\ref{4.9}), enables the relaxation of fundamental micromechanical assumptions, including the ellipsoidal symmetry hypothesis and the EFH {\bf H1a}.
As a result, novel physical effects have been identified that could not be captured within the classical (first) theoretical framework of micromechanics.

Equations (\ref{4.4}) and (\ref{4.9}) are derived for general nonlinear cases (see, e.g. 
(\ref{2.18})-(\ref{2.20})), incorporating both state-based (\ref{2.8}) and bond-based (\ref{2.15}) peridynamic models (PM). In the special case of a linear bond-based PM (\ref{2.24}), these equations simplify previously established results (see \cite{Buryachenko2023a}).

\subsection{Effective constitutive law and  effective dataset}

Equation (\ref{4.9}) is reformulated in terms of displacement fields ($\bfx\in v_i$)
\BBEQ
\label{4.10}
\overline{\bfu}^{[n+1]}(\bfx) \!\!&=&\!\!\!
\bfu^{b(0)} ({\bf x})
+\int \bfcL^{uu}_q(\bfx-\bfx_q,\overline{\bfu}^{[n]}\vert; v_i,\bfx_i)
\nonumber\\
&\times&\!\! 
\varphi (v_q,{\bf x}_q\vert; v_i,{\bf x}_i)
d{\bf x}_q.
\EEEQ
The initial approximation is defined by the driving term, corresponding to the zero-order approximation, given as $ \overline{\bfu}^{[0]}(\bfx)=\bfu^{b(0)} ({\bf x})$, which itself is expressed by Equation (\ref{3.15}).
The iterative scheme in Eq. (\ref{4.10}) constructs a Neumann series representation for the solution
($\bfx\in v_i^l$)
\BB
\label{4.11}
\lle\overline{\bfu}\rle_i(\bfx) := \lim_{n\to \infty}\lle\overline{\bfu}^{[n+1]}\rle_i(\bfx) =\widehat{\bfcD}_i^{ub}(\bfb,\bfx).
\EE
This leads to the formulation of the statistical averages of the field within the fixed extended inclusion, specifically the conditional averages for $\bfx\in v_i^l$
\BBEQ
\label{4.12}
\!\!\!\!\!\!\!\!\!\!\lle{\bfu}\rle_i(\bfx) \!\!&=&\!\! \widehat{\bfcD}_i^{ub}(\bfb,\bfx) +
\bfcL_i^{uu}(\bfx-\bfx_i, \widehat{\bfcD}_i^{ub}(\bfb,\bfx)),\\
\label{4.13}
\!\!\!\!\!\!\!\!\!\!\!\!\lle{\bfsi}\rle_i(\bfx) \!\!&=&\!\!\bfcL^{\sigma}(\lle{\bfu}\rle_i)(\bfx).
\EEEQ
The tensor $\widehat{\bfcD}_q^{ub}(\bfb,\bfy)$ (\ref{4.11})
is an inhomogeneous function of coordinates of the fixed inclusion $\bfx\in v_i^l$ depending on all interecting inclusions (at least at $|\bfx_q|\leq a^{\delta}+l_{\delta}$) whereas in the dilute approximation the corresponding tensor is $\bfI\bfb(\bfy)$.

Equation (\ref{4.12}) provides a means to compute the expected value of the displacement field over the macroscopic region $\bfX$
\BB
\label{4.14}
\langle {\bfu}\rle(\bfX) :=c^{l(0)}\lle\bfu\rle^{l(0)}(\bfX)+
c^{l(1)}\lle\bfu\rle^{l(1)}(\bfX)
\EE
The statistical averages of the displacement field in the macropoint are expressed through the ensemble averages of displacements at the point $\bfX$ within the matrix, denoted as $\lle\bfu\rle^{l(0)}(\bfX)$, and within the inclusions, represented as $\lle\bfu\rle^{l(1)}(\bfX)$.
To establish this framework, we introduce an auxiliary domain $v ^1_i (\bfX)$, which is characterized by the indicator function $V^1_i(\bfX)$ and is bounded by $\partial v ^1_i (\bfX)$, the locus of the centers of translated ellipsoidal regions $v_q({\bf 0})$ around the fixed macroscopic point point $\bfX$. The domain $v ^1_i (\bfX)$ is then obtained as the limiting form of a sequence $v^ 0_{ki}\to v_q^1(\bfX) $,
where $v^ 0_{ki}\to v_q^1(\bfX) $ as a given ellipsoid $v_k$ contracts to the point $\bfX$.
{\color{black} Then $\lle\bfu\rle^{l(1)}(\bfX)$ can be estimated as ($\bfy\in v^l_q)$
\BBEQ
\label{4.15}
\!\!\!\!\!\!\!\!\!\!\!\!c^{l(1)}\lle\bfu\rle^{l(1)}(\bfX)\!&=&\!
c^{l(1)}\bfu^{b(0)}(\bfX)+
\!\!\int_{v^1_i({\bf X})} \!\!\!\!\!\!\!n^{(1)}(\bfx_q)
\nonumber\\
\!\!&\times&\!\!\bfcL^{uu}_q(\bfX\!-\bfx_q,\bfcD_q^u(\bfb,\bfy))
d{\bf x}_q.
\EEEQ
Consequently, the statistical expectation of the displacement field, given by Eq. (\ref{4.14}), can be represented in terms of the volume force density distribution within the region $v_q^l$, where $\bfy\in v^l_q$
\BBEQ
\label{4.16}
\!\!\!\!\!\!\!\!\!\!\lle\bfu\rle(\bfX)\!\!\!&=&\!\!\!
\bfu^{b(0)}(\bfX)+ c^{(0)}
\int \bfcL^{uu}_q(\bfX-\bfx_q,\bfcD_q^{ub}(\bfb,\bfy))
\nonumber\\
\!\!&\times&\!\!
\varphi (v_q,{\bf x}_q\vert; \bfX) d{\bf x}_q %\nonumber\\
%\!\!\!&+&\!\!\! 
+\int_{v^1_i({\bf X})} n^{(1)}(\bfx_q)
\nonumber\\
\!\!&\times&\!\!
\bfcL^{uu}_q(\bfX-\bfx_q,\bfcD_q^{ub}(\bfb,\bfy))
d{\bf x}_q,
\EEEQ
In Eq. (\ref{4.16}), the first and second integrals correspond respectively to the first and second terms on the right-hand side of Eq. (\ref{4.14}). Additionally, the function $\varphi (v_q,{\bf x}_q\vert; \bfX)$ vanishes, i.e., $\varphi (v_q,{\bf x}_q\vert; \bfX)=0$, whenever the point $\bfx_q$ belongs to the domain $v^1_i(\bfX)$.

The macroscopic stress field in the effective constitutive law is expressed as
\BB
\label{4.17}
\lle\bfsi\rle (\bfx)=\lle\bfcL^{\sigma(0)}(\bfu)\rle(\bfx)+\lle\bftau\rle(\bfx)
\EE
where the first term on the right-hand side simplifies for the linear matrix and depends on the statistical averages $\lle\bfu\rle(\bfx)$:
$\lle\bfcL^{\sigma(0)}(\bfu)\rle(\bfx)=\bfcL^{\sigma(0)}(\lle\bfu\rle)(\bfx)$
($\bfx\in \mathbb{R}^d$) (\ref{4.16}).
To estimate the average $\lle\bftau\rle(\bfx)$, we first consider the fixed inclusion
$v_q$ centered at $\bfx_q$,
which is estimated using Eq. (\ref{2.17}).

This inclusion yields the value
$\bfcL^{\sigma}_1(\bfx-\bfx_q,\bfu)=\bfcL^{\sigma}(\widetilde{\bftau})$ which is estimated using Equation (\ref{2.17}) with the substitution $\bff^{(0)}\to \bff_1$ .
The statistical average of the local polarization tensor $\lle\bftau\rle(\bfX)$ is then obtained by averaging over the domain
$v_i^1$
\BBEQ
\label{4.18}
\!\!\!\!\!\!\!\!\!\!\!\!\!\!\!\!\lle\bftau\rle(\bfX)\!\!&=&\!\!\int_{v^1_i({\bf X})} n^{(1)}
\bfcL^{\sigma}(\bfC_1\bfeta^D)(\bfy-\bfX)d\bfy,%
\\
\label{4.19}
\!\!\!\!\!\!\!\!\!\!\!\!\!\!\!\!\! \bfeta^D(\bfx,\bfy)\!\!&=&\!\!\Big[\bfcD_i^{ub}(\bfb,\bfx)+\bfcL_i^{uu}(\bfx-\bfx_i,
\bfcD_i^{ub}(\bfb,\bfx) \nonumber\\
\!\!&-&\!\! \bfcD_i^{ub}(\bfb,\bfy)-\bfcL_i^{uu}(\bfy-\bfx_i,
\bfcD_i^{ub}(\bfb,\bfy))\Big],
\EEEQ
Here, $v_q^1(\bfx)$ is defined by Equation (\ref{2.41}), and the statistical average of the displacement field within the inclusion $\bfx\in v_q$ is represented by $\lle\bfu\rle_q(\bfx)$, as shown in Equation (\ref{4.15}).
The estimation of the effective micromodulus \cite{Yanget2024}, which involves averaging the strain energy of the peridynamic material model under homogeneous boundary conditions, raises certain concerns regarding its validity.

The new nonlocal effective constitutive law, expressed in Equation (\ref{4.17}), bears a strong resemblance to the previously established nonlocal effective constitutive law
\BB
\label{4.20}
\lle\bfsi\rle(\bfx) ={^L}\!\bfL^{(0)}\lle\bfep\rle(\bfx) +\lle ^L\!\bftau\rle(\bfy), \ \ \ ^L\!\bftau(\bfy):={^L}\!\bfL_1(\bfy)\bfep(\bfy).
\EE
This is analogous to the locally elastic constitutive law (\ref{2.2}) at the microlevel. However, the first term
${^L}\!\bfL^{(0)}\lle\bfep\rle(\bfx) $
on the right-hand side of Eq. (\ref{4.20}) corresponds to a general inhomogeneous strain field, where $\lle\bfep\rle(\bfx)\not ={\bf 0}$. On the other hand, the term $\lle\bfcL^{\sigma(0)}(\bfu)\rle(\bfx)$ from Eq. (\ref{4.17}) simplifies to $\!\bfL^{(0)}\lle\bfep\rle(\bfx) $ only when the strain field $\lle\bfep\rle(\bfx)$ is homogeneous, i.e., $\lle\bfep\rle(\bfx)\equiv$cons.

Numerical analysis was conducted for a 1D statistically homogeneous random structure bar \cite{{Buryachenko2023},{Buryachenko2023a}} under a self-equilibrated body force condition (\ref{2.3}), where $\bfb(\bfx) = 0$ for
$|\bfx| > B^b$ and $\bfb(\bfx)=-\bfb(-\bfx)$ for $|\bfx| \leq B^b$; $c^b=0$. The engineering approach, utilizing the scale separation hypothesis (\ref{2.38}$_2$), follows this scheme: first, estimate $\bfL^*$ by CAM (see Comment 3.6 and a generalized EFM, \cite{Buryachenko2022a}) under homogeneous boundary conditions (\ref{2.40}), and then evaluate $\lle\bfu^{\rm EA}\rle(\bfx)$ using Eqs. (\ref{2.1})–(\ref{2.3}) with the substitution
$\bfL\to \bfL^*$. This leads to a monotonically increasing displacement field $\lle\bfu^{\rm EA}\rle(\bfx)$ (represented by the solid curve 4 in Fig. 1). 

\vspace{-1.mm}% \noindent
%\hspace{30mm}
\parbox{8.0cm}{%\hspace{-10mm}
\centering \epsfig{figure=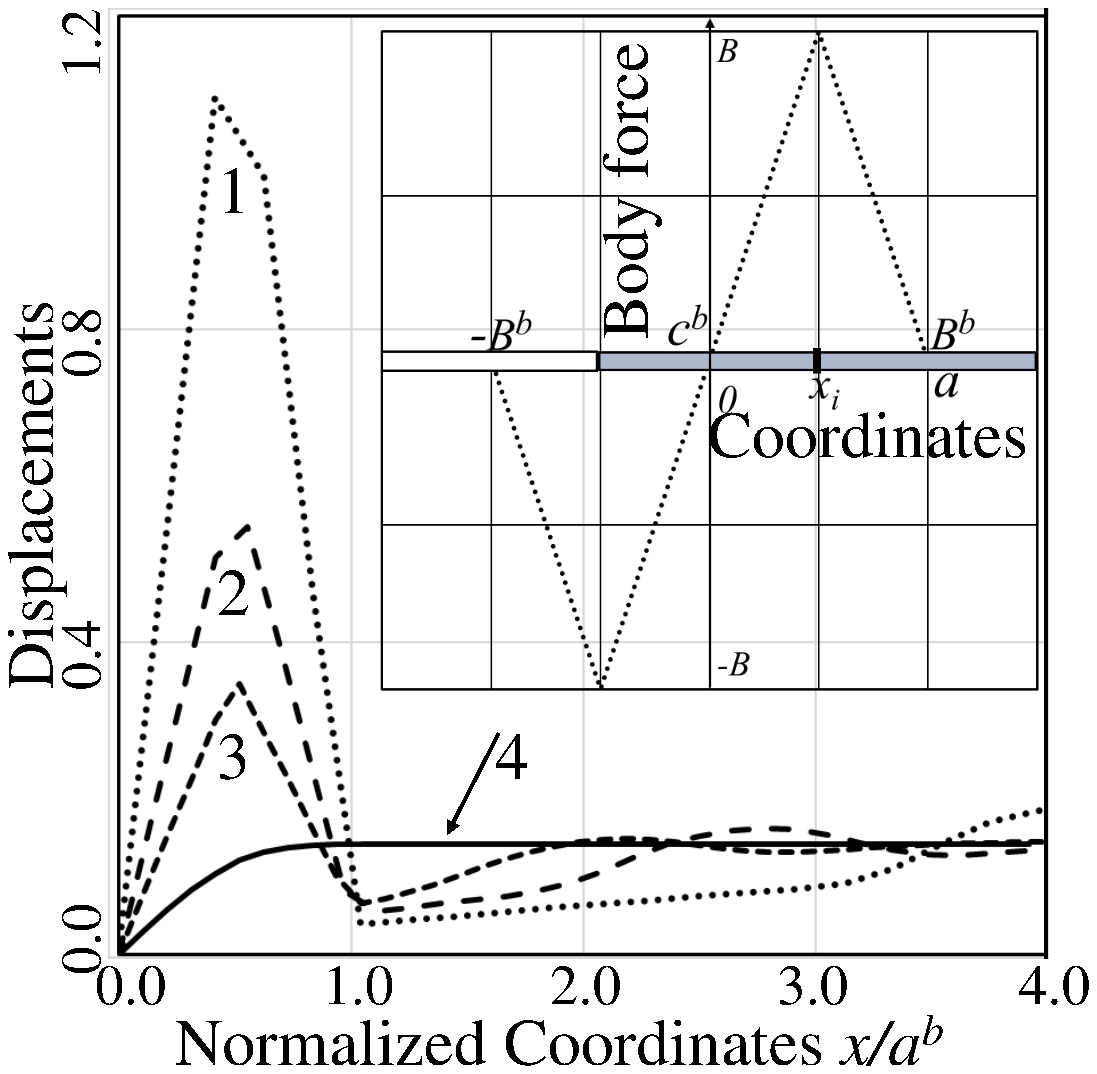, width=7.0cm}      \vspace{-107.mm}
\vspace{80.mm}
%\vspace{-89.mm} 
\tenrm \baselineskip=8pt
{\hspace{1.mm}{\sc Fig. 1} Average displacement $\lle\bfu\rle(\bfx)$ vs $x/a^b$}}
%\vspace{80.mm}

\noindent In Fig. 1, the parameters $l_{\delta}/a=1$ and $c^{(1)}=0.5$ are fixed, while $ B^b/a$ takes values 
of $0.25,\ 0.375,\ 0.5$ (shown in curves 1, 2, and 3, respectively).
As observed in Fig. 1, the simultaneous inclusion of the peridynamic constitutive equation (\ref{2.5}) and the inhomogeneity of the body force $\bfb(\bfx)$ results in strongly nonmonotonic distributions of the displacement field $\lle\bfu\rle(\bfx)$, which differ from $\lle\bfu^{\rm EA}\rle(\bfx)$ by a factor of 9 for curve 1. 
%\noindent 
Due to nonlocal effects, the range of long-range action is restricted not only by $|\bfx|\leq B^b$ , but by a domain $|\bfx|\leq a^{\rm l-r}$ $(a^{\rm l-r}\approx 3 B^b$), meaning that the problem domain
$\bfx\in R^1$ is effectively reduced to a finite domain, avoiding the inconsistencies previously mentioned in \cite{Buryachenko2022b}.
This implies that the domain size $|\bfx|\leq a^{\rm l-r}\approx 3 B^b$ (at the considered scale ratios $a/a^{b}/l_{\delta}$) requires stabilization of the displacement fields $\lle\bfu\rle(\bfx)\approx$const. (for $|\bfx|>a^{\rm l-r}$), which significantly exceeds the combined scale of the body force $ B^b$ and the horizon $l_{\delta}$ (a synergistic effect).
The parameter $a^{\rm l-r}$ should be further learned.
Although the term ``RVE" ( see for details Subsection 8.3) was not directly employed, the domain $|\bfx|\leq a^{\rm l-r}$ in fact represents the RVE for $\lle\bfu\rle(\bfx)$ and this RVE depends on the scale ratios $a/a^{b}/l_{\delta}$.
In generalizing this phenomenon for an arbitrary body force $\bfb(\bfx)$ as given in Eq. (\ref{2.4}), we can define the RVE as the region for which there exists a characteristic size $ B^{\rm RVE}$ such that
\BB
\label{4.21}
\lle\bfu\rle(\bfx)=\bfu^{\infty}\equiv {\rm const}., \ \ {\rm for}\ \ \forall \bfx\geq B^{\rm RVE},
\EE
where this equality holds within a prescribed tolerance.

For the linear bond-based model of constituents, the governing equations (\ref{4.14})–(\ref{4.17}) reduce to the corresponding relations derived in \cite{{Buryachenko2023},{Buryachenko2023a}}.
For statistically homogeneous composite materials with locally elastic constituent phases, the nonlocal effective operator derived in \cite{LucianoW2001} incorporates microstructural information via the Green's function of the reference matrix medium (\ref{2.5}), allowing for a rigorous treatment of the effective response under arbitrary body force distributions. However, the extension of this methodology to PM, where nonlocal interactions play a fundamental role, is not straightforward due to the absence of classical differential operators.
Potential applications of micromechanical modeling of composites subjected to body forces with compact support—such as those induced by localized thermal loading (e.g., laser heating, as discussed in \cite{{Yanget2019},{Isakariet2017}}, and \cite{Yilbas2013})--{\color{black}were analyzed in \cite{Buryachenko2025c} for the LM of CMs. 
When laser-based techniques are employed, the resulting thermal and stress fields typically vary over spatial scales comparable to the microstructural heterogeneities (e.g., particle sizes \cite{Yanget2019}). This fact emphasizes the need to account for nonlocal effects to accurately describe the thermomechanical behavior of even locally elastic CMs under such localized thermal excitations.
The situation becomes even more complex for composites with peridynamic phase properties, which require a detailed consideration of peridynamic thermal expansion and heat transfer phenomena (see, e.g., \cite{OterkusO2024}, \cite{Oterkuset2014}). These aspects, however, lie beyond the scope of the present study.
} 

For an arbitrary body force $\bfb(\bfx)$ with compact support, we have derived formal representations for the effective macroscopic and microscopic fields in a composite material (CM) with a random microstructure. Specifically, we consider the macroscopic displacement
$\lle\bfu\rle(\bfx)$ and stress $\lle\bfsi\rle(\bfx)$, as well as their corresponding microscopic counterparts $\lle\bfu\rle_i(\bfz,\bfx) \lle\bfsi\rle_i(\bfz,\bfx)$ and $\lle\bfsi\rle_i(\bfz,\bfx)$, where the subscript $i$ denotes quantities evaluated within a representative inclusion $v_i$.
For computational implementation and subsequent applications, we construct an  effective dataset of effective parameters for various realizations of the applied body force $\bfb_k(\bfx)$, expressed as ($\bfx\in R^d)$
\BBEQ
\label{4.22}
\!\!\!\!\!\!\!\!\!\!\!{\bfcD}^{\rm r}&\!=\!&\{\bfcD^{\rm r}_k\}_{k=1}^N, \ \ \ {\bfcD}^{\rm r}_k=\{
\lle{\bfu}_k\rle(\bfb_k,\bfx), \lle\bfsi_k\rle(\bfb_k,\bfx), 
\nonumber \\
\!\!\!\!\!\!\!\!\!\!\!&&\!\!\! \lle{\bfu}_{ik}\rle(\bfb_k,\bfz,\bfx),
\!\lle{\bfsi}_{ik}\rle(\bfb_k,\bfz,\bfx), \bfb_k(\bfx)
\}.
\EEEQ
Here, each realization $k$th corresponds to a specific body force $\bfb_k$, for which the macroscopic displacement $\lle{\bfu}_k\rle(\bfx)$ and stress $\lle{\bfsi}_k\rle(\bfx)$ fields are computed, along with the local inclusion-scale displacement
$\lle\bfu_{ik}\rle(\bfz,\bfx):=\lle\bfu_{k}\rle_i(\bfz)$ and stress $\lle{\bfsi}_{ik}\rle(\bfb_k,\bfz,\bfx):=\lle\bfsi_{k}\rle_i(\bfz)$.
The macroscopic coordinate $\bfx\in R^d$ spans the domain of the homogenized CM, while the local coordinate $\bfz\in v_i$ resides within the representative inclusion $v_i$.
This  effective dataset
${\bfcD}^{\rm r}$ forms a basis for data-driven modeling, enabling efficient retrieval and interpolation of effective response functions for new realizations of $\bfb_k$ in a computationally efficient manner.

%\section{Periodic structure CMs}

\section{Periodic structure CMs}

\setcounter{equation}{0}
\renewcommand{\theequation}{5.\arabic{equation}}

{\color{black} In Section 5, a new translation averaging technique is introduced for periodic systems under BFCS loading. This approach yields a new type of effective dataset, which idoes not deliver for further manipulation any detailed information about the microstructure or phase properties. }  

Equations (\ref{3.23}) and (\ref{3.26}) were derived under the general assumption of arbitrary probability densities $\varphi (v_j,{\bf x}_j)$ and the conditional probability density $\varphi (v_j,{\bf x}_j\vert; v_1,{\bf x}_1)$, which characterize the statistical distribution of the microstructural inclusions within the CM. However, in subsequent developments, we specialize these representations to periodic structure CMs, where the probability densities $\varphi (v_j,{\bf x}_j)$ and $\varphi (v_j,{\bf x}_j\vert; v_1,{\bf x}_1)$ take deterministic forms, given explicitly as sums of Dirac delta functions (\ref{2.36}) localized at the periodic grid points $\bfx_{\alpha}\in \bfLa$, corresponding to the structured arrangement of inclusions.

Consequently, the integral relations expressed in Eqs. (\ref{4.11})–(\ref{4.13}) undergo a transformation to a discrete summation form reflecting the periodic nature of the material structure
\BBEQ
\label{5.1}
\!\!\!\!\!\!\!\!\overline{\bfu}(\bfx) \!\!&:=&\!\! \lim_{n\to \infty}\overline{\bfu}^{[n+1]}(\bfx) =\widehat{\bfcD}_i^{ub}(\bfb,\bfx),\\
\label{5.2}
!\!\!\!\!\!\!\!\!{\bfu}(\bfx) \!\!&=&\!\! \widehat{\bfcD}_i^{ub}(\bfb,\bfx) +
\bfcL_i^{uu}(\bfx-\bfx_i, \widehat{\bfcD}_i^{ub}(\bfb,\bfx)),\\
\label{5.3}
\!\!\!\!\!\!\!{\bfsi}(\bfx) \!\!&=&\!\!\bfcL^{\sigma}({\bfu})(\bfx).
\EEEQ
Despite their formal similarity in appearance, these transformed equations (\ref{5.1})–(\ref{5.3}) and the corresponding set of equations (\ref{4.11})–(\ref{4.13}) are conceptually distinct. Equations (\ref{4.11})–(\ref{4.13}) inherently involve ensemble-averaged effective parameters.
Equations (\ref{5.1})–(\ref{5.3}), though similar in structure, pertain specifically to periodic composites and thus rely on a deterministic framework where the microstructure $\bfx_{\alpha}\in \bfLa$ is predefined.

To perform the statistical averaging of Eqs. (\ref{5.1})–(\ref{5.3}), we employ the translated averaging technique (see
\cite{{Buryachenko2023a},{Buryachenko2024a},{Buryachenko2025}} for details),
which is based on shifting the periodic grid $\bfLa_0$ with corresponding inclusion centers by a translation vector $\bfchi$, generating a new grid
$\bfLa_{\bf \chi}$. The body force field $\bfb(\bfx)$ remains fixed throughout this transformation. Mathematically, we consider a % continuous translation of the initial lattice $\bfLa_0$} by the vector
$\bfchi$, leading to a corresponding transformation in the material properties, the characteristic functions of inclusions, and the displacement field $(V_i(\bfx,\bfchi)=V_{i0}(\bfx-\bfchi))$:
\BBEQ
\label{5.4}
\bfC(\bfx,\bfq,\bfchi)&=&\bfC_0(\bfx-\bfchi, \bfq-\bfchi), \ \ \\
\label{5.5}
\bfb(\bfx,\bfchi)&=&\bfb_0(\bfx),\ \ \bfu(\bfx,\bfchi)\not\equiv \bfu_0(\bfx-\bfchi),
\EEEQ
where the inequality (\ref{5.5}$_2$) holds because $\bfb(\bfx)$ is fixed.
For each translation $\bfchi\in \cV_{\rm \bf x}$, the corresponding displacement field $\bfu(\bfx,\bfchi)$ is determined by solving the governing equations (\ref{2.5}) or (\ref{2.6}). This ensemble of solutions allows us to define the macroscopic (or effective) displacement field by averaging over all possible translations $\bfchi$, ensuring a statistical homogenization of the material response across different realizations of the periodic microstructure. The macroscopic displacement field is thus given by:
\BBEQ
\label{5.6}
\!\!\!\!\!\!\!\!\!\!\!\!\!\lle\{\cdot\}\rle(\bfx)\!\!&=&\!\!{1\over\overline
{\cV}_{\bf x}}\int_{{\cal V}_{\rm \bf x}}\{\cdot,\bfchi\}~d\bfchi, \nonumber\\
\lle\{\cdot\}\rle^{l(1)}(\bfx)\!\!&=&\!\! {1\over\overline{\cV}_{\bf x}}\int_{{\cal V}_{\rm \bf x}}\{\cdot,\bfchi\}V_i(\bfx,\bfchi)~d\bfchi.
\EEEQ
The averages in Eqs. (\ref{5.6}) represent ensemble averaging over all possible translated realizations of a periodic microstructure, where translations $\bfchi$ are uniformly distributed over the periodicity cell. This translated averaging applies to periodic CMs with any phase constitutive laws and any inhomogeneous loading.
A special case, similar to (\ref{5.6}$_1$), was introduced in asymptotic homogenization by \cite{SmyshlyaevC2000} and \cite{Ameenet2018} for periodic media under periodic loading. Interestingly, the statistical formulations in Eq. (\ref{5.6}$_2$) resemble a student probability problem: inclusion $v_i$ is randomly ``dropped" onto a fixed point $\bfx\in R^d$. However, in (\ref{5.6}$_2$), the inclusion belongs to a randomly translated periodic grid $\bfLa_{\chi}$.

Applying the translated averaging operation (\ref{5.6}) to Eq. (\ref{5.2}) and Eq. (\ref{4.11}) yields the following expressions
($\bfy\in v^l_q)$
\BBEQ
\label{5.7}
\!\!\!\!\!\!\!\!\!\!\!\!\!\!\!\!\!\!\!\lle\bfu\rle^{l(1)}(\bfX)\!\!&=&\!\!
\bfu^{b(0)}(\bfX)+
{1\over\overline{\cV}_{\bf x}}\int_{{\cal V}_{\rm \bf x}} V(\bfx_q)
\nonumber\\
\!\!\!\!\!\!\!\!\!\!\!\!\!\!\!\!\!\!\!\!&\times&\!\!
\bfcL^{uu}_q(\bfX-\bfx_q,\widehat\bfcD_q^{ub}(\bfb,\bfx_q))
d{\bf x}_q,\\
\label{5.8}
\!\!\!\!\!\!\!\!\!\!\!\!\!\!\!\!\!\!\!\!\!\!\!\!\!\!\!\!\!\!\!\!\!\!\!\!\!\!\!\!\lle\bfu\rle(\bfX)\!&=&\!
\bfu^{b(0)}(\bfX)+ {c^{l(0)}\over\overline{\cV}_{\bf x}}\int_{{\cal V}_{\rm \bf x}} V^{l(0)}(\bfx_q)
\nonumber\\
\!\!\!\!\!\!\!\!\!\!\!\!\!\!\!\!\!\!\!\!\!\!\!\!\!\!&\times&\!\!
\bfcL^{uu}_q(\bfX-\bfx_q,\widehat\bfcD_q^{ub}(\bfb,\bfx_q))
d{\bf x}_q,+{c^{l(1)}\over\overline{\cV}_{\bf x}}\int_{{\cal V}_{\rm \bf x}}\nonumber\\
\!\!\!\!\!\!\!\!\!\!\!\!\!\!\!\!\!\!\!\!\!\!\!\!\!&\times&\! V^{l(1)}(\bfx_q)
\bfcL^{uu}_q(\bfX-\bfx_q,\widehat\bfcD_q^{ub}(\bfb,\bfx_q))
d{\bf x}_q,
\EEEQ
which correspond to Eqs. (\ref{4.15}) and (\ref{4.16}), respectively,
for the general case of the conditional probability density
$\varphi (v_q,{\bf x}_q\vert; \bfX)$.

The averages $\lle\bfu\rle^{l(1)}(\bfX)$ (\ref{5.7}) and $\lle\bfu\rle(\bfX)$ (\ref{5.8})
depend on the macrocoordinates $\bfX$ and are not tied to a specific periodic grid $\bfLa$.
In a similar manner, the effective fields $\lle\bfsi\rle^{l(1)}(\bfX)$ and $\lle\bfsi\rle(\bfX)$ can be found.
Translation averaging of constitutive lau (\ref{3.4}) with linear matrix (\ref{3.1}) will lead to formally identical Eq. (\ref{4.17})
\BB
\label{5.9}
\lle\bfsi\rle (\bfx)=\bfcL^{\sigma(0)}(\lle\bfu\rle)(\bfx)+\lle\bftau\rle(\bfx)
\EE
with
$\lle\bfcL^{\sigma(0)}(\bfu)\rle(\bfx)=\bfcL^{\sigma(0)}(\lle\bfu\rle)(\bfx)$
($\bfx\in \mathbb{R}^d$) (\ref{5.8}).
I.e., the term $\lle\bfcL^{\sigma}(\widetilde{\bftau}^{(0)})\rle(\bfx)$ (\ref{4.13}) is equivalent to the stresses in a pure matrix
$\bfC^{(0)}(\bfx,\hat{\bfx})$ produced by the total displacement $\lle\bfu\rle(\bfx)$ (\ref{5.8}).

For estimation of an average $\lle\bftau\rle(\bfx)=\lle{\bfcL}^{\sigma}(\bfC_1,\bfu)\rle(\bfx)$ (\ref{3.6}), we consider, at first, the fixed inclusion
$v_q$ with the center $\bfx_q$. This inclusion produces value
${\bfcL}^{\sigma}(\bfC_1,\bfu)(\bfx-\bfx_q)$.
Then
{\color{black}a statistical average of the local polarization tensor $\lle\bftau\rle(\bfX)$ is also obtained by averaging over
the domain $v_i^1$
\BBEQ
\label{5.10}
\!\!\!\!\!\!\!\!\!\!\!\!\!\!\!\!\lle\bftau\rle(\bfX)\!\!\!&=&\!\!\!{1\over\overline{\cV}_{\bf x}}\!\!\int_{{\cal V}_{\rm \bf x}}\!\!
V^{l(1)}(\bfy)
\bfcL^{\sigma}(\bfC_1\bfeta^D)(\bfX-\bfy)d\bfy,%
\\
\label{5.11}
\!\!\!\!\!\!\!\!\!\!\!\!\!\!\!\!\! \bfeta^D(\bfx,\bfy)\!\!\!&=&\!\!\! \Big[\widehat\bfcD_i^{ub}(\bfb,\bfx)+\bfcL_i^{uu}(\bfx-\bfx_i,
\widehat\bfcD_i^{ub}(\bfb,\bfx) \nonumber\\
& &\widehat\bfcD_i^{ub}(\bfb,\bfy)-\bfcL_i^{uu}(\bfy-\bfx_i,
\widehat\bfcD_i^{ub}(\bfb,\bfy))\Big],
\EEEQ
where one used the representation for the displacement field in
the inclusion $\bfx\in v_q$: $\bfu(\bfx)$ (\ref{5.2}).}

Analogously to the  effective  dataset $\bfcD^{\rm r}$ (\ref{4.22}) for random structure CMs, we can define a dataset $\bfcD^{\rm p}$ for periodic structure CMs
\BBEQ
\label{5.12}
\!\!\!\!\!\!\!\!\!\!\!{\bfcD}^{\rm p}&\!=\!&\{\bfcD^{\rm p}_k\}_{k=1}^N, \ \ \ {\bfcD}^{\rm p}_k=\{\lle{\bfu}_k\rle(\bfb_k,\bfx),\!\lle\bfsi_k\rle(\bfb_k,\bfx),\nonumber\\
\!\!\!\!\!\!\!\!\!\!\!\!\!\!&\!\!& \lle{\bfu}_{k}\rle^{l(1)}(\bfb_k,\bfx),
\!\lle{\bfsi}_{k}\rle^{(1)}(\bfb_k,\bfx), \bfb_k(\bfx)
\},
\EEEQ
where each effective parameter is computed for a given $\bfb_k$ using Eqs. (\ref{5.6}) during the offline stage for the $k$-th realization. The macrocoordinates $\bfX\in R^d$ define an  effective  dataset $\bfcD^{\rm p}$ (\ref{5.12}), which contains less detailed information than $\bfcD^{\rm r}$ (\ref{4.22}). The latter depends not only on the macrocoordinates $\bfX\in R^d$ but also on the local coordinates $\bfz\in v_i$ within the representative inclusion $v_i$, providing finer-scale resolution.

Thus, the macroscopic stresses in the effective constitutive law (\ref{5.9})
depends on the statistical average $\lle\bfu\rle(\bfX)$ ($\bfx\in R^d$) (\ref{5.8}) and
$\bfu(\bfy)$ ($\bfy\in v^{l}_q$) (\ref{5.2}), respectively. However, these displacement fields
$\lle\bfu\rle(\bfX)$ (\ref{5.8}) and
$\bfu(\bfy)$ ($\bfy\in v^{l}_q$) (\ref{5.2}) are not the prime variables and depend on the body force $\bfb(\bfX)$.
The estimation of effective constitutive law in the next subsection is based on the solution of the AGIE (\ref{3.26}) rather than GIE (\ref{3.27}). %exploited for obtaining of Eq. (\ref{4.16}).

\sffamily
\noindent{\bf Comment 5.1.}
To the best of the author's knowledge, the method of reducing DNS to smoothed (effective) parameters (\ref{5.12}) is novel. Furthermore, a critical aspect in the analysis of periodic structure CMs is the appropriate selection of periodic boundary conditions (PBC) (\ref{2.39}), (\ref{2.40}) and volumetric periodic boundary conditions (VPBC) (\ref{2.37}), (\ref{2.38}) at the interface of unit cells (UCs), which link the field distributions between adjacent UCs.
The established PBC (\ref{2.39}) and VPBC (\ref{2.37}) definitively correspond to both homogeneous remote loading (\ref{2.28}) or (\ref{2.29}) and the zero body force condition $\mathbf{b} (\mathbf{x}) \equiv \mathbf{0}$ (\ref{2.4}). However, for a general body force case (\ref{2.4}), both PBC (\ref{2.39}) and VPBC (\ref{2.37}) become incorrect (see Subsection 2.3). Nevertheless, enforcing any PBC (\ref{2.39}) and VPBC (\ref{2.37}) is unnecessary if DNS is utilized to estimate Eqs. (\ref{5.6}) on a representative volume element (RVE, see Section 7) (possibly consisting of multiple UCs) rather than on a single UC.
The size of the RVE (see Comment 4.13) acts as a controlled parameter that must be determined with a prescribed tolerance. Although the integrands in Eqs. (\ref{5.6}) are computed through DNS within CMic, the combination of Eqs. (\ref{5.6}) with the RVE concept facilitates the construction of the  effective  dataset $\bfcD^{\rm p}$. Consequently, the method (\ref{5.6}) is referred to as CAM's version for periodic structure CMs.

\noindent{\bf Comment 5.2.}
In Sections 3 and 4, we analyzed matrix random structure composites (CMs) with GIE (\ref{3.24}) and AGIE (\ref{3.23}), which apply to the first group, where a continuous matrix phase contains isolated inhomogeneities. For periodic structure CMs, translation averaging (\ref{5.6}) is used without microtopology restrictions. The second group \cite{GibsonA1998} includes skeletal, percolated, or laminated composites, where at least two phases form a monolithic frame, such as metallic foams or cellulose-reinforced polymers.
Cellulose-reinforced polymers
\cite{GilorminiB1999}, metallic wool,
and glass wool have both a network connecting them to the penetrated structure, and
also an incomplete network with only one connection to the network tree.
The third group involves composites where the phases are contiguous but not interconnected, like polycrystalline metals with different crystallographic orientations, forming heterogeneous materials. Adjusting the constituent concentration can move a composite from one group to another. All periodic structure CMs discussed enable the creation of the  effective dataset $\bfcD^{\rm p}$ (\ref{5.12}).

\noindent{\bf Comment 5.3.} The effective  datasets, $\bfcD^{\rm r}$ and $\bfcD^{\rm p}$, corresponding to the scales $B^b$ (\ref{2.3}) and $|\Omega_{00}|$, respectively, are functionally independent of the microstructural details of the composite material. Specifically, they do not encode explicit dependencies on the two-point probability distribution $\varphi(v_q,\bfx_q\vert; v_i,\bfx_i)$ or the grid $\bfLa$, nor do they retain information regarding the methods used to evaluate the reference fields in Eqs. (\ref{4.11})–(\ref{4.19}) and (\ref{5.1})–(\ref{5.11}).
Furthermore, Eq. (\ref{3.24}) is obtained from Eq. (\ref{3.23}) via a centering transformation, thereby making it a particularized form of the latter. However, Eq. (\ref{3.23}) holds a significant computational advantage over Eq. (\ref{3.24}) as it provides the foundation for constructing the  effective dataset $\bfcD^{\rm r}$, which is instrumental in the development of machine learning (ML) and neural network (NN) methodologies—particularly surrogate operators (see Subsection 7). These ML-based techniques facilitate the generation of surrogate operators for arbitrary macroscopic loading conditions, whether specified in terms of the macroscopic displacement field $\lle\bfu\rle(\bfx)$ or the body force distribution $\bfb(\bfx)$. Importantly, this approach circumvents the need to explicitly solve the micro-scale problem dictated by Eq. (\ref{3.24}), thereby significantly reducing computational complexity.
\rmfamily

\section{Estimation of field fluctuations and effective energy-based criteria}

\setcounter{equation}{0}
\renewcommand{\theequation}{6.\arabic{equation}}

{\color {black} Section 6 discusses extensions and applications of AGIE to the field fluctuation estimations used in nonlinear phenomena; this section may be skipped on a first reading.}

In nonlinear LM problems, a key challenge is estimating the second moment of phase fields ($\bfpi=\bfep,\ \bfsi$; $\bfx\in v^{(i)}$):
\BBEQ
\label{6.1}
%\!\!\!\!\!\!\!\!
\!\!\!\!\!\!\!\!\langle \bfpi \otimes \bfpi \rangle _{i}({\bf x}) =\langle \bfpi \rangle _{i}({\bf x})\otimes\langle \bfpi \rangle _{i}({\bf x})
+\Delta^{\pi 2}_i(\bfx),
\EEEQ
where $\Delta^{\pi 2}_i(\bfx)$ is called the dispersion (or the fluctuation field).
Two methods--exact differential analysis and integral equations--yield similar numerical results ( \cite{{Buryachenko2007},{Buryachenko2022a}}). However, using the simplified engineering assumption ⟨$\langle \bfpi \otimes \bfpi \rangle _{i}({\bf x}) =\langle \bfpi \rangle _{i}({\bf x})\otimes\langle \bfpi \rangle _{i}({\bf x})$ can result in infinite errors under hydrostatic loading of porous isotropic media.

The integral equation method in LM allows for its extension to PM. Specifically, by considering only binary inclusion interactions, new approximations for the second moment can be derived:
\BBEQ
\label{6.2}
\!\!\!\!\!\!\!\!\!\!\!\langle \bfthe \!\!&\otimes&\!\! \bfthe \rangle _{i}({\bf
x}) = \langle \bfthe \rangle _i({\bf x})
\otimes \langle \bfthe \rangle _{i}({\bf x})
+\!\!\int\!\! \bfcA^{\theta\theta}_i
\nonumber\\
\!\!&*&\!\!\lle\bfcL_p^{\theta\zeta}(\bfx,\overline{\bfzeta})\vert;v_i,\bfx_i\rle
%\nonumber\\
\otimes \bfcA^{\theta\theta}_i
\nonumber\\
\!\!&*&\!\!\lle \!\bfcL_p^{\theta\zeta}(\bfx,\overline{\bfzeta})
\vert ;v_{i},{\bf x}_{i} \rangle
%\nonumber\\
%\!\!&\times&\!\!
\varphi (v_{p},{\bf x}_p \mid ;v_{i},{\bf x}_{i})d{\bf x}_p.
\EEEQ

\sffamily
\noindent{\bf Comment 6.1.}
The new equation (\ref{6.3}) is derived for nonlinear PM problems (see, e.g.
(\ref{2.18})-(\ref{2.20})) involving the operators $\bfcA^{\theta\theta}_i$ (\ref{3.10}$_2$) and $\bfcL_p^{\theta\zeta}(\bfx,\overline{\bfzeta})$ (\ref{3.10}$_1$). Notably, it exactly matches the equation for a medium with remote BCs (\ref{6.3}), (\ref{3.24}) rather than body force loading (\ref{2.4}), (\ref{3.26}). These equations extend previous results from linear bond-based and state-based models \cite{{Buryachenko2023c},{Buryachenko2023e}} and unify them with existing LM formulations \cite{{Buryachenko2007},{Buryachenko2022a}}. This marks a rare instance where PM methods for second-moment fields (nonlinear problems) inspire new LM techniques for nonlinear analysis. Furthermore, the formulations can be extended, as in \cite{Buryachenko2023e}, to include triple-point probability densities $\varphi (v_{p},{\bf x}_{p},v_q,{\bf x}_{q}|;v_{i},{\bf x}_i)$ for greater accuracy.

\rmfamily

Equation (\ref{6.3}) can be simplified for linear operators $\bfcA^{\theta\theta}_i$ (\ref{3.10}$_2$) and $\bfcL_p^{\theta\zeta}(\bfx,\overline{\bfzeta})$ (\ref{3.10}$_1$) within the EFH framework {\bf H1a} (\ref{3.8}), enabling its application to derive effective energy-based criteria:
\BBEQ
\label{6.3}
\!\!\!\!\!\!\!\!\!\!\langle \bfeta \otimes \bfeta \rangle _{i}({\bfz}) &=& \langle \bfeta \rangle _i({\bf z})
\otimes \langle \bfeta \rangle _{i}({\bf z})+\Delta^{\eta 2}_i(\bfz),
\EEEQ
where
\BBEQ
\label{6.4}
\!\!\!\!\!\! \Delta^{\eta 2}_i(\bfz)&=& \int [\bfbA_{i,j}(\bfz)\lle\bfep\rle]%\nonumber\\
\otimes[\bfbA_{i,j}(\bfz)\lle\bfep\rle]
\nonumber\\
\!\!&\times&\!\!
\varphi (v_{j},{\bf x}_j \mid ;v_{i},{\bf x}_{i})~d\bfx_j,
\\
\label{6.5}
\!\!\!\!\!\! \bfbA_{i,j}(\bfz)&=&\bfA^{\eta\varepsilon}_i(\bfz)[\bfJ^{I\varepsilon\varepsilon}_{i,j}\bfD^0_i
+\bfJ^{J\varepsilon\varepsilon}_{i,j} (\bfy)\bfD^0_j].
\EEEQ
Here $\bfA^{\eta\varepsilon}_i(\bfz)=\underline{\bfA}_i^{u\varepsilon}[\bfx]\lle\bfxi\rle$ ($\bfxi=\hat\bfx-\bfx)$ whereas
$\bfD^0_i$ is the effective field concentration factor $\lle\overline{\bfep}\rle_i=\bfD^0_i\lle\bfep\rle$ (see for details \cite{Buryachenko2023e}).

Equation (\ref{6.3}) serves as a ``elementary block" for developing various effective energy-based criteria. For instance, in the peridynamic approach to fatigue cracking,{Nguyenet2021} estimated the energy release rate by analyzing the micropotentials governing interactions between material points. The state-based formulation of this effective energy-based criterion is as follows.
$( \bfxi,\bfze\in {\cal H}_{\bf x }\subset v_i)$
\BBEQ
\label{6.6}
%\!\!\!\!\!\!\!\!
\!\!\!\!\!\!\!\!\!\!\!\!\!\!\langle \underline\bfU^{\top}\underline\bfbK\underline\bfU \rangle _{i}(\bfxi,\bfze) \!\!&=&\!\!
\langle \underline\bfU^{\top}\rle_i\underline\bfbK\lle\underline\bfU \rangle _{i}(\bfxi,\bfze)
+\Delta^{\rm UKU}(\bfxi,\bfze),
\\
\label{6.7}
\!\!\!\!\!\!\!\!\!\!\!\!\Delta^{\rm UKU}(\bfxi,\bfze)\!\!&=&\!\!\int [ \bfbA_{ij}(\bfxi)\lle\bfep\rle]^{\top}
%\nonumber\\&\times&
\underline\bfbK(\bfxi,\bfze) [ \bfbA_{ij}(\bfze)\lle\bfep\rle]%\nonumber\\
\nonumber\\
\!\!&\times&\!\!
\varphi (v_{j},{\bf x}_j \mid ;v_{i},{\bf x}_{i}) ~d\bfx_j,
\EEEQ
where the statistical average of the displacement state is defined as
$\lle\underline\bfU \rangle _{i}[\bfx]\lle\bfxi\rle =\underline\bfA^{u\epsilon*}_i[\bfx]\lle\bfxi\rle\lle\bfep\rle=
\bfA^{\eta\epsilon*}_i(\bfz)\lle\bfxi\rle\lle\bfep\rle$ (\ref{4.17}); $\bfbK$ is the modulus state (\ref{2.13}) \cite{Sillinget2007}.
The energy-based failure model \cite{Jafarzadehet2022} evaluates the local strain energy density in bond-based peridynamics. From this, an effective criterion for the state-based model can be derived
\BBEQ
\label{6.8}
%\!\!\!\!\!\!\!\!
\!\!\!\!\!\!\!\!\!\!4\lle{\cal W}\rle_i(\bfx)
&\!\!=\!\!&\langle \underline \bfU^{\top}\bullet\underline\bfbK\bullet\underline \bfU\rangle_{i}({\bf
x}) \nonumber\\
\!\!\!\!\!\!\!\!\!\!\!\!\!\!&\!\!=\!\!&\langle \underline \bfU^{\top}\rle_i\bullet\underline\bfbK\bullet\lle\underline \bfU\rangle_{i}({\bf
x})+\Delta^{\rm UKU}(\bfx),\\
\label{6.9}
\!\!\!\!\!\!\!\!\!\!\Delta^{\rm UKU}(\bfx)&=& \!\int\!\! \Big\langle
[ \bfbA_{ij}(\bfxi)\lle\bfep\rle]^{\top}
%\nonumber\\&\times&
\big\langle\underline \bfbK(\bfxi,\bfze) [ \bfbA_{ij}(\bfze)\lle\bfep\rle]\nonumber\\
\!\!\!\!\!\!\!\!\!\!\!\!\!\!\!\!\!\!\!\!&& \big\rangle^{\cal H_{\bf x} }
\Big\rangle^{\cal H_{\bf x} }\!\!(\bfx) %nonumber\\
\varphi (v_{j},{\bf x}_j \mid ;v_{i},{\bf x}_{i}) ~d\bfx_j
\EEEQ
averaging over the horizon region, ${\cal H}_{\bf x}\subset v_i$
($ \bfy\in v_j$).

Seemingly, a second moment of the force vector state $\underline\bfT=\underline\bfbK\bullet\underline\bfU$ (\ref{2.10}), (\ref{2.11}) can be obtained
$( \bfxi, \bfze\in {\cal H}_{\bf x}\subset v_i$)
\BBEQ
\label{6.10}
\!\!\!\!\!\!\!\!\!\!\!\!\!\!\!\!\lle\underline\bfT\otimes\underline\bfT\rle_i^{\cal H_{\bf x} }(\bfx)\!\!&=&\!\!
\Big\langle(\underline\bfbK\bullet\lle\underline\bfU\rle_i)\otimes
(\underline\bfbK\bullet\lle\underline\bfU\rle_i)\Big\rangle^{\cal H_{\bf x} }(\bfx)
\nonumber\\
\!\!&+&\!\!\Delta^{\rm T2}(\bfx),\\
\label{6.11}
\!\!\!\!\!\!\!\!\!\!\!\!\!\!\!\!\!\!\!\!\!\!\!\!\!\!
\Delta^{\rm T2}(\bfx)\!\!&=&\!\! \int\!\! \Big\langle \big\langle\underline\bfbK(\bfxi,\bfze)
[ \bfbA_{ij}(\bfze)\lle\bfep\rle]\big\rangle^{\cal H_{\bf x} }\otimes\big\langle\underline\bfbK(\bfxi,\bfze)\nonumber\\
\!\!\!\!\!\!\!\!\!\!\!\!\!\!\!\!&\times&\!\!\!\!
[ \bfbA_{ij}(\bfze)\lle\bfep\rle]\big\rangle^{\cal H_{\bf x} }
\Big\rangle^{\cal H_{\bf x} }(\bfx)
\nonumber\\ \!\!&\times&\!\!
\varphi (v_{j},{\bf x}_j \mid ;v_{i},{\bf x}_{i}) ~d\bfx_j.
\EEEQ
From Eq. (\ref{6.10}), we can determine $\lle\underline\bfT\cdot\underline\bfT\rle_i^{\cal H_{\bf x} }(\bfx)$
($\bfx\in{\cal H}_{\bf x}\subset v_i$), providing an estimate for the yield function. Specifically, the criterion based on the co-deviatoric force state is given by:
$\psi(\bfx)=\big\langle \underline t^{\rm d}[x]\lle\xi\rle$ $\times\underline t^{\rm d}[x]\lle\xi\rle \rangle^{\cal H_{\bf x} }(\bfx)/2$.
This corresponds to the second invariant of the deviatoric stress tensor in classical local theory, where $\underline t^{\rm d}[x]\lle\xi\rle$ represents the co-deviatoric component of the scalar force state $\underline t$ (see \cite{{Mitchell2011},{Sillinget2007}} for details).

A failure criterion for bond breakage between material points is based on energy dissipation, ${\cal W}_{\bf \xi}(\bfx)$ (see \cite{{Fosteret2011},{SunF2021},{Sunet2020},{WangW2023}}). It depends on the stress state $\underline\bfT=\underline\bfbK\bullet\underline\bfU$ (\ref{2.10}), (\ref{2.11}) and the displacement state $\underline\bfU$ between two points ($\bfxi\in {\cal H}_{\bf x}\subset v_i)$
\BBEQ
\label{6.12}
\!\!\!\!\!\!\!\!\!\!\!\!\!\!\!\!\!\!\!\!{\cal W}_{\bf \xi}(\bfx)&=&\big \langle (\underline\bfbK\bullet \underline\bfU)\rangle_i\langle \underline \bfU\big\rangle_i(\bfx)
\nonumber\\ \!\!&=&\!\!
(\underline\bfbK^{(i)}\bullet \langle \underline\bfU\rangle_i) \langle\underline \bfU\rangle_i(\bfx)+\Delta^{\rm TU2}(\bfxi,\bfx),
\\
\label{6.13}
\!\!\!\!\!\!\!\!\!\!\!\!\!\!\Delta^{\rm TU2}(\bfxi,\bfx)&=\!\!&\int \!\!\big\langle\underline\bfbK(\bfxi,\bfze)
[ \bfbA_{ij}(\bfze)\lle\bfep\rle]\big\rangle^{\cal H_{\bf x} }
\nonumber\\ \!\!&\times&\!\!
[ \bfbA_{ij}(\bfxi)\lle\bfep\rle](\bfx)
%\nonumber\\
%\!\!\!\!\!\!\!\!&\times&
\varphi (v_{j},\!{\bf x}_j \!\mid ;v_{i},{\bf x}_{i}) d\bfx_j.
\EEEQ
Equations (\ref{6.6})-(\ref{6.13}) for the linear operators $\bfcA^{\theta\theta}_i$ (\ref{3.10}$_2$) and $\bfcL_p^{\theta\zeta}(\bfx,\overline{\bfzeta})$ (\ref{3.10}$_1$) simplify to the corresponding equations for both the linear state-based \cite{Buryachenko2023e} and bond-based \cite{Buryachenko2023c} models.

\sffamily
\noindent {\bf Comment 6.2.}
All effective energy-based criteria in (\ref{6.6})-(\ref{6.13}) are derived from a single fundamental component: the second moment of relative displacement $\lle\bfeta\otimes\bfeta\rle_i(\bfz)$ (\ref{6.3}) (or, that is the same, $\lle\underline\bfU[\bfx]\lle\bfxi\rle\otimes\underline\bfU[\bfx]\lle\bfxi\rle\rle_i(\bfx)$ ($\bfx\in v_i$, $\bfxi\in {\cal H}_{\bf x}\subset v_i$)).
These criteria are defined on
either one bond $\bfxi\in {\cal H}_{\bf x}\subset v_i$, or two bonds $
\bfxi,\bfze\in {\cal H}_{\bf x}\subset v_i$ with the field fluctuations, either
$\Delta^{\rm UKU}(\bfxi,\bfze), \Delta^{\rm UKU}(\bfx)$, $\Delta^{\rm T2}(\bfx)$, or
$\Delta^{\rm TU2}(\bfxi, \bfx)$. All the field fluctuations can be incorporated into the effective dataset $\bfcD^{\rm r}$ (\ref{4.22}).
%The estimation process for these criteria using this fundamental component is illustrated in %Fig. 8.

\noindent {\bf Comment 6.3.}
Effective energy-based criteria using field second moments were developed for the linear bond-based \cite{Buryachenko2023e} and state-based \cite{Buryachenko2023c} models, while Eq. (\ref{6.3}) applies to nonlinear constitutive laws \cite{Buryachenko2023k}. A promising approach would be to estimate effective strain energy directly, as in \cite{PCastanedaS1998}, without preliminary computing field second moments.

\rmfamily

\section {CMs with other constitutive laws of phases}

{\color {black} Section 7 discusses extensions and applications of AGIE to CMs with plural constitutive laws. This section may be skipped on a first reading.}

\setcounter{equation}{0}
\renewcommand{\theequation}{7.\arabic{equation}}

\subsection {Locally elastic}

We will consider the local basic equations of thermoelastostatics
(\ref{2.1})-(\ref{2.3}) of composites. Locally elastic counterpart of Eqs. (\ref{3.17}) and (\ref{4.6}) are
\BBEQ
\label{7.1}
\!\!\!\!\!\!\!\!\!\!\!\!\!\!\!\!\!\!\!\!^L\!\bfze(\bfx)-^L\!\overline{\bfze}(\bfx)=^L\!\!\bfcL^{\zeta\zeta}_i(\bfx,^L\!\overline{\bfze}), \
^L\!\bfze(\bfx)=^L\!\!\bfcA^{\zeta\zeta}(^L\!\overline\bfze)(\bfx),
\EEEQ
where we introduce the substitutions $(\bfep,\bfsi)\leftrightarrow \,^L\!\bfze$.
The most straightforward method is reduced to the solution of the integral Eq. (\ref{3.12}) using the Green function $\bfG^{(0)}$
and replacing $\bfep^{w\Gamma}
\rightarrow\, ^L\!\overline{\bfze}(\bfx)$. For a homogeneous ellipsoidal inclusion under a uniform effective field, this yields Eshelby’s solution \cite{Eshelby1957}. When the effective field is inhomogeneous, the multipole expansion method \cite{Kushch2020} is effective.
For an infinite medium, the volume integral equation (VIE) \cite{{Buryachenko2022a},{LeeH2020}} applies, allowing discretization of inclusions only, unlike FEA (which is effectively used at $^L\!\overline{\bfze}(\bfx)\equiv$const.). The boundary integral equation (BIE) methods \cite{{Ballaset1989},{HsiaoW2008},{MukherjeeL2013},{Steinbach2008}} are widely used for homogeneous elasticity but face challenges like 3D meshing and singular integral computation. Alternative meshless methods \cite{Fasshauer2016}, such as the local boundary integral equation, boundary knot
method, boundary collocation method, non-dimensional dynamic influence functions
method, and the method of fundamental solutions (MFS) \cite{KupradzeA1964} (see also \cite{Chenet2008}), offer advantages. In Amic (\ref{3.9}), MFS \cite{Buryachenko2017c} adapted to CAM is most effective. Each method has its strengths and limitations, requiring careful selection based on application.

We also consider a case of an imperfect interface
\BBEQ
\label{7.2}
[[\bfu(\bfx)]]\not={\bf 0},\ \ [[\bfsi(\bfx)]]\cdot\bfn(\bfx)\not={\bf 0},\
\EEEQ
with the jumps (\ref{7.2}$_1$) and (\ref{7.2}$_2$) at the
$\bfx\in \Gamma_i^u$ and $\bfx\in \Gamma_i^{\sigma}$
($\Gamma_i^u, \Gamma_i^{\sigma}\subset\Gamma_i$), respectively, described by the different models of interface imperfections.
As such, on the right-hand side of Eq. (\ref{3.12}) with the volume integral, we need to add the surface integrals
with some Green functions kernels and integrands (\ref{7.2}$_1$) and (\ref{7.2}$_2$) (see for details \cite{Buryachenko2017c}).
The first kind of model can be
referred to as interface models in which the traction is continuous across the interface $\bfx\in \Gamma_i^u$ while the displacement is in general discontinuities (\ref{7.2}$_1$) at $\bfx\in \Gamma_i^u$ such as, e.g., in {\it linear spring model} (LSM, see, e.g.,
\cite{{Hashin1991a},{Hashin1991b},{Hashin2002},{DvorakB1992a},{Huanget1993},{ZhongM1997}}). The LSM was generalized to a {\it cohesive zone model} (CZM),
where the traction vector assumed to be continuous is a non-linear
function of the displacement jump (bilinear, trapezoidal, exponential, and polynomial
cohesive laws). The cohesive model originated by Barenblatt \cite{Barenblatt1962} in fracture mechanics (see also
\cite{{Needleman1990},{OrtizP1999}}) has received wide development in micromechanics of CM (see, e.g.,
\cite{{Othmaniet2011},{Tanet2007a},{Tanet2007b}}).

The second kind of interface model, the Gurtin–Murdoch interface stress model (ISM) \cite{{GurtinM1975},{Gurtinet1998}}
(see also \cite{{Ibach1997},{MarangantiS2007},{Povstenko1993},{Wanget2011}}), also known as the coherent interface model,  is dual to the linear spring-layer model. It enforces displacement continuity across $\bfs\in\Gamma_i^{\sigma}\subset \Gamma_i$ while allowing discontinuities in stress and strain. Steigmann and Ogden \cite{SteigmannO1999} extended ISM to account for surface resistance to stretching and bending by incorporating surface membrane strain and curvature tensors. ISM has been applied to nanoscale structures \cite{{MillerS2000},{Shenoy2002}} and adapted for Eshelby-type inclusion problems \cite{{Chenet2007a},{Chenet2007b},{Duanet2005},{HeL2006},{SharmaG2004},{SharmaW2007}},
see also \cite{{Firoozet2020},{FiroozJ2019}}. All imperfect interface models can be incorporated into the perturbator $^L\!\!\bfcL^{\zeta\zeta}_i(\bfx,^L\!\overline{\bfze})$ (\ref{7.1}) \cite{Buryachenko2017c}.

Replacing the PM perturbator (\ref{3.17}) with the LM perturbator (\ref{7.1}) transforms AGIE (\ref{3.23}) into the LM-based AGIE:
\BBEQ
\label{7.3}
%\!\!\!\!\!\!\!\!\!\!\!\!\!\!\!\!\!\!\!\!
\langle ^L\!{\bfze} \rle_i (\bfx)&=& \bfze^{b(0)} ({\bf x})+
\int {^L\!\bfcL}^{\zeta\zeta}_j(\bfx-\bfx_j,^L\!\overline{\bfze})
\nonumber\\
\!\!\!\!\!\!\!\!\!\!\!\!\!\!&\times&
\varphi (v_j,{\bf x}_j\vert v_1,{\bf x}_1)d{\bf x}_j
\EEEQ
where $\bfze^{b(0)} ({\bf x})$ represents the deterministic field induced by the body force $\bfb(\bfx)$ in an infinite homogeneous matrix.
When the perturbator $^L\!\bfcL^{\zeta\zeta}_j$ is expressed via the Green function (as in Eq. (\ref{3.12})), the left-hand side and integral terms in Eqs. (\ref{3.12}) and (\ref{7.3}) match. However, Eq. (\ref{7.3})—derived for body forces (\ref{2.4})—is exact, whereas Eq. (\ref{3.12}), formulated for remote boundary conditions (\ref{2.30}), is incorrect.

Similarly to Eq. (\ref{7.3}), the second field moment within an inclusion $\bfx\in v_i$ is given by
\BBEQ
\label{7.4}
\!\!\!\!\!\!\!\!\!\!\!\!\!\!\!\!\langle ^L\! \bfze \otimes ^L\!\bfze \rangle _{i}({\bf
x}) &=& \langle ^L\!\bfze \rangle _i({\bf x})
\otimes \langle ^L\!\bfze \rangle _{i}({\bf x})
+\!\!\int\!\! \varphi (v_{p},{\bf x}_p \mid ;v_{i},{\bf x}_{i})
\nonumber\\
\!\!\!\!\!\!\!\!\!\!\!\!\!\!&\times&\!\!^L\!\bfcA^{\zeta\zeta}_i*\lle ^L\!\bfcL_p^{\zeta\zeta}(\bfx,^L\!\overline{\zeta})\vert;v_i,\bfx_i\rle
%\nonumber\\
\otimes ^L\!\bfcA^{\theta\theta}_i
\nonumber\\
\!\!\!\!\!\!\!\!\!\!\!\!\!\!\!\!\!\!&*&\!\!
\lle ^L\! \bfcL_p^{\zeta\zeta}(\bfx,^L\!\overline{\bfzeta})
\vert ;v_{i},{\bf x}_{i} \rangle %\nonumber\\
%\!\!&\times&\!\!
d{\bf x}_p.
\EEEQ
As expected, this equation coincides with its counterpart in LM under remote boundary conditions (\ref{2.30}), as derived in \cite{Buryachenko2022a}.

Solution of Eq. (\ref{7.3}) is performed by repeating step-by-step of solution of Eq. (\ref{3.23}), see Eqs. (\ref{4.5})-(\ref{4.22}). Finally, we obtain a dataset
($\bfx\in R^d)$
\BBEQ
\label{7.5}
\!\!\!\!\!\!\!\!\!\!\!^L\!{\bfcD}^{\rm r}&\!=\!&\{^L\!\bfcD^{\rm r}_k\}_{k=1}^N, \ \ ^L\!{\bfcD}^{\rm r}_k\!=\!\{\lle{\bfep}_k\rle(\bfb_k,\bfx),\!\lle\bfsi_k\rle(\bfb_k,\bfx),\nonumber\\
\!\!\!\!\!\!\!\!\!\!\!\!&&\!\!\!\!\!\!\lle{\bfep}_{ik}\rle(\bfb_k,\bfz,\bfx),
\!\lle{\bfsi}_{ik}\rle(\bfb_k,\bfz,\bfx), \bfb_k(\bfx)
\},
\EEEQ
which looks as the effective dataset ${\bfcD}^{\rm r}$ (\ref{4.22}). Analysis of periodic structure CM is reduced to the replacement of probability densities $\varphi (v_i,{\bf x}_i )$ and $\varphi (v_i,{\bf x}_i\vert; v_j,\bfx_j)$ by their $\delta$ function representations
(\ref{2.36}) that leads to an effective dataset $^L\!{\bfcD}^{\rm p}$ similar to ${\bfcD}^{\rm p}$ (\ref{5.12}).

\subsection {Strongly nonlocal (strain type) model}

Following Rogula \cite{Rogula1982} and Eringen \cite{Eringen2002}, the stress-strain relationship in nonlocal linear thermoelasticity is governed by the integral equation:
\BBEQ
\label{7.6}
\!\!\!\!\!\!\!\!\!\!\!\!\!\bfsi(\bfx)=\bfcL*\bfep(\bfx)+\bfal(\bfx),\ \bfep(\bfx)=\bfcM*\bfsi(\bfx)+\bfbe(\bfx)
\EEEQ
where the integral response operators $\bfcK=\bfcL, \bfcM; \ \bfga=\bfal,\bfbe$) and transformation field $\bfga=\bfal,\bfbe$
are defined as
\BB
\label{7.7}
\bfcK*(\cdot)=\!\!\int \!\!\bfcK(\bfx,\bfy)(\cdot)(\bfy)d\bfy,\ \bfga(\bfx)=\!\!\int\!\!\bfm^{\gamma}(\bfx,\bfy)\Delta T(\bfy)d\bfy. %(2.9)
\EE
Here, $\bfcK(\bfx,\bfy)$ is the kernel governing nonlocal interactions, while $\bfga(\bfx)$ represents the transformation field, assuming a uniform temperature difference $\Delta T(\bfy)\equiv (T-T^0)$. This formulation modifies only the stress-strain relation, keeping equilibrium and compatibility equations unchanged, as in linear isotropic elasticity \cite{{BazantJ2002},{EdelenL1971},{Eringen1999},{Eringen2002},{Kroner1967}}.

The response and inverse response parameters in Eq. (\ref{7.6}) satisfy the relationships:
\BB
\label{7.8}
\bfcL\!*\!\bfcM=\bfcM\!*\!\bfcL=\bfI, \ \ \ \bfcM\!*\!\bfal=-\bfbe,\ \ \ \bfcL\!*\!\bfbe=-\bfal,
\EE
with material parameters subjected to the following symmetry regulations:
$\bfcK=\bfcL, \bfcM; \ \bfga=\bfal,\bfbe$):
${\cal K}_{ijkl}={\cal K}_{ijlk}={\cal K}_{jikl}$, ${\cal K}_{ijkl}(\bfx,\bfy)={\cal K}_{klij}(\bfy,\bfx)$, $\gamma_{ij}=\gamma_{ji}$ \cite{{Kunin1967},{Kroner1970}}.
The kernel $\bfcK(\bfx,\bfy)$ is given by:
\BB
\bfcK(\bfx,\bfy) = \bfK(\bfx) \lambda(\bfx,\bfy), \label{7.9}
\EE
where $\lambda(\bfx,\bfy)$ is a nonlocal weight function, often distance-dependent $\lambda(\bfx,\bfy)=\lambda_{\infty}(|\bfx-\bfy|)$ \cite{Polizzotto2001}, and normalized as:
\BB
\int_{w}\lambda(\bfx,\bfy)d\bfy=1, \ \ \forall \bfx\in w= R^d \label{7.10}
\EE
This ensures $\bfcK$ represents stiffness or compliance in uniform straining and stressing. In the limit $\lambda_{\infty}(\bfx)\to \delta(\bfx)$, the nonlocal operator reduces to the local case $\bfcK(\bfx,\bfy)\to \bfK(\bfx)\delta(\bfx-\bfy)$, recovering the classical formulation (\ref{2.2}).

Micromechanical models begin with analyzing a single heterogeneity in an infinite matrix under remote loading \cite{KumasakaH1996} (see comprehensive reviews in \cite{{Buryachenko2011c},{Gutkin2006},{MarangantiS2007}}).
For various nonlocal models, involving spatial integrals or field gradients, the internal field within a homogeneous ellipsoidal inclusion remains non-uniform even under homogeneous remote loading. The strongly nonlocal strain-based counterparts of Eqs. (\ref{3.17}) and (\ref{4.6}) are:
\BBEQ
\label{7.11}
\!\!\!\!\!\!\!\!\!\!\!\!\!\!^S\!\bfze(\bfx)-^S\!\overline{\bfze}(\bfx)=^S\!\!\bfcL^{\zeta\zeta}_i(\bfx,^S\!\overline{\bfze}), \
^S\!\bfze(\bfx)=^S\!\!\bfcA^{\zeta\zeta}(^S\!\overline\bfze)(\bfx),
\EEEQ
where $(\bfep,\bfsi)\leftrightarrow \,^S\!\bfze$. Unlike MEF and MTM (see Comment 3.6), which focus on perturbators inside the inclusion
[see e.g. \cite{{SharmaD2002},{Xunet2004},{ZhangS2005}}], CAM requires evaluating $^S\!\bfcL^{\zeta\zeta}_i$
outside $\bfx\not \in v_i$ (see \cite{Buryachenko2024b}). This was addressed in \cite{Buryachenko2011c} using an iterative solution of the volume integral equation.

Substituting the PM perturbator (\ref{3.17}) with the strongly nonlocal strain-based perturbator (\ref{7.11}) modifies AGIE (\ref{3.23}) into the following nonlocal AGIE formulation:
\BBEQ
\label{7.12}
\!\!\!\!\!\!\!\langle ^S\! \bfze \rle_i (\bfx)&=& \bfze^{b(0)} ({\bf x})+
\!\!\int \!\!{^S\!\bfcL}^{\zeta\zeta}_j(\bfx-\bfx_j,^S\!\overline{\bfze})\nonumber\\
\!\!&\times&\!\!\varphi (v_j,{\bf x}_j\vert v_1,{\bf x}_1)d{\bf x}_j
\EEEQ
where $\bfze^{b(0)} ({\bf x})$ denotes the deterministic field generated by the body force $\bfb(\bfx)$ in an infinite homogeneous matrix.
By centering Eq. (\ref{7.12}), we obtain a GIE analogous to (\ref{3.24}), which was applied in \cite{Buryachenko2011c} to study CMs reinforced with circular inclusions. It was shown that the local statistical average stress $\lle\bfsi\rle_i(\bfx)$ ($\bfx\in v_i$) depends on both the radial distribution functions (RDFs) and the excluded volume $v_i^0$, potentially leading to stress values that differ in sign from those predicted by classical methods such as the MEF and MTM. However, the differences in the effective moduli $\bfL^*$ estimated using GIEs—nonlocal counterparts to Eqs. (\ref{3.13}) and (\ref{3.14})—were found to be relatively minor.

To estimate the second field moment $\langle ^S\! \bfze \otimes ^S\!\bfze \rangle _{i}({\bf
x}) $ within an inclusion $\bfx\in v_i$, one simply needs to replace the superscript $^L$ with $^S$ in Eq. (\ref{7.3}). Likewise, the dataset $^D\!\bfcD$ can be derived from Eq. (\ref{7.5}) by substituting the superscript $^L$ with $^S$.

\subsection {Coupled problems of composites}

We formulate the Additive GIEs for coupled problems using thermoelectroelasticity as an example. To maintain consistency, elastic and electric variables are treated equivalently. Following the notation from \cite{BarnettL1975} and referencing works like \cite{{Maugin1988},{PartonK1988},{KanaunL2008}}, the local linear constitutive relations are recast in a unified form, leading to a basic equation analogous to Eqs. (\ref{2.1})–(\ref{2.3}).
\BBEQ
\label{7.13}
\bfcD\bfSi &=& \bfzir,\ \bfcE=\bfcD\bfcU,
\\
\label{7.14}
\bfcE&=& \bfbM\bfSi+\bfLa,\ \
\bfSi=\bfbL(\bfcE-\bfLa),
\EEEQ
where
\BBEQ
\!\!\!\!\!\!\!\!\!\!\!\!\bfcE&=&
\left\|\begin{array}{c}
\bfep\\
\bfE
\end{array}\right\|,\
\bfSi=
\left\|\begin{array}{c}
\bfsi\\
\bfD
\end{array}\right\|,\
\bfcU=
\left\|\begin{array}{c}
\bfu\\
\phi
\end{array}\right\|,\
%\label{7.15}
\nonumber
\\
\bfbL&=&
\left\| \begin{array}{cc}
\bfL\ \ \ \ \bfe^{\top}\\
\bfe \ \ -\bfk
\end{array} \right\|,\ \
\ \ \bfbM=
\left\| \begin{array}{cc}
\bfM\ \ \ \ \bfd^{\top}\\
\bfd \ \ -\bfb
\end{array} \right\|,\ \
\label{7.15}
\\
\bfcD&=&
\left\| \begin{array}{cc}
{\rm def} \ \ \ \ 0 \\
0 \ \ {\rm grad}
\end{array} \right\|,
% \bfLal=
% \left\| \begin{array}{c}
% \bfal^{\top}\\
% \bfp\theta
% \end{array} \right\|, \ \
\bfLa=
\left\| \begin{array}{c}
\bfbe\\
\bfq\theta
\end{array} \right\|, \ \
\label{7.16}
\EEEQ
here $\bfD$ and $\bfE$ are the vectors of induction and
electric field intensity, $\theta$ is a deviation of
a stationary temperature field from a given value,
$\bfk$ and $\bfb$ are the tensors of dielectric permeability and impermeability,
$\bfq$ is the pyroelectric coefficient,
$\bfe$ and $\bfd$ are the piezoelectric moduli, and $\phi$ is the electric potential.
To obtain a symmetric matrix of coefficients, we replaced the electric
field $\bfE$ by $-\bfE$, and the tensors $\bfk$ and $\bfb$ by $-\bfk$ and $-\bfb$
on the right-hand sides of (\ref{7.14}). It is assumed that the properties of both the comparison medium and the
matrix coincides. Some background representations for micromechanics of thermoelectroelasticity such as generalized Hill's conditions, effective energy functions, and phase-averaged first and second moments are considered in \cite{Buryachenko2007} (see also \cite{QinY2008}).

These and other methods with the classical background defined by both the hypothesis {\bf H1} and GIEs counterpart of (\ref{3.13}) (see, e.g. \cite{{DinzartS2017}, {Levinet2011}} and referenced in \cite{Buryachenko2013})
are based on the estimation of the average field and polarization tensors inside the heterogeneities rather than outside ones.
Exploiting the new background (\ref{3.26}) and (\ref{3.27}) is accomplished by a straightforward generalization of the approaches of Subsections 4.1 and 4.2 taking
the unified notations (\ref{7.13})-(\ref{7.16}) into account.

The coupled electroelasticity counterparts of Eqs. (\ref{3.17}) and (\ref{4.6}) are given by $(\bfx\in R^d$):
\BBEQ
\label{7.17}
\!\!\!\! \!\!\!\! \!\! \!\! \!\! \bfcE(\bfx)-\overline{\bfcE}(\bfx)=^{EE}\!\!\!{\bfcL}^{EE}_i(\bfx,\overline{\bfcE}), \
\bfcE(\bfx)=^{EE}\!\!\!{\bfcA}^{EE}(\overline\bfcE)(\bfx),
\EEEQ
Unlike classical methods such as MEF and MTM, which rely on hypothesis {\bf H1a} (\ref{3.8}) and the GIE counterpart of (\ref{3.13}) (see \cite{{DinzartS2017},{DinzartS2018},{DunnT1993a},{DunnT1993b},{Dumontet2020},{Guet2015},{Li2000b},{Li2004},{LiD1999},{KhoroshunD2004},{ShermergorY1993}}), the CAM approach (both Eqs. (\ref{3.23}) and (\ref{3.24})) requires evaluating $^{EE}\!\!\!{\bfcL}^{EE}_i$ (\ref{7.17}) outside the inclusion region ($\bfx \not\in v_i$), rather than just inside ($\bfx \in v_i$). For inclusions of arbitrary shape and non-uniform effective fields ($\overline{\bfcE}(\bfx) \not\equiv$ const.), numerical methods mentioned in Subsections
4.1 and 4.2 must also be adapted accordingly.

Replacing the PM perturbator (\ref{3.17}) with the coupled electroelasticity perturbator (\ref{7.17}) transforms AGIE (\ref{3.23}) into:
\BBEQ
\label{7.18}
\langle \bfcE\rle_i (\bfx)\!\! &=&\!\! \bfcE^{b(0)} ({\bf x})+
\int {^{EE}}\!\!\bfcL^{EE}_j(\bfx-\bfx_j,\overline{\bfcE})\nonumber \\
\!\! &\times&\!\!
\varphi (v_j,{\bf x}_j\vert v_1,{\bf x}_1)d{\bf x}_j.
\EEEQ
Here, $\bfcE^{b(0)} ({\bf x})$ represents the deterministic field produced by the coupled body force $\bfb(\bfx)$ in an infinite homogeneous matrix.
By centering Eq. (\ref{7.18}), we derive a GIE analogous to (\ref{3.24}), which was introduced in \cite{Buryachenko2022a} for analyzing random-structured piezoelectric composite materials.

To estimate the second field moment $\langle \bfcE \otimes \bfcE\rangle _{i}({\bf x})$ within an inclusion $\bfx \in v_i$, one only needs to replace the superscript $^L$ with $^{EE}$ in Eq. (\ref{7.3}). Similarly, the effective dataset $^{EE}\!\bfcD^{\rm r}$ can be obtained in the same way as $\bfcD^{\rm r}$ (\ref{7.5}).

\subsection{ First Strain Gradient Medium}

Subsections 6.3 and 6.4 provide an excerpt from \cite{Buryachenko2022a}.

In the linearized theory of Mindlin’s form-II first strain gradient elasticity ( \cite{{Mindlin1964},{MindlinE1968}}, see also \cite{{Lazaret2020},{Poet2019},{Solyaevet2020}}), the strain energy density of a homogeneous centrosymmetric material depends quadratically on both the strain $\bfep$ and its gradient $\bfka = \nabla\bfep$:
\BBEQ
\label{7.19}
\!\!\!\!\!\!\!\!\!\!\!\!\!\!W(\bfep,\nabla\bfep)\!\! &=&\!\! {1\over 2}\bfep:\bfL:\bfep+{1\over 2}\bfka\therefore\bfC\therefore\bfka\nonumber\\
\!\!\!\!\!\!\!\!\!\!\!\!\!\! &=&\!\! {1\over 2}L_{ijkl}\varepsilon_{ij}\varepsilon_{kl}+{1\over 2}C_{ijmkln}\partial_m\varepsilon_{ij}\partial_n\varepsilon_{kl},
\EEEQ
where $\bfL$ is the rank-four elasticity tensor, and $\bfC$ is the rank-six strain gradient elasticity tensor, both possessing specific symmetries ( \cite{Buryachenko2022a}). The operator $\therefore$ denotes contraction over the last three indices of the left operand and the first three of the right.

Using the Euler–Lagrange equations, the corresponding static governing equation for displacement takes the Navier-like form:
\BBEQ
\label{7.20}
\!\!\!\bfbL^{\rm M}\bfu\!\! &=&\!\!-\bfb, \ \ {\bfbL}^{\rm M}={\bfbL}-{\bfbC}, \ \ {\bfbL}_{ik}=L _{ijkl}\partial_j\partial_l,
\nonumber\\
\!\!\! {\bfbC}_{ik}\!\! &=&\!\!C_{ijmklm}
\partial_j\partial_l\partial_m\partial_n.
\EEEQ
Thus, the Mindlin operator $\bfbL^{\rm M}$ incorporates both standard elasticity and strain gradient effects, ensuring a symmetric total stress tensor in the first strain-gradient model \cite{AskesA2011}.
Some well-known simplified strain gradient theories, such as the couple stress theory,
Aifantis, Kleinert, Mindlin, and Wei \& Hutchinson’s models \cite{{Aifantis2019},{AltanA1997},{AskesA2011},{Kleinert1989},{Lamet2003},{Mindlin1964},{Zhouet2016}} have similar forms of governing equations as that of the general isotropic second gradient theory (see their connection in \cite{Buryachenko2022a}).

Consider an inhomogeneity $v_i$ centered at $\bfx_i$with material parameters$\bfL^{(1)}(\bfx)$ and $\bfC^{(1)}(\bfx)$ ({\ref{7.18}) embedded in an infinite homogeneous matrix $w$ characterized by $\bfL^{(0)}$ and $\bfC^{(0)}$. Under generalized traction-free remote boundary conditions and body force $\bfb(\bfx)$ ({\ref{2.4}), the resulting fields in the homogeneous matrix (without $v_i$) are $\overline {\bfep}^b(\bfx)$ and $\overline{\bfka}^b(\bfx)$

($\bfx\in w$). The equilibrium equation takes the form:
\BB
\label{7.21}
\bfbL^{\rm M (0)}\bfu=-\bfbL^{\rm M}_1\bfu-\bfb,
\EE
where $\bfL_1(\bfx):=\bfL(\bfx)-\bfL^{(0)}$ and $\bfC_1(\bfx):=\bfC(\bfx)-\bfC^{(0)}$ vanish outside $v_i$.

Following \cite{SmyshlyaevF1994}, we define the stress and double-stress polarization tensors:
\BBEQ
\label{7.22}
\bfta(\bfx)\!\! &=&\!\!\bfsi(\bfx)-\bfL^{(0)}:\bfep(\bfx), \nonumber\\
\bfpi(\bfx)\!\! &=&\!\!\bfmu(\bfx)-\bfC^{(0)}\therefore\bfka(\bfx),
\EEEQ
which are zero outside $v_i$. Substituting these into the equilibrium equation ({\ref{7.21}) results in:
\BB
\label{7.23}
\bfbL^{\rm M (0)}\bfu=-\nabla[\bfta-\nabla\bfpi].
\EE
For convenience, we introduce the notation:
\BB
\label{7.24}
\hat{\bfta}=\begin{pmatrix}
\bfta \\
\bfpi
\end{pmatrix},\ \ \
\hat{\bfep}=\begin{pmatrix}
\bfep \\
\bfka
\end{pmatrix},
\ \ \
\hat{\overline{\bfep}}=\begin{pmatrix}
\overline{\bfep} \\
\overline{\bfka}
\end{pmatrix},\ \ \ \hat{\bfU}=\begin{pmatrix}
\bfU^{\epsilon\tau}&\bfU^{\epsilon\pi } \\
\bfU^{\kappa \tau}&\bfU^{\kappa \pi}
\end{pmatrix},
\EE
Here, $\hat{\overline{\bfep}}=(\overline{\bfep},
\overline{\bfka})^{\top}$ represents the field in a homogeneous medium with $\bfL^{(0)}$, $\bfC^{(0)}$ under given BC (like ({\ref{2.4})). The Green functions and their derivatives, $\bfU^{\epsilon\tau}=\nabla\nabla\bfG^{\rm M},$ $\bfU^{\epsilon\pi }=-\nabla\nabla\nabla\bfG^{\rm M}$,
$\bfU^{\kappa \tau}=\nabla \nabla\nabla\bfG^{\rm M}$, $\bfU^{\kappa \tau}=$ $-\nabla\nabla\nabla\nabla\bfG^{\rm M}$, relevant for both isotropic and anisotropic media, are detailed in \cite{LazarP2018} and \cite{Poet2019}.

By applying the Green function ({\ref{7.24}) and integrating by parts, the equilibrium equation ({\ref{7.24}) is transformed into the convolution form
\BB
\label{7.25}
\hat{\bfep}=\hat{\overline{\bfep}}+\hat{\bfU}*\hat{\bfta},
\EE
which corresponds to Eq. ({\ref{3.11}) up to notation differences.

The integral equation ({\ref{7.25}) with a nonsingular convolution kernel $\hat{\bfU}$ can be rewritten in operator form as
\BBEQ
\label{7.26}
\!\!\!\!\!\!\!\!\!\!\!\!\hat{\bfep}(\bfx)-\hat{\overline{\bfep}}(\bfx)&:=&^{\epsilon}\!\bfcL_i(\bfx-\bfx_i,\hat{\bfta})=
^{\epsilon}\!\!\bfcL_i(\bfx-\bfx_i,\hat{\overline{\bfep}}), \nonumber\\
\!\!\!\!\!\!\!\!\!\!\!\!\!\!\hat{\bfep}(\bfx)&=&^{\epsilon}\!\bfcA_i(\bfx-\bfx_i,\hat{\overline{\bfep}}),
\EEEQ
Particular case of homogeneous effective field $\hat{\overline{\bfep}}(\bfx)\equiv$const. and homogeneous ellipsoidal inclusion corresponds to the Eshelby solution considered in \cite{{GaoM2010},{Maet2018},{ZhangS2005},{ZhengZ2004}}.
In the general case of the inclusion shape and inhomogeneity of the effective field $\hat{\overline{\bfep}}(\bfx)\not \equiv$const.,
the solutions $^{\epsilon}\!\bfcL_i(\bfx-\bfx_i,\hat{\bfta})$ and
$^{\epsilon}\bfcA_i(\bfx-\bfx_i,\hat{\overline{\bfep}})$ can be obtained via the volume integral equation method (Subsection 6.1) or finite element analysis (e.g.,\cite{{AskesA2011},{DorganV2006},{VoyiadjisS2020}}).

Replacing the PM perturbator (\ref{3.17}) with the strain gradient perturbator (\ref{7.26}) transforms AGIE (\ref{3.23}) into:
\BBEQ
\label{7.27}
\langle \hat{\overline\bfep}\rle_i (\bfx)\!\!&=&\!\! \hat\bfep^{b(0)} ({\bf x})+
\int {^{\epsilon}}\! \bfcL_j(\bfx-\bfx_j,\hat{\overline\bfep})\nonumber\\
\!\!&\times&\!\!\varphi (v_j,{\bf x}_j\vert v_1,{\bf x}_1)d{\bf x}_j.
\EEEQ
Here, $\hat\bfep^{b(0)} ({\bf x})$ represents the deterministic field produced by the body force $\bfb(\bfx)$ in an infinite homogeneous matrix.
By centering Eq. (\ref{7.27}), we derive a GIE analogous to (\ref{3.24}), which was introduced in \cite{Buryachenko2022a} for analyzing random-structured CMs with strain gradient properties of phases.

To estimate the second field moment $\langle \hat{\bfep} \otimes \hat{\bfep}\rangle _{i}({\bf x})$ within an inclusion $\bfx \in v_i$, one only needs to replace the superscript $^L$ with $^{\epsilon}$ in Eq. (\ref{7.4}). Similarly, the n effective dataset $^{\epsilon}\bfcD^{\rm r}$ can be obtained in the same way as $\bfcD^{\rm r}$ (\ref{7.5}).

\subsection {Stress-Gradient Elasticity Model}}

In this subsection, we briefly summarize the stress-gradient elasticity framework, introduced by Forest and Sab \cite{ForestS2012} and further developed in \cite{{Brisard2017},{Sabet2016},{Tranet2018}}, adapted for obtaining AGIE (as discussed in Section 3 for PM). In the strain-gradient model (\ref{7.19}), the strain-energy density $W$ depends on the strain $ \bfep $ and its gradient $\nabla\bfep$. Similarly, in the stress-gradient model of Forest and Sab \cite{ForestS2012} (see also \cite{Buryachenko2022a}), the complementary strain-energy density
$W^c$
of a linear homogeneous centrosymmetric material
depends on the stress $\bfsi$
and its first gradient $\nabla\bfsi$
\BB
\label{7.28}
W^c(\bfsi,\bfR)={1\over 2}\bfsi:\bfM:\bfsi+{1\over 2}\bfR\therefore\bfS\therefore\bfR,
\EE
where $\bfM$ and $\bfS$ are the classical and generalized compliances, which are tensors with minor
and major symmetries. The trace-free part $\bfR$ of the stress-gradient $\nabla \bfsi$ is the orthogonal projection of $\nabla \bfsi$
onto the space of third-rank, trace-free tensors, and we write $\bfR=\bfI'_6\therefore\nabla\bfsi $, where the sixth-rank
tensor $I'_{6|ijkpqr}=I_{ijpq}\delta_{kr}-(I_{pq(ir}\delta_{j)k})/2$ \cite{Tranet2018} is defined as the orthogonal projection (in the sense of the $\therefore$ scalar product).
Being a projector, $\bfI'_p$ enjoys the classical property $\bfI'_p\therefore \bfI'_p=\bfI'_p$.

The strain measures, the energy-conjugate to the stress $\bfsi$ and trace-free variable $\bfR$, are
\BB
\label{7.29}
\bfe=\bfM:\bfsi={\partial W^c\over\partial \bfsi}, \ \ \ \bfphi=\bfS\therefore\bfR= {\partial W^c\over\partial \bfR},
\EE
where the total strain $\bfe$ is not necessarily the symmetric gradient of the displacement $\bfu$.

Equations (\ref{7.28}) can be considered as an inversion of Eq. (\ref{7.29})
\BB
\label{7.30}
\bfsi=\bfL:\bfe, \ \ \ \bfR=\bfC\therefore\bfphi,
\EE
where $\bfL=(\bf M)^{-1}$ is a classical stiffness whereas the generalized compliance $\bfS$ and stiffness $\bfC$ are inverse to each other
$\bfC\therefore\bfS=\bfS\therefore\bfC=\bfI'_6$.

Analogously to ({\ref{7.22}), we introduce the stress and stress gradient polarization tensors
\BBEQ
\label{7.31}
\bfta(\bfx)\!\!&=&\!\!\bfsi(\bfx)-\bfL^{(0)}:\bfe(\bfx), \nonumber\\
\bfpi(\bfx)\!\!&=&\!\!\bfR(\bfx)-\bfC^{(0)}\therefore\bfphi(\bfx),
\EEEQ
vanishing in the matrix $\bfx\in w\setminus v_i$. Similarly to Eq. ({\ref{7.25}) (see also \cite{Tranet2018}), we get an equation for one inhomogeneity in the infinite matrix
\BBEQ
\label{7.32}
\hat{\bfe}=\hat{\overline{\bfe}}+^{\bf e}\!\hat{\bfU}*\hat{\bfta},
\EEEQ
where
\BB
\label{7.33}
\hat{\bfta}=\begin{pmatrix}
\bfta \\
\bfpi
\end{pmatrix},\ \ \
\hat{\bfe}=\begin{pmatrix}
\bfe \\
\bfphi
\end{pmatrix},
\ \ \
\hat{\overline{\bfe}}=\begin{pmatrix}
\overline{\bfe} \\
\overline{\bfphi}
\end{pmatrix},\ \ \ ^{\bf e}\!\hat{\bfU}=\begin{pmatrix}
^{\bf e}\!\bfU^{e \tau}&^{\bf e}\!\bfU^{e \pi} \\
^{\bf e}\!\bfU^{\phi\tau}&^{\bf e}\!\bfU^{\phi\pi}
\end{pmatrix},
\EE
where $\hat{\overline{\bfe}}=(\overline{\bfe}, \overline{\bfpi})$ is a field that would exist in the medium with homogeneous properties $\bfM^{(0)}$, $\bfS^{(0)}$, and
appropriate remote boundary condition. Green operators $^{\bf e}\!\bfU^{e \pi}$,
$^{\bf e}\!\bfU^{\phi\tau}$, and $^{\bf e}\!\bfU^{\phi\pi}$ can be derived from the single operator $^{\bf e}\!{\bfU}^{e \tau}$,
as obtained by Tran \cite{Tran2016} for an isotropic simplified model.

The integral Eq. ({\ref{7.32}) with nonsingular convolution kernel $\hat{\bfU}$ and their solution can be presented in an operator form
({\ref{3.17}) and ({\ref{4.6})
\BBEQ
\label{7.34}
\!\!\!\!\!\!\!\!\!\!\hat{\bfe}(\bfx)-\hat{\overline{\bfe}}(\bfx)&:=&^{\bf e}\!\bfcL^{\tau}_i(\bfx-\bfx_i,\hat{\bfta})=
^{\bf e}\!\!\bfcL^{e}_i(\bfx-\bfx_i,\hat{\overline{\bfe}}), \nonumber\\
\!\!\!\!\!\!\!\!\!\!\!\!\bar v_i\hat{\bfta}(\bfx)&=&^{\bf e}\!\bfcR^{e}_i(\bfx-\bfx_i,\hat{\overline{\bfe}}),
\EEEQ
where the solutions $^{\bf e}\!\bfcL^{e}_i(\bfx-\bfx_i,\hat{\bfe})$ and $^{\bf e}\!\bfcR^{e}_i(\bfx-\bfx_i,\hat{\overline{\bfe}})$ can be found, e.g. either by the volume integral equation method (see Subsection 7.1)
by finite element analysis (see, e.g., \cite{{AskesA2011},{DorganV2006},{VoyiadjisS2020}}).

\section {Representative volume element (EVE)}
{\color{black} The second major achievement is the introduction of a new {\it RVE} concept, emerging from the AGIE framework (involving BFCS loading), which is developed in Subsections 8.3 and 8.4 and employed for constructing novel {\it effective datasets}. The new RVE concept is fundamentally differ from classical RVE considered in Subsections 8.1 and 8.2 introduce through the passing to the limit of the sample. A summary of AGIE applications in both analytical and computational micromechanics is provided in Subsection 8.5. The discussion of FFT-based solution methods for AGIE in Subsection 8.6 may be skipped on a first reading.}

\subsection {RVE for CMs subjected to remote homogeneous loading}
\setcounter{equation}{0}
\renewcommand{\theequation}{8.\arabic{equation}}
\rmfamily
The Representative Volume Element (RVE), introduced by Hill \cite{Hill1963}, has a complex history (see \cite{Ostojaet2016}). To accurately convey its definition and significance, we cite Hill \cite{Hill1963}, which provides its rigorous foundation.

\noindent{\bf Definition 8.1.} {\it Representative volume element (RVE)
(a) is structurally entirely typical of the whole mixture on average, and
(b) contains a sufficient number of inclusions for the apparent overall moduli ({\scshape and statistically averaged field distributions in the phases} \rmfamily) to
be effectively independent of the surface values of traction and displacement, so
long as these values are ‘macroscopically uniform'.... The contribution of this surface layer to any average can be negligible by taking the sample large enough.}

In Hill’s original definition \cite{Hill1963}, a {\scshape small caps} font fragment was included and is explained later. We highlight key RVE concepts reflecting this definition (see for details \cite{Buryachenko2025}):
1. The RVE applies to statistically homogeneous or periodic CMs under remote homogeneous boundary conditions (BCs) (\ref{2.28}) or (\ref{2.29}). While Hill \cite{Hill1963} did not use terms like "statistically homogeneous" or "periodic structure CMs," the concept aligns with them. Additionally, "functionally graded structures" (a later term) do not fit within the RVE framework.

2. The RVE estimates effective moduli $\bfL^*$ by averaging phase fields. For periodic structure, CMs, RVE is equivalent to a unit cell (RVE$\equiv$UC), and remote homogeneous BCs (\ref{2.28}) or (\ref{2.29}) reduce to periodic BCs (\ref{2.39}). For statistically homogeneous CMs, no specific geometric representation is required. However, in practice, RVE sampling via computational  
{\color{black}(with``arbitrary simulated fields", see, e.g., \cite{{Ongaroet2022},{Wildmanet2017},{Yanget2024},{Yanget2023}})
or image-based methods (modeling ``real" structures, see, e.g., \cite{{Ahmadiet2022},{AnbarlooieH2024},{Talamadupulaet2020}})} is common, and verifying finite-size BC replacements is crucial to avoid distortions in property estimation.

3. Hill \cite{Hill1963} applied his definition to linear locally elastic CMs, but Definition 8.1 omits the term “constitutive law.” It generalizes to peridynamic CMs by replacing “surface” (\ref{2.42}), (\ref{2.43}) with “volumetric” (\ref{2.40}), (\ref{2.41}).

The inaccuracies of intuitive RVEs in PM have long been addressed in LM (see the Introduction for references, see also
\cite{{Galliet2012},{Harperet2012},{Kanitet2003},{Moumenet2021},{Saroukhaniet2015},{Temizeret2012}}). In LM, a sample response is estimated from DNS of microstructural volume elements (MVEs) simulated or extracted (e.g., from micro-CT). The homogenized properties $\bfL^{\rm A}_{\rm KUBC}$ and $\bfM^{\rm A}_{\rm SUBC}$ for KUBC (\ref{2.41}) and SUBC (\ref{2.42}) are different. The convergence of $\bfL^{\rm A}_{\rm KUBC} - (\bfM^{\rm A}_{\rm SUBC})^{-1}$ 
\BB
\label{8.1}
\bfL^{\rm A}_{\rm KUBC} - (\bfM^{\rm A}_{\rm SUBC})^{-1}\to {\bf 0}
\EE
as MVE size increases help approximate the RVE size and estimate the effective moduli $\bfL^*$, requiring large (theoretically infinite) material domains (see \cite{Ostojaet2016} and \cite{Buryachenko2022a}). However, even with convergence in the number of realizations (e.g.,\cite{Ongaroet2022}), boundary layer and edge scale effects may persist. 

Another well-known inconsistency in LM is the replacement of PBC by homogeneous BC (see \cite{Buryachenko2007}, p. 508). The same issues are expected in PM when modeling infinite (e.g., periodic) media with finite domains:
see, e.g.,
\cite{{Ahmadiet2022},
{Askariet2006},
{Askariet2008},
{Askariet2015},
{Chenget2024},
{DeckleverS2016},
{Jenabidehkordiet2020},
{Yanget2023b},
{Zhanet2021},
{ZhangQ2021}}.
Using body force with compact support (e.g., (\ref{2.4})) as a training parameter helps eliminate sample size, boundary layer, and edge effects.

In the context of AMic, the RVE interpretation for random structure CMs avoids issues like sample size and edge effects, as well as the need for ``surface values." 
 The RVE size is now fully determined by the micromechanical model used, making the concept of ``taking the sample large enough" irrelevant.
For both random and periodic structure CMs, the effective moduli $\bfL^*$ and field concentration factors $\bfA^*(\bfz)$ ($\bfz\in v_i$) are related by mutual coupling (see {\scshape small caps} in Definition 8.1), see equations
\BB
\label{8.2}
^L\!\bfL^*=^L\!\bfL^{(0)}+^L\!\bfR^*, \ \ \ \lle\bfep\rle_i(\bfz)=^L\!\bfA^*(\bfz)\lle \bfep\rle.
\EE
In random structure CMs, the RVE size depends on the micromechanical estimation method. For one-particle methods (like EFM, MTM, and EMM, see Comment 3.6), the RVE is defined as $v_i$, and the interaction size $a^{\rm int}=a$. In the multiparticle EFM, considering binary interactions increases $a^{\rm int}=6a$, thus increasing the RVE size.
In PM, methods for one-particle approaches yield different $\bfL^*$ estimations with the same RVE size. This eliminates the need for sample size control in Amic methods. For periodic structure CMs, the RVE coincides with the unit cell (UC). Therefore, under the hypothesis (\ref{2.33}$_2$) in LM, the equality $\rm RVE^{\bf L^*}\equiv RVE^{\lle\bf \epsilon\rle_i({\bf z})}$ holds for both random and periodic structure CMs.

\subsection {RVE for CMs subjected to inhomogeneous loading}

The RVE concept is rigorously valid when the field scale (inhomogeneity) vastly exceeds the material scale (microstructural variations, see (\ref{2.33}$_2$), (\ref{2.34}$_2$). When these scales are comparable, nonlocal material behavior emerges, requiring stress-strain descriptions via statistical averages weighted by a tensorial kernel. Nonlocal models range from strong nonlocality (strain type and displacement type, peridynamics) to weakly nonlocality (strain-gradient, stress-gradient, and higher-order models).

\noindent{\bf Definition 8.2.} {\it RVE (a) is structurally entirely typical of the whole mixture in $R^d$ in a statistical average sense, and
(b) contains a sufficient number of inclusions for the apparent effective nonlocal operator (and statistically averaged field distributions in the phases) to be effectively independent of the applied inhomogeneous field.}

Definition 8.2 omits the notion of a “large enough sample” and instead considers a heterogeneous medium (statistically homogeneous or periodic) in full space $R^d$. Rather than estimating effective moduli $\bfL^*$, it focuses on effective nonlocal operators of a given structure.
If $ \langle \bfep \rangle ({\bf y})$ varies slowly near ${\bf y}$, it allows for a Taylor expansion of its statistical average and the use of Fourier transforms. The most common effective nonlocal operator is a second-order differential operator
\BB
\label{8.3}
\lle\bfsi\rle(\bfx)=\bfL^*\lle\bfep\rle(\bfx)+\bfcL^*\nabla\otimes\nabla\lle\bfep\rle(\bfx).
\EE
The quasi-crystalline approximation by Lax \cite{Lax1952} (\ref{4.2}) is commonly used to truncate integral equation hierarchies, enabling explicit derivations of the nonlocal differential operator via methods like the effective field method \cite{KanaunL1994} or conditional moments \cite{Khoroshun1996}. A key advantage of the rigorous approach \cite{{Drugan2000},{Drugan2003},{DruganW1996}} is its foundation in variational principles, providing approximation bounds \cite{DruganW1996}.
By combining MEFM with the Fourier transform method \cite{DruganW1996}, explicit expressions for the nonlocal overall operator as a second-order differential operator (\ref{8.3}) were derived
\cite{{Buryachenko1998},{BuryachenkoR1998a}}.
For MEFM, the RVE size is $ B^{\rm RVE{\bf L}*} = 6a$ (see Subsection 8.1), while for the corresponding nonlocal operator, it is
$ B^{\rm RVE{\cal L}*} = 18a$ \cite{{Buryachenko1998},{Buryachenko1999b}}. This approach, initially developed within the first background of micromechanics in LM \cite{Buryachenko2007}, was later extended to the second background in \cite{Buryachenko2022a}.

We analyze periodic structure CMs in LM using CMic methods. For a fixed periodic body force $\bfb(\bfx)$ using variational and asymptotic expansion techniques (see Block 1 in (\ref{3.10})),
the approach in \cite{SmyshlyaevC2000} constructs elliptic higher-order homogenized equations rigorously
(for the particular case of translation averaging solution (\ref{5.6})). 
However, in \cite{SmyshlyaevC2000}, neither numerical examples nor the definition of a representative volume element (RVE) were discussed.
A quantitative assessment was performed in \cite{Ameenet2018} on a two-phase composite under periodic anti-plane shear. Another method, second-order computational homogenization \cite{{Kouznetsovaet2004a},{Kouznetsovaet2004b}}, incorporates macroscopic deformation gradients and higher-order stress measures, resulting in a full second gradient continuum.
Analytical second-order homogenization of linear elastic CMs leads to the second gradient Mindlin continuum (see \cite{Buryachenko2022a}), where the macroscopic length scale parameter depends on UC size. This approach highlights the impact of physical and geometrical nonlinearities on the relation between RVE size and macroscopic response.

The approach in \cite{{Buryachenko1998},{BuryachenkoR1998a}} was later extended to PM in \cite{Buryachenko2024b}. Equation (\ref{3.24}) was analyzed under Hypothesis {\bf H1a} (\ref{3.8}) for linear operators $\bfcJ^{\theta\zeta}_{i,j}$, which decompose into tensors when applied to constant effective fields (\ref{3.8}).
The most general method, an iterative scheme (see LM counterpart in \cite{Buryachenko2022a}), derives a nonlocal integral operator without assuming a predefined form. However, a major drawback is that solving the micromechanical problem (\ref{3.24}) must be repeated for each different external loading $\langle \bfthe\rangle ({\bf z})$.
If the field $\lle\overline{\bfze}_{j}\rle$ is smooth, it can be approximated using a Taylor expansion (see \cite{Eskin1981}), reducing the integral operator to a differential one (\ref{8.3}). This method \cite{Buryachenko2024b} can be adapted for periodic
structure CMs \cite{Buryachenko2025a}, cite{Buryachenko2025e} by replacing the conditional probability density $\varphi(v_q,\bfx_k\vert v_i,\bfx_i)$ by its $\delta$ function form (\ref{2.36}) in Eq. (\ref{3.24}).

In Subsection 8.2, we identified two main shortcomings in the discussed methods.
The first issue is that the approaches for both random and periodic structure CMs (see \cite{{Ameenet2018}, {Buryachenko1998},{BuryachenkoR1998a},{Buryachenko1998},{BuryachenkoR1998a}, {Drugan2000},{Drugan2003},{DruganW1996},
{KanaunL1994},{Khoroshun1996},{SmyshlyaevC2000},{Kouznetsovaet2004a},{Kouznetsovaet2004b}}) are designed specifically for estimating the second-order gradient model effective operator (\ref{8.3}). Their applicability to constructing a more general (e.g., integral) effective operator remains uncertain.
The second issue, related to \cite{{Buryachenko1999b},{Buryachenko2024c}}, is that these methods rely on the EFH (\ref{3.8}) and assume linear phase properties. Additionally, when evaluating the effective response $\lle\bfsi\rle$ under changing overall strains $\lle\bfep\rle$, the entire micromechanical problem must be solved again from scratch.
In the next subsection, we introduce a universal method for estimating any form of the effective (surrogate) nonlocal operator for both random and periodic structure CMs. Its computational complexity matches that of the original approach in Subsection 86.1. Thus, in terms of generality and applicability (see Subsection 8.3 and Section 7), the methods in Subsection 8.2 are a dead end.

\subsection{RVE for CMs subjected to body force with compact support}

Not all  effective datasets $\bfcD^{\rm r}$ and $\bfcD^{\rm p}$ are suitable for ML and NN-based surrogate operators. A generalized RVE concept, extending Hill’s \cite{Hill1963} definition, serves as a threshold to filter out inappropriate sub-datasets, ensuring compatibility with PM under self-equilibrated forcing $\bfb(\bfx)$ (\ref{2.4}).

{\bf Definition 8.3.} {\it RVE is structurally entirely typical of the whole CM area which is sufficient for all apparent effective parameters $\bfcD^{I}$ ($I={\rm r,p})$ to be effectively stabilized (vanised for stresses and strains) outside RVE (as, e.g. in Eq. (\ref{4.12})) in the infinite random or
periodic structure CMs
with any elastic constitutive laws of phases.}

We analyze the region $\overline {\rm RVE}=R^d\setminus {\rm RVE}$, where the  effective dataset
$\bfcD^{\rm r}$ (or $\bfcD^{\rm p}$) stabilizes. Ensuring stabilization and selecting the correct 
$\overline {\rm RVE}=R^d\setminus {\rm RVE}$ guarantees that all effective parameters between $|\bfx|=B^{\rm RVE}$ and $|\bfx|=B^{\rm RVE}+B^{b}/2$ remain consistent within a given tolerance (i.e. $\bfcD^I\to \bfcD^{I{\rm RVE}}$, $I=r,p$). When this holds, the region $|\bfx|>B^{\rm RVE}+B^{b}/2$ can be disregarded, allowing an infinite medium to be approximated by a finite sample. A properly chosen RVE eliminates edge effects, typically confined to a boundary layer of thickness $6|\Omega|$ if $\lle\bfep\rle^{\Omega} (\bfx)$ (though this is not the case here; refer to p. 129 in \cite{Buryachenko2007}). Conversely, an undersized $B^{\rm RVE}$ leads to numerical errors due to sample size limitations and edge effects (see Subsection 8.1).
In this framework,  effective datasets $\bfcD^{\rm r}$ and $\bfcD^{\rm p}$ are computed via micromechanical methods like CAM. Under the scale separation hypothesis (\ref{2.33}$_2$) in LM, Definition 8.3 reduces to Hill’s classical definition \cite{Hill1963} for apparent effective moduli $^L\!\bfL^*$.
{\color{black} 

A fundamental distinction should be noted between RVE Definitions 8.1 and 8.3. Specifically, RVE 8.1 is formulated for a large (theoretically infinite) sample, whereas the purpose of RVE 8.3 is to introduce a preliminary finite-size RVE (to be discussed later). Despite this conceptual difference, there exists a technical similarity between the two definitions.
In particular, RVE 8.1 involves taking the limit in Eq. (\ref{8.1}) when estimating the effective material parameter (moduli or operator) as defined in Eq. (\ref{8.2}$_1$). Similarly, RVE 8.3 is defined through the limiting process $\bfcD^I \to \bfcD^{I{\rm RVE}}$, where $\bfcD^I$ ($I = r, p$) denotes the effective field parameters (generalized analog of Eq. (\ref{8.2}$_2$)).
Importantly, both approaches—Definitions 8.1 and 8.3—do not encounter complications associated with the analysis of stress singularities (in local models) or displacement discontinuities (in peridynamic models) in particular realizations of random or periodic fields. These difficulties are circumvented through the use of volume or statistical averaging of the relevant fields.
}

The RVE concept is even more critical in PM, where three types of nonlocal effects emerge: inhomogeneous applied fields $B^b$, Eq. (\ref{2.4}), material nonlocality $l_{\delta}$, and inclusion interactions $a^{int}$, all of which interact synergistically. In contrast, in LM \cite{Hill1963}, under the scale separation hypothesis (\ref{2.33}$_2$), the nonlocal effects of $a^{int}$ and $a$ reduce to a constant.
For a self-equilibrated forcing term (\ref{2.4}), statistical displacements $\lle\bfu\rle(\bfx)$ in a random heterogeneous bar ($\bfx\in R^1$) were analyzed in \cite{Buryachenko2023} using the AMic framework (see Fig. 1) . Results showed that for $|\bfx|\geq a^{\rm l-r}$,
the solution $\lle\bfu\rle(\bfx)\approx$const.; i.e. the domain
$|\bfx|\leq a^{\rm l-r}$ is RVE$^{\lle\bf u\rle(\bf x)}$ for $\lle\bfu\rle(\bfx)$, depending on scale ratios $a/B^{b}/l_{\delta}$. These studies, though not explicitly naming RVE, pioneered its estimation in PM for random composites.
Additionally, RVEs must be assessed for other effective parameters, such as $\bfcD^{\rm r}$ (\ref{4.22}) and $\bfcD^{\rm p}$ (\ref{5.12}), across different scale ratios $a/B^{b}/l^{(1)}_{\delta}/
l^{(0)}_{\delta}$. Unlike LM (\ref{8.2}), where $\rm RVE^{\bf L^*}\equiv RVE^{\lle\bf \epsilon\rle_i({\bf x})}$., these RVEs may differ for each scale ratio.

In peridynamic CMic, the term RVE is commonly used in both Block 3 (finite inclusion samples) and Block 2 (periodic structure CMs), as referenced in (\ref{3.9}). The general limitations of Block 3 were discussed in Subsection 8.1.
For infinite periodic media, RVE has primarily been used in Block 2 as a shorthand for ``unit cell (UC) of CM under remote homogeneous loading (\ref{2.28}) at scale separations $\Lambda/L=\infty$, $\Lambda/|\Omega_{00}|=\infty$, and $L/|\Omega_{00}|=\infty$", as seen in studies like
\cite{{Madenciet2017},{Madenciet2018},{Diyarogluet2019a},{Diyarogluet2019b},
{Galadimaet2023},{Galadimaet2023b},
{Galadimaet2023c},{Galadimaet2024},{Huet2022},
{Liet2022b},{Qiet2024},{Xiaet2020},{Xiaet2019},{Xiaet2021a},{Xiaet2021b},{WangQet2024}}.
While this shortcut is valid in LM under the scale separation hypothesis (\ref{2.33}$_2$)
and has been widely accepted for decades, its direct application in PM requires caution.
In particular, using periodic boundary conditions (PBC, (\ref{2.39}))—which are correct in LM—can be problematic in PM (see Subsection 2.3 and Comment 5.1). Instead, the variable periodic boundary conditions (VPBC) originally proposed in \cite{Buryachenko2023} should be further generalized to accommodate arbitrary scale ratios $|\Omega_{00}|/a/l^{(1)}_{\delta}/l^{(0)}_{\delta}$ and unit cell geometries (see also Subsection 2.4).

This implies that the RVE for the $k$-th
implementation serves as a crucial parameter (or prerequisite) for building the  effective dataset required for the surrogate operator (refer to Subsection 8.2). Specifically, if RVE$\subset w$, then $\bfcD^{\rm r}_k$
is incorporated into $\bfcD^{\rm r}$. Otherwise, these elements
are excluded from further analysis of the surrogate operator.
The dimensions $B^{\rm RVE}_k$ of the RVE may vary across different implementations indexed by $k$-th. The primary role of the RVE is to ensure the stabilization of the field variables within the domain $\overline{\rm RVE}_k$, rather than merely serving as a mechanism to determine the size of a subdomain for modeling the medium. In this context, the RVE is not strictly a tool for selecting a characteristic size of subdomain (see, e.g.,{Hill1963}) of the micro inhomogeneous medium but is instead a means of achieving convergence in the statistical or spatial distribution of field quantities.
This does not imply that the final size of the RVE, $B^{\rm RVE}_{\rm final}$, is equal to $\max_k
B^{\rm RVE}_k$. Once the contribution $\bfcD^{\rm r}_k$ has been accepted into the overall  effective dataset $\bfcD^{\rm r}$, the specific value of $B^{\rm RVE}_k$ associated with that particular $ k$-th implementation may no longer be retained or considered.
Furthermore, it is well-established in LM that increasing the sample size effectively mitigates the sample size effect (see, e.g.,\cite{{Kanitet2003},{TeradaK2001}}). However, even resolving the sample size effect does not address the boundary layer and edge effects associated with nonlocal phenomena. The primary advantage of the new RVE concept (as defined) lies in its ability to completely eliminate boundary layer and edge effects for random and periodic structured composite materials, irrespective of the phases' elastic properties—whether local or nonlocal, linear or nonlinear.

For estimating any effective nonlocal operator, including surrogate models in ML or NN
(see Section 9), analyzing field distributions and determining the appropriate RVE size for a finite scale ratio $B^b/|\Omega_{00}|\not =\infty$ is crucial. To the author's knowledge, the term RVE has not been explicitly used in LM or PM concerning the forcing term (\ref{2.4}).
The RVE depends on scale ratios $a/B^{\rm RVE}/l^{(1)}_{\delta}/l^{(0)}_{\delta}$ and varies with the gradient $\nabla \bfb(\bfx)$. This means two functions $\bfb(\bfx)$ and $\bfb'(\bfx)$ with identical scale ratios may yield different RVEs. Additionally, satisfying $b({\bf 0}, B^b)\subset \Omega_{00}$ does not guarantee RVE$\subset \Omega_{00}$;
in fact, the RVE may span multiple unit cells.
Despite this complexity, the RVE remains a key benchmark: if the solution domain $w$ of Eqs.
({4.10})-({4.20}) fully contains the RVE (RVE$\subset w$), the computed elements $\bfcD^{\rm I}_k$ ($I={\rm r,p},\ k=1,\ldots,N$)
contribute to the surrogate operator dataset (Subsection 8.2).
Otherwise, $w$ must be expanded or $|\nabla\bfb(\bfx)|$ reduced.
If the RVE size RVE$\subset w$, then $\bfcD^{\rm I}_k$ varies across implementations, ensuring field stabilization rather than merely defining a subdomain for modeling.
Unlike traditional LM approaches, where increasing the sample size mitigates finite-size effects \cite{{Kanitet2003},{TeradaK2001}}, this novel RVE concept eliminates boundary layer and edge effects in random or periodic composites, regardless of phase properties—local or nonlocal, linear or nonlinear.
\sffamily

\noindent{\bf Comment 8.1.} Special attention is given to statistically inhomogeneous (or functionally graded, FG) composites, where the arrangement of elements (such as concentration and orientation, see \cite{Buryachenko2022a}) depends on position, as described in Eq. (\ref{2.35}). FG composites were studied using the multiparticle effective field method (MEFM) within the framework of the EFH (\ref{3.1}), where the exact GIE ({3.14}) is simplified to Eq. ({3.13}). The area where Eq. (\ref{3.14}) should be solved is defined as $w^{\rm FG}=B^{\lle\bfep\rle}\oplus b(0,a^{\rm int})$; here, $B^{\lle\bfep\rle}$ represents a domain where $\lle\bfep\rle(\bfx) \neq \text{const}$, and $a^{\rm int}$ is the scale of long-range interactions between inclusions (\ref{2.33}). However, a new rigorous Definition 8.3 is immediately beneficial for FG composites with random structures (\ref{2.35}).
In statistically homogeneous media, the RVE is invariant to parallel translation, meaning that if $\bfx \to \bfy$ ($\bfy = \bfx + \bfz$), then RVE($\bfx$) = RVE($\bfy$) for any $\bfz$. However, in statistically inhomogeneous (functionally graded) media (\ref{2.35}), this invariance is lost, so RVE($\bfx$)$\not=$RVE($\bfy$). As a result, the AGIE ({3.23}) should be solved for a set of uncoupled solutions of RVE($\bfx$) instead of a less effective coupled solution over the domain $w^{\rm FG}$. The resulting  effective datasets $\bfcD^{\rm r}(\bfx)$, depending on body force $\bfb$, are not invariant (in contrast to $\bfcD^{\rm r}$ (\ref{4.22})) under parallel translation along the cross-section of FG composites.

{\color{black} \noindent{\bf Comment 8.2.} 
Let us compare the similarities and differences between the effective moduli $^L\!\bfL^*$ and the strain concentrator factor $^L\!\bfA^*(\bfz)$ (\ref{8.2}), on one hand, and the effective dataset $\bfcD^{\rm r}$  (\ref{4.22}), on the other.
None of the quantities $^L\!\bfL$, $^L\!\bfA^*(\bfz)$,  , or $\bfcD^{\rm r}$ depend explicitly on microstructural geometric descriptors (e.g., $c^{(1)}$ or $\varphi(v_i,\bfx_i|v_j,\bfx_j)$ or phase properties. Moreover, $\bfcD^{\rm r}$ is applicable to the analysis of FGMs and does not depend on overall properties, unlike
$^L\!\bfL$ and  $^L\!\bfA^*(\bfz)$.
While $^L\!\bfL$, $^L\!\bfA^*(\bfz)$ )are independent of the homogeneous boundary conditions (\ref{2.30}) or (\ref{2.31}), the effective dataset $\bfcD^{\rm r}$  (\ref{4.22}) depends explicitly on the BFCS (\ref{2.4}).
The key distinction, however, is that the effective moduli $^L\!\bfL^*$ form a single constant tensor, whereas the effective dataset
$\bfcD^{\rm r}=\{\bfcD^{\rm r}_k\}_{k=1}^N$
can contain an arbitrary number of elements $\bfcD^{\rm r}_k$  with potentially nonlinear, nonlocal, or otherwise complex relationships among multiple effective parameters. As a result, $\bfcD^{\rm r}$ is far more informative than the constant tensor  $^L\!\bfL^*$, while also eliminating both sample-size and boundary-layer effects. These properties make the dataset $\bfcD^{\rm r}$ (\ref{4.22}) ideally suited for subsequent use in standard ML\&NN methodologies (see Section 9).

}

\rmfamily

\subsection {RVE for deterministic structure CMs}

Random or periodic structure CMs are characterized by general probability densities ($\varphi (v_i,{\bf x}_i)$ and $\varphi (v_i,{\bf x}_i\vert; v_j,\bfx_j)$, see Subsection 2.3) or specific cases (\ref{2.36}) with a periodic grid $\bfLa$. In contrast, deterministic structure CMs do not exhibit randomness or periodicity. For example,
from microcomputer tomography (micro-CT) or scanning electron microscopy (SEM)
images of composite samples, and typically consist of
several hundreds or thousands of inclusions \cite{{Bellenset2024},{Echlinet2014}}.
Notably, CM images obtained through CT techniques can be viewed as observational snapshots (``window of observation") of deterministic structure CMs. The observation window $w$ with boundary conditions (either (\ref{2.28}) or (\ref{2.29})) is shown
in Fig. 2.

As discussed in Subsection 8.2, even with a sufficiently large window $w$ for estimating $\bfL^*$ and the application of highly precise numerical methods in DNS, the boundary layer effect persists and cannot be fully eliminated. The difficulties in using finite-size samples to approximate effective behavior are well recognized in LM (see, e.g., \cite{Buryachenko2022a}, p. 593; \cite{Buryachenko2007}, pp. 226-229).
It is known that determining the overall behavior of a sample $w$ using DNS, such as, e.g., in Eq. (4.22), $(\bfx\in w$)

\vspace{1.mm} \noindent %\hspace{30mm} 
\parbox{8.8cm}{`
\centering \epsfig{figure=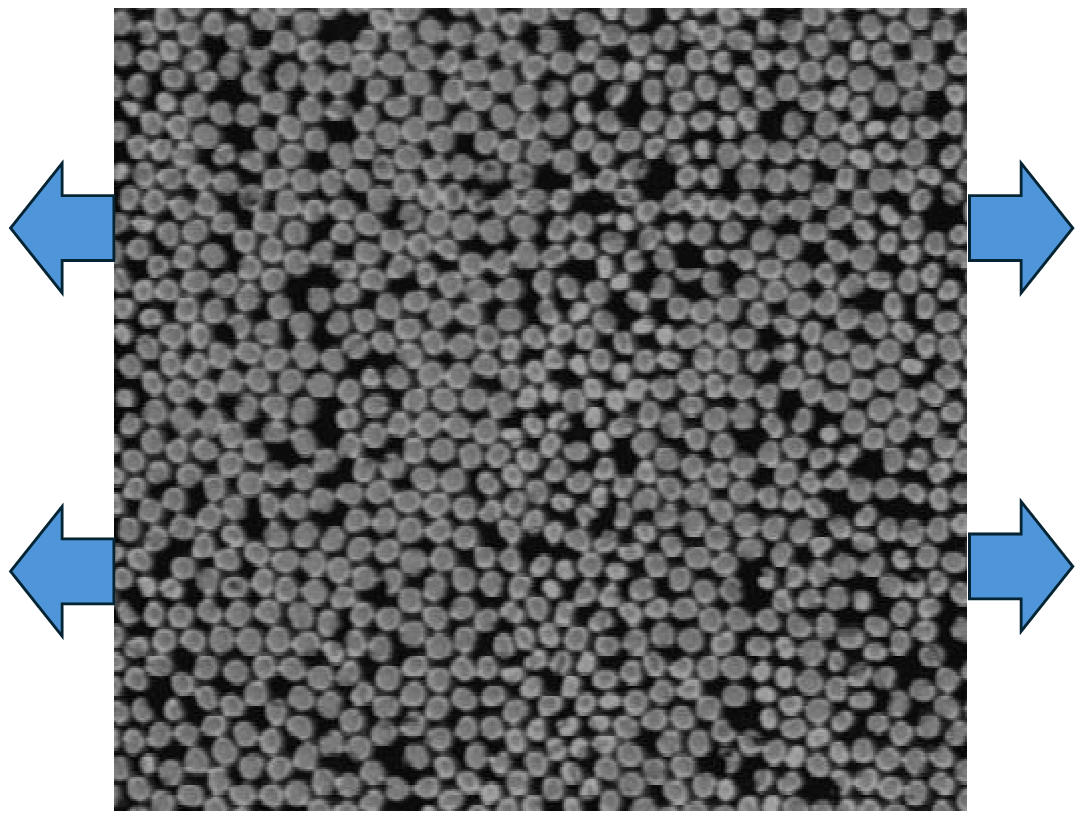, width=7.8cm}\\ \vspace{-50.mm}
\vspace{122.mm}
\vspace{-117.mm} \tenrm \baselineskip=8pt
{{\sc Fig. 2:} CT image with BC}}
%\vspace{2.mm}

\BBEQ
\label{8.4}
\!\!\!\!\!\!\!\!\!{\bfcD}^{\rm DNS}=\{\bfcD^{\rm DNS}_k\}_{k=1}^N, \ \ 
\bfcD^{\rm DNS}_k=\{\bfu(\bfb_k,\bfx),\bfb_k(\bfx)\}, 
\EEEQ
results in the complete loss of information related to boundary layer effects;
Here, the applied body force
$\bfb(\bfx)$ does not necessarily conform to Eq. (\ref{2.4}) (see, e.g.,{Silling2021} with oscillatory forcing).
For example, Fig. 3a (Fig. 3 is reproduced from in \cite{Youet2022b}) presents an experimental 
%\noindent 
counterpart of Fig. 2, showing a biological tissue specimen under biaxial stretching, with the digital image correlation (DIC) displacement tracking grid marked in green. A schematic of a specimen subjected to Dirichlet-type boundary conditions is illustrated in Fig. 3b. Additionally, a neural network-based learning model was utilized to develop a surrogate operator capable of predicting overall displacement fields for previously unseen loading conditions in finite samples of soft biological tissue (see Fig. 3c).
Figures 3a-3c from \cite{Youet2022b} have been reproduced in several key studies on physics-informed neural operators (e.g., 
\cite{{Gosmaniet2022},{Jafarzadehet2024},{Jinet2023},{Youet2022}}, see also \cite{Jafarzadehet2024b}). However, how the sample size and boundary layer effects influence these estimations remains an unresolved issue.

\vspace{2.mm}% \noindent
\hspace{-19mm} 
\parbox{11.8cm}{%\hspace{-10mm}
\centering \epsfig{figure=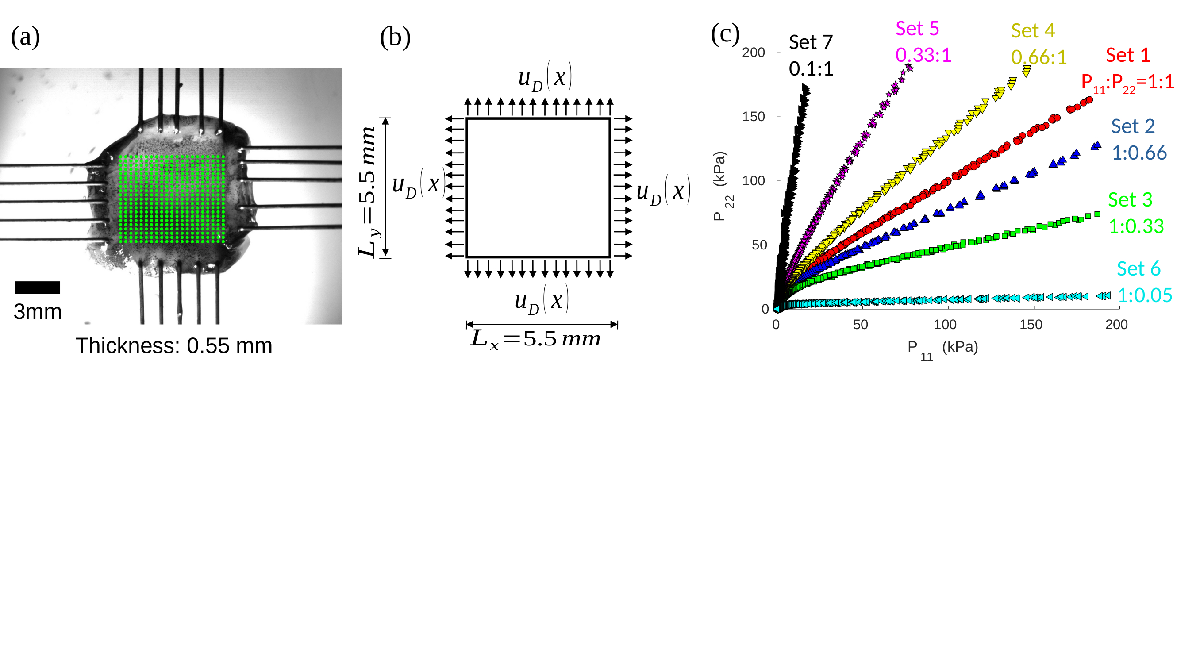, width=8.4cm}\\ \vspace{-18.mm}
\vspace{80.mm}
\vspace{-84.mm} \tenrm \baselineskip=8pt
{\hspace{10.mm}{\sc Fig. 3:} Biological tissue specimen}}
\vspace{5.mm}

A refined version of Maxwell’s approach \cite{Maxwell1873}, often referred to as its “second birth,” represents an infinite statistically homogeneous (or periodic) CM as an inclusion cloud embedded within an infinite matrix \cite{Sevostianovet2019}. Unlike the classical Maxwell model \cite{Maxwell1873}, these generalized approaches account for the interactions between inclusions within the cloud.
However, Buryachenko \cite{Buryachenko2022b} demonstrated that 23 key papers in this field \cite{Sevostianovet2019} are fundamentally flawed when using homogeneous remote BC (\ref{2.28}). To address this, BC (\ref{2.28}) can be replaced with freeloading at infinity, supplemented by a body force (\ref{2.4}). This allows the inclusion cloud (of learnable size) to estimate the RVE per Definition 8.3. Through Monte Carlo simulations of ``random" inclusion sets, we can determine $\bfcD^{\rm p}$ similarly to (\ref{5.6}), replacing uniform $\bfchi$ in (\ref{5.6}) with random inclusion sets. At this stage, the microstructure of the cloud and its computation method (e.g., multipole expansion \cite{Sevostianovet2019}) become irrelevant. The final step involves using ML or NN techniques (see Section 9) to derive a surrogate nonlocal operator for modeling infinite statistically homogeneous media. This marks the “third birth” of Maxwell’s approach, proving highly effective.

The concept of the ``inclusion cloud" in Maxwell’s approach \cite{Maxwell1873} can be reformulated by replacing the remote BC
(\ref{2.28}) (or (\ref{2.29})),
with a body force $\bfb(\bfx)$ (\ref{2.4}) that has compact support.
Under this modification, the domain of interest $\bfx\in w$ (\ref{8.4}), effectively reduces to the $\bfx\in$RVE in the sense of Definition 8.3. The DNS data can then be represented as ($\bfx\in {\rm RVE}$)
\BBEQ
\label{8.5}
\!\!\!\!\!\!\!\!\!\!\!{\bfcD}^{\rm DNSd}=\{\bfcD^{\rm DNSd}_k\}_{k=1}^N, \ 
\bfcD^{\rm DNSd}_k=\{\bfu(\bfb_k,\bfx),\bfb_k(\bfx)\}.
\EEEQ
The loading configuration of a deterministic structure within the domain $\bfx \in w$ (see Fig. 2), subjected to a body force $\bfb(\bfx)$ defined by (\ref{2.4}), is illustrated in Fig. 4.
In this figure, the localized force application region $b(\bfx_i,B^b)\subset {\rm RVE}$ is explicitly shown.
This highlights that the applied force is confined within a subset of the RVE, ensuring the stabilization of effective parameters $\bfcD^
{\rm DNSd}$ outside RVE. In this approach, the dataset $\bfcD^{\rm DNSd}$ is constructed from various realizations of the body force $\bfb(\bfx)$ and the microstructural configurations (e.g., derived from distinct CT images). Generations of deterministic structures at the fixed $\bfb_k(\bfx)$ allow the analog of dataset (\ref{5.12}).
The generation of deterministic structures while keeping the body force $\bfb_k(\bfx)$ fixed enables the construction of an analogous  effective dataset to Eq. (\ref{5.12})
\BBEQ
\label{8.6}
\!\!\!\!\!\!\!\!\!\!\!{\bfcD}^{\rm d}&\!=\!& \{\bfcD^{\rm d}_k\}_{k=1}^N,\ \ {\bfcD}^{\rm d}_k=\{\lle{\bfu}_k\rle(\bfb_k,\bfx),\lle\bfsi_k\rle(\bfb_k,\bfx), \nonumber\\
\!\!\!\!\!\!&&\!\!\!\!\!\!\!\lle{\bfu}_{k}\rle^{(1)}(\bfb_k,\bfx),
\!\lle{\bfsi}_{k}\rle^{(1)}(\bfb_k,\bfx), \bfb_k(\bfx)
\},
\EEEQ
where the statistical averages $\lle(\cdot)\rle$ are computed by averaging over the ensemble of configurations, ensuring a statistical representation of the mechanical response for the given loading conditions. 
Naturally, there are no restrictions on the shapes of the domains $b({\bfx}_i,B^b)$ [Eq. (\ref{2.4})] and the RVE (see Definition 8.2), nor on the size ratio between them. The choice of spherical geometries for these regions, as illustrated in Fig. 4, is made purely for convenience and does not reflect any fundamental requirement.

In the framework of computational micromechanics applied to 2D specific realizations of statistically homogeneous random media (i.e., deterministic microstructures as defined in Eq. (\ref{3.26})), Silling {\it et al.} \cite{Sillinget2024} introduced a coarse-graining approach for the homogenization of mechanical properties of CMs.
This model is based on Monte Carlo-simulated inclusion placements ($\approx$900 inclusions) within a finite-sized square domain $w$, which is subjected to a self-equilibrated BFCS $\bfb(\bfx)$ (\ref{2.4}).
The body force $\bfb(\bfx)$ corresponds to the region of long-range interaction, as described in \cite{Buryachenko2023}. The geometric scale separation is characterized by the ratio
${\rm dist}(\partial {\rm RVE}, \partial w)\approx 10a=100\lambda$,
where $\lambda$ represents the lattice spacing.
This scale separation satisfies the conditions of an RVE according to Definition 8.3.
The validity of this RVE assumption is supported by the numerical observation that the effective strain field $\lle\bfep\rle(\bfx)$
vanishes in the vicinity of $\partial w$, as illustrated in colored Fig. 12 of \cite{Sillinget2024}, which is consistent
with Fig. 4. Notably, the term ``RVE" was not explicitly used in 
\cite{Sillinget2024}.

\vspace{1.mm} \noindent %\hspace{30mm} 
\parbox{8.8cm}{
\centering \epsfig{figure=Fel4.eps, width=7.8cm}\\ \vspace{-50.mm}
\vspace{122.mm}
\vspace{-117.mm} \tenrm \baselineskip=8pt
{{\sc Fig. 4:} Scheme of CT image with $b({\bfx}_i,B^b)\subset{\rm RVE}\subset w$}}
\vspace{0.mm}

The purple regions, where $\lle\bfep\rle(\bfx)\approx {\bf 0}$, and the blue regions in Fig. 5 correspond to the areas outside $w\setminus$RVE and inside RVE, respectively, as shown in Fig. 4. Fig. 5 is the best illustration of the new RVE concept, although the term ``RVE" was not used in \cite{Sillinget2024} (because this term was at the first time introduced later in \cite{Buryachenko2025}).
The proposed upscaled peridynamic model allows for a significantly coarser discretization compared to the original fine-scale model, which relies on DNS. This coarse-graining approach enables large-scale simulations to be conducted efficiently, through the estimation of quantities such as
$\lle\bfu\rle_i(\bfz,\bfx),\
\lle\bfsi\rle_i(\bfz,\bfx)$, $\bfz\in v_i, \ \bfx\in R^d$
remains outside the scope of the study.

\vspace{1.mm} \noindent %\hspace{30.mm} 
\parbox{8.8cm}{
\centering \epsfig{figure=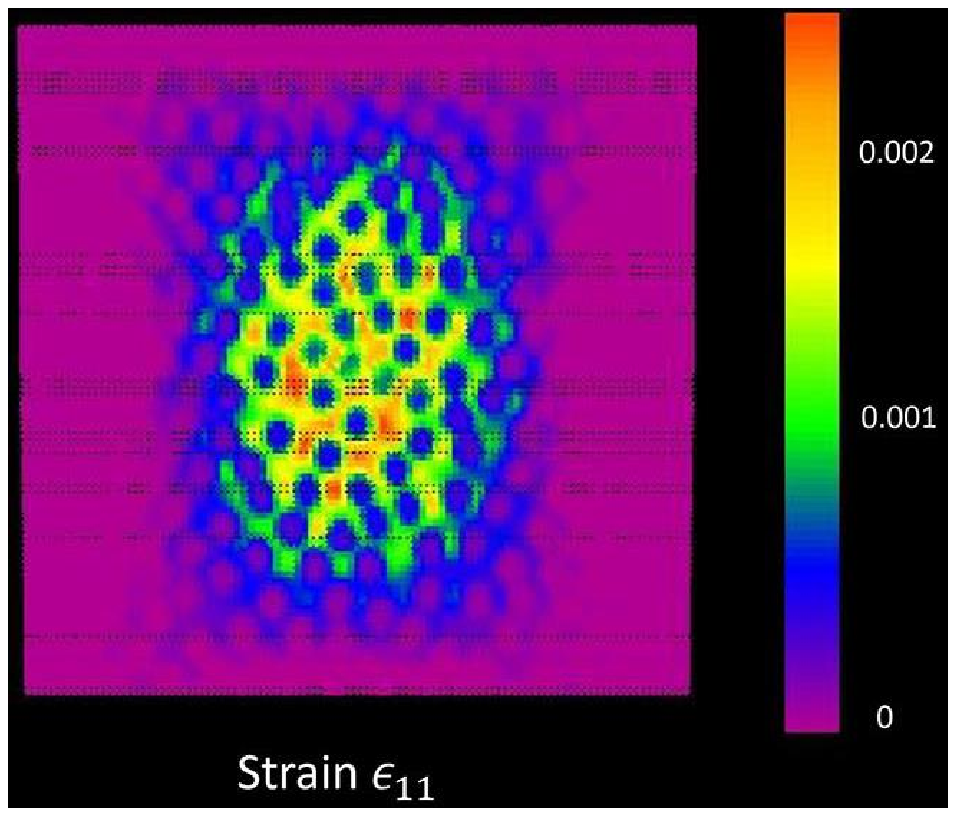, width=7.8cm}\\ \vspace{-50.mm}
\vspace{122.mm}
\vspace{-117.mm} \tenrm \baselineskip=8pt
{{\sc Fig. 5:} Strain $\lle\bfep\rle(\bfx)$ in a sample $w$}}
\vspace{2.mm}

{\color {black} 
Considerable advancements in the nonlocal homogenization of random microstructures were achieved in \cite{Fanet2023}, \cite{Fanet2024a}, \cite{Silling2021}, and \cite{Youet2024}. In the case of 1D systems, the random inclusion configurations were generated using uniform distribution functions. Subsequently, a DNS solver was employed to compute the wave velocity fields, thereby producing deterministic realizations (i.e., specific structures in the sense of Eq. (\ref{3.26})) corresponding to a prescribed forcing term $\bfb(\bfx,t)$ along with given boundary and initial conditions.
Surrogate nonlocal operators were developed using several regression techniques, including Nonlocal Operator Regression (NOR) \cite{Youet2024}, Bayesian Nonlocal Operator Regression (BNOR) \cite{Fanet2023}, and Embedded Nonlocal Operator Regression (ENOR) \cite{Fanet2024a}. To eliminate both the effects of sample size and boundary layers, these approaches can be straightforwardly adapted for the BFCS loading conditions described in Eq. (\ref{2.4}).
}

\subsection {Classification and schematic representation of CAM}
We begin by classifying PM in relation to the homogeneous volume boundary condition (VBC) (\ref{2.28}), starting with the Amic branch. The classification of the LM method (\ref{3.9}) also applies to PM.
Model methods (Gr1) include simplified approaches like mixture theory \cite{{Askariet2006},{Askariet2008},{Askariet2015},{Chenget2024},{Huet2011},{Huet2012a},{Mehrmashhadiet2019},{WuC2023},{Wuet2021}}, which have been widely used for laminated structures \cite{{Basogluet2022},{Diyarogluet2016},{Ghajariet2014},{Huet2014},{MadenciO2014},{Madenciet2021},{Madenciet2023},{Renet2022},{Xuet2008}}).
The CAM of peridynamic CM, a part of {\it Analytical Micromechanics} (Amic), does not rely on DNS,
although 
%\noindent 
nothing in n Eq. (\ref{3.24}) and their solution (see { \cite{Buryachenko2024a}) is analytical. 
The effective medium method (EFM), a Gr1 approach in LM, extends naturally to PM as a peridynamic inclusion in a local effective medium.
Perturbation methods (Gr2) relate to the dilute approximation in PM \cite{Buryachenko2020b}, while variational methods (Gr3) are discussed in \cite{Buryachenko2020c}. The original CAM formulation in PM (Gr4) was introduced in \cite{Buryachenko2020} for linear bond-based properties and later generalized to nonlocal elastic properties \cite{Buryachenko2023k}, \cite{Buryachenko2022a}.
Additionally, the classification of {\it Computational Mechanics} using DNS (Eq. (\ref{3.10})) in LM applies to PM as well. Asymptotic homogenization methods (Block 1 in (\ref{3.10})) for PM are explored in \cite{{AllaliL2012},{Duet2016},{Duet2020}}).
The generalization of classical computational homogenization approaches (Block 2 in (\ref{3.10})) has been explored in studies such as \cite{{Buryachenko2018},{Buryachenko2018b},{Diyarogluet2019a},{Diyarogluet2019b},{Galadimaet2023},{Galadimaet2023b},{Sillinget2023},{Qiet2024},{Xiaet2020},{Xuet2021},{XuF2020},{Youet2022}}. Additionally, numerous works address problems involving single or multiple inclusions (or cracks) within a sample (Block 3 in (\ref{3.10})):
\cite{{Agwaiet2011},{Askariet2006},
{Askariet2006},
{Askariet2008},
{Askariet2015},
{Azdoudet2013},
{Birneret2023},
{BobaruH2011},
{Breitenfeldet2014},
{Dipasqualeet2022},
{HaB2010},
{Huet2012b},
{Jenabidehkordiet2020},
{KilicM2010},
{KilicM2010b},
{Laurienet2023},
{Leet2014},
{MacekS2007},
{MadenciG2015},
{Mousaviet2021},
{Nguyenet2021},
{Renet2017},
{Saregoet2016},
{Wanget2018},
{Wen et2023},
{Yanget2023b},
{Zhanet2021},
{ZhouW2021}}.

Amic functions like a solar system, with GIE (\ref{3.24}) as its gravitational core, guiding all related methodologies Gr1)-Gr4). In contrast, the Cmic branch orbits independently, employing approaches that do not rely on GIE (\ref{3.24}). Now, we turn to an entirely different centralized framework, where AGIE (\ref{3.23}) becomes the new focal point, standing on its own without any dependence on GIE (\ref{3.24})

The scheme of the AGIE-CAM (\ref{3.17}), (\ref{3.19}), (\ref{4.5})-(\ref{4.7}) and GIE-CAM (\ref{3.17}), (\ref{3.19}), (\ref{3.24})
are represented in a block diagram in Fig. 6a and 6b, respectively. To simplify notation, the scheme corresponds to the iterative solution of Eq. (\ref{4.5})-(\ref{4.7}).
At its core, the process revolves around Block 3 Perturbator and
%\noindent 
Block 4 Micromechanics (see Fig. 6a), which forms a central feedback loop. Block 3 computes the perturbator $\bfcL^{\theta\zeta}_i(\bfz,\overline{\bfze})$ (\ref{3.17}). The preprocessing Block 1 Input provides Block 3 with essential data, including the microstructural geometry of an inclusion $v_i(\bfx)$ within an infinite homogeneous matrix (or a large sample), material properties $\bfbC(\bfx)$(such as micromodulus $\bfC^{\rm bond}$), and a square mesh $\Omega^{\rm sq}:=\{(x_1, . . . , x_d )^{\top}| x_i = hp_i \}$ $(\bfp=(p_1,\ldots,p_d)\in Z^d)$. Similarly, Block 2 Input supplies Block 4 with geometrical information, specifically the same square mesh $\Omega^{\rm sq}$
and the spatial distribution of inclusions $X$. For random structures, this distribution is characterized by probability densities $\varphi (v_j,{\bf x}_j)$ and conditional densities $\varphi (v_j,{\bf x}_j\vert; v_i,{\bf x}_i)$. For the periodic structures $X=\bfLa$, whereas $X$ is a deterministic field for deterministic structure CMs.
The use of the same mesh $\Omega^{\rm sq}$ in Block 3 and Block 4 is primarily for ease of integration between these two components (see the case of LM on p. 428 in \cite{Buryachenko2022a}). In the iterative cycle [Block 3]$\rightleftharpoons$[Block 4] (\ref{4.5}), Block 3 takes as input $\lle\overline{\bfze}\rle_i(\bfx)$
and outputs $\bfcL^{\theta\zeta}_i(\bfz,\lle\overline{\bfze}\rle_i)$, where $\bfx,\bfz\in\Omega^{\rm sq}$.
However, the mesh $\Omega^{\rm sq}$ is used solely to maintain consistency between Block 3 and Block 4 and does not determine the internal solution method within Block 3 (see Eqs. (\ref{3.17})). Importantly, the internal workings of Block 3 are irrelevant to Block 4, and vice versa. As a result, these blocks function as independent ``Black Boxes" that interact solely through their inputs and outputs, as indicated in Fig. 6a by their black shading. Output Block 5 contains the datasets either $\bfcD^{\rm r}$, $\bfcD^{\rm p}$, or $\bfcD^{\rm d}$ for CMs with either random, periodic, or deterministic structures, respectively.
It should be mentioned that particular cases of problems indicated in Block 3 in Eq. (\ref{3.10}) can be considered as a Block 3 Perturbator (see Fig. 6a) for one inclusion (or crack) in a big sample.

For comparison, the structure of the GIE-CAM, as defined in Eqs. (\ref{3.17}), (\ref{3.19}), and (\ref{3.24}), is illustrated schematically in the block diagram shown in Fig. 6b, which is described in detail in \cite{Buryachenko2024b}. Interestingly, Block 3—representing the Perturbators $\bfcL^{\theta\zeta}_i(\bfz,\overline{\bfze})$ (see Eq. (\ref{3.17}))—is identical in both Fig. 6a and Fig. 6b. However, Block 4 differs between the two 
%\noindent 
diagrams. Specifically, Block 4 in Fig. 6a (refer to Eq. (\ref{3.23})) is structurally simpler than its counterpart in Fig. 6b (see Eq. (\ref{3.24})).
In particular, the integral term in Eq. (\ref{3.23}) lacks the renormalizing term $\lle\bfcL^{\theta\zeta}_j(\bfz-\bfx_j,\overline{\bfze})\rle(\bfx_j)$ that appears in Eq. (\ref{3.24}). Additionally, the free term $\bfep^{b(0)}(\bfz)$ in Eq. (\ref{3.23}) is deterministic, being defined by the BFCS loading (\ref{2.4}) applied to an infinite homogeneous matrix. In contrast, the free term $\lle \bfthe\rle ({\bf z})$ in Eq. (\ref{3.24}) represents an a priori unknown statistical average of the solution.
Furthermore, Block 2 (Input) in Fig. 6a includes an additional BFCS term as given in Eq. (\ref{2.4}). The distinction continues in Block 5 (Output): in Fig. 6b, it provides an estimate of the effective moduli $\bfC^*$, whereas in Fig. 6a, it yields  effective datasets such as $\bfcD^{\rm r}$, $\bfcD^{\rm p}$, or $\bfcD^{\rm d}$ corresponding to composite materials (CMs) with random, periodic, or deterministic microstructures, respectively.

\vspace{1.mm} \noindent\hspace{-10mm} 
\parbox{10.2cm}{
\centering \epsfig{figure=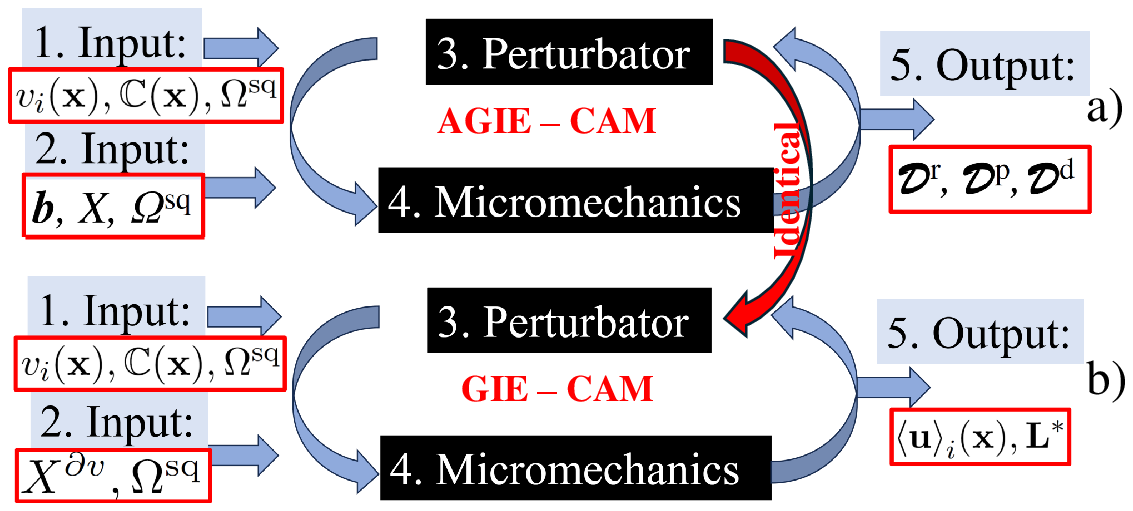, width=9.8cm}\\ \vspace{-122.mm}
\vspace{122.mm}
\vspace{-85.mm} \tenrm \baselineskip=8pt
{{\sc Fig. 6:} The iteration scheme of the AGIE-CAM and GIE-CAM}}
\vspace{3.mm}

The AGIE (\ref{3.23}) plays a fundamental role in both Amic and Cmic within the PM framework. Sections 3 and 4 focus on the formulation of AGIE (\ref{3.23}) and its solution (\ref{4.1})–(\ref{4.20}), establishing it as the core component of the self-consistent method  in Amic (see Fig. 7). 
{\color {black}  
The proposed approach (\ref{4.1})–(\ref{4.20}) is analogous to the corresponding solution of Gr4) (\ref{3.9}), with the essential distinction arising from the difference between the AGIE (\ref{3.23}) and the GIE (\ref{3.24}).
Similarly, the analog of Gr2) derived from Gr4) for the AGIE-CAM formulation, by taking the limit $c^{(1)} \to 0$, differs from the Gr2) solution (\ref{3.9}) obtained from the GIE.}
Furthermore, the model methods Gr1) corresponding to (\ref{3.10})
(referenced in the first paragraph of this subsection) implicitly or explicitly rely on EFH (\ref{3.8}), which is eliminated under body force loading (\ref{2.4}). Consequently, the physical interpretation of Gr1) methods under such loading conditions remains unclear.
Variational methods, widely used in LM (see \cite{Buryachenko2022a} for references), have been extended to PM in \cite{{Buryachenko2022a},{Buryachenko2020c}} for statistically homogeneous materials and field parameters. However, even for statistically homogeneous media subjected to body force loading (\ref{2.4}), the statistical homogeneity of the field is disrupted, making the formulation of variational methods for this loading nontrivial. 
{\color{black} Thus, the methods corresponding to Blocks Gr1)–Gr4) (\ref{3.9}) are inapplicable to the loading configuration defined by (\ref{2.4}) and are therefore indicated with red cross marks in Fig. 7.}
In the Cmic
%\noindent 
branch, for periodic structured CMs, the equations (\ref{5.1})–(\ref{5.3}) derived from CAM (\ref{3.26}) formally resemble the equations (\ref{4.11})–(\ref{4.13}) for random structured CMs. Additionally, the translation averaging (\ref{5.6}) applied to periodic structures leads to equations (\ref{5.7})–(\ref{5.11}), which are structurally similar to (\ref{4.15})–(\ref{4.19}) for random structures. This process generates the  effective dataset $\bfcD^{\rm p}$ (\ref{5.12}), which is analogous to $\bfcD^{\rm r}$ (\ref{4.22}).
Notably, periodic boundary conditions (\ref{2.39}), (\ref{2.40}) (or VPBC (\ref{2.39}), (\ref{2.40})) are effectively eliminated, raising questions about the generalization of methods from Blocks 1 and 2 (see (\ref{3.10})) for body force loading (\ref{2.4}) in both LM and PM. 
As highlighted in the opening paragraph of this subsection, these classes of methods have produced significant theoretical developments and a substantial corpus of influential results for the periodic BC (\ref{2.39}), (\ref{2.40})—or, equivalently, for the VPBC formulation. However, the methods categorized under Blocks 1) and 2) are entirely unsuitable for the loading condition specified in (\ref{2.4}) and must therefore be excluded from the present analysis.
As a clear indication of their limitation, these Blocks are marked with red cross lines in Fig. 7. 
This implies that the previously well-developed theories corresponding to Blocks 1) and 2) (as outlined in the Introduction) are not applicable and therefore will not be employed in the context of BFCS loading (\ref{2.4}).
Ultimately, the AGIE-CAM approach simplifies the analysis of periodically structured CMs to the examination of a deterministic finite set of inclusions within the RVE, thereby aligning it with the methodology of the modified Block 3 (see Fig. 7), while differing from the original Block 3 formulation (\ref{3.10}). This suggests that DNSs within Cmic are systematically reorganized within Amic to construct  effective datasets $\bfcD^{\rm p}$ (\ref{5.12}) and $\bfcD^{\rm d}$ (\ref{8.6}), both structurally analogous to $\bfcD^{\rm r}$ (\ref{4.22}).
{\color{black} 
Thus, presenting the known classifications of both Amic (\ref{3.9}) and Cmic (\ref{3.10}) within the LM framework (generalized also to the PM in \cite{Buryachenko2024a}) serves not merely as a comprehensive background review.
More importantly, all these previously developed methods are {\it inapplicable} see Fig. 7) to the proposed AGIE-CAM approach, which is specifically designed to establish the computational chain
AGIE$\to$ new RVE$\to$ effective dataset $\bfcD^{\rm r}$ (\ref{4.22}) (or alternatively, $\bfcD^{\rm p}$ (\ref{5.12}) or $\bfcD^{\rm d}$ (\ref{8.6})).

\vspace{1.mm} \noindent \hspace{-8mm} 
\parbox{10.2cm}{
\centering \epsfig{figure=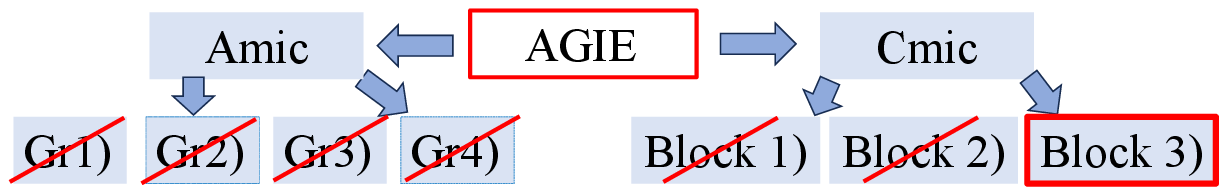, width=9.6cm}\\ \vspace{-122.mm}
\vspace{122.mm}
\vspace{-100.mm} \tenrm \baselineskip=8pt
{{\sc Fig. 7:} AGIE as a central governing equation of PM}}
\vspace{3.mm}

Despite advances in efficient DNS in Cmic at the microscale, applying computational homogenization to realistic heterogeneous structures remains costly. Hence, reduced-order models (ROMs) have been developed to significantly accelerate microscale evaluations while maintaining accuracy.
Although Transformation Field Analysis by Dvorak \cite{Dvorak1992} (see also for references \cite{Buryachenko2023d}, \cite{Buryachenko2023g}) was not labeled as a ROM at that time, the TFA can, in retrospective analyses, be considered as an ignition spark for a wide class of ROMs, such as: 
Nonuniform Plastic Strain Analysis (NPSA, \cite{MichelS2003}, \cite{MichelS2004}), 
Proper Orthogonal Decomposition (POD, \cite{Berkoozet1993}),
Proper Generalized Decomposition \cite{Ladevezeet2010},
Reduced Basis Method \cite{Boyaval2008}, 
High-Performance Reduced-Order Model \cite{Hernandezet2014},
Empirical Cubature Method \cite{Hernandezet2017},
and Wavelet-Reduced-Order Model \cite{vanTuijlet2019}. These methods proposed for linear and nonlinear, local and nonlocal problems are usually applied to the BC (either (\ref{2.30}), (\ref{2.31}), or (\ref{2.39}), (\ref{2.40})) or VBC
(either (\ref{2.28}), (\ref{2.29}), or (\ref{2.37}), (\ref{2.38})). 
The most suitable POD application in PD for reformulation in the framework of the current review is presented in \cite{Donget2023} (see also \cite{Białeckiet2005}), where the authors developed an adaptive partitioned ROM for
fast solution of the PD model involving fracture problems. In constructing ROMs, the global equilibrium equations are typically reformulated in block form involving displacement vectors $\mathbb{U}_b$ and $\mathbb{U}_i$ for boundary layer and internal points, along with their respective load vectors $\mathbb{B}_b$ and $\mathbb{B}_i$. For analysis under BFCS (\ref{2.4}) instead of conventional boundary conditions (e.g., VBC in (\ref{2.28}), (\ref{2.29}) or (\ref{2.37}), (\ref{2.38})), Buryachenko \cite{Buryachenko2024a} proposed setting $\mathbb{B}_b \equiv \mathbf{0}$ and using a compactly supported $\mathbb{B}_i$. A similar modification of existing ROMs to incorporate BFCS (\ref{2.4}) in place of standard BCs would be an interesting direction.}
}

It should be noted that the methods Amic (\ref{3.9}) and Cmic (\ref{3.10}), corresponding to the boundary conditions (\ref{2.28})–(\ref{2.30}) or (\ref{2.37})–(\ref{2.40}), are fundamentally distinct in their mathematical formulation and underlying assumptions (see Introduction). However, when the forcing term exhibits compact support (\ref{2.4}), both methods, as highlighted by the dashed blue boxes in Fig. 7, are governed by the same integral equation-AGIE (\ref{3.23}). This equation serves as a unifying operator, establishing a common functional framework for both approaches. In other words, AGIE (\ref{3.23}) acts as a projection operator that maps both methods onto a shared solution space, effectively reducing their conceptual disparity under these specific conditions of the forcing term loading (\ref{2.4}).

\subsection{Fast Fourier transform methods in micromechanics of periodic and deterministic structure CMs} 
{\color{black} It is worth highlighting the Fast Fourier Transform (FFT) method as a highly effective tool for analyzing a broad class of Cmic problems described by Eq. (\ref{3.10}) (Block 2). This family of methods was originally introduced for linear elastic CMs with periodic microstructures in \cite{{MoulinecS1994}, {MoulinecS1998}}. In this context, the implicit Lippmann–Schwinger (L–S) Eq. (\ref{8.7}) arising in locally elastic micromechanics can be transformed into a simple multiplication operation in the Fourier space, as shown in Eq. (\ref{8.8})
\BBEQ
\label{8.7}
\bfep(\bfx)&=&\bfep^{w_\Gamma}+^L\!\bfGa^{(0)}*^L\!\bftau(\bfx), \\
\label{8.8}
\widehat\bfep(\bfk)&=&^L\!\widehat\bfGa^{(0)}(\bfk)
^L\!\widehat\bftau(\bfk) \ \ \ (\bfk\not={\bf 0}), \widehat{\bfep}({\bf 0})={\bfep}^{w_\Gamma}.
\EEEQ
The convolution $*$ involves the Eshelby–Green operator $^L\!\bfGa^{(0)}={\rm div}^L\!\bfG^{(0)}\nabla^s $, where $\nabla^s$ is the symmetrized gradient and $^L\!\bfG^{(0)}$ is the Green function of Eq. (\ref{2.5}).
This operator is periodic in the unit cell $\Omega_{00}$ and explicitly known in Fourier space for an infinite isotropic matrix \cite{{MoulinecS1994}, {Mura1987}}.

FFT methods exploit this by transforming the implicit Lippmann–Schwinger equation into a product in Fourier space (\ref{8.8}), reducing computational cost to $O(N log N)$ compared to $O(N^2)$ for FEM. The original scheme by Moulinec and Suquet \cite{MoulinecS1994}, based on fixed-point (Picard) iterations, can, in a retrospective sense, be considered as an ignition spark for
a wide range of FFT-based methods for CMs with high contrast, nonlinear \cite{{Lucariniet2022}, {Schneider2021}}, and nonlocal \cite{Buryachenko2023j} properties; 
it was considered either matrix or polycrystal \cite{Seguradoet2018} structures with either strain, polarization or displacement 
fields as the primary unknown variables. 
Modern FFT solvers are highly optimized for both memory usage and computational performance. Furthermore, they are well-suited for parallel computing, where the FFT grid can be efficiently distributed across multiple processors.

Equation (\ref{8.7})
is formulated under the periodic boundary conditions (PBC) given by (\ref{2.39}) and (\ref{2.40}). For the peridynamic (PD) model described by Eq. (\ref{2.6}), these should be replaced by their nonlocal counterparts--the volume periodic boundary conditions (VPBC) defined in (\ref{2.37}) and (\ref{2.38}).
Furthermore, if the PBC are replaced with the BFCS loading (\ref{2.4}), the classical formulation given by Eqs. (\ref{8.7}) and (\ref{8.8}) transforms accordingly into a new pair of equations that reflects this loading framework (see for details \cite{Buryachenko2025b})
\BBEQ
\label{8.9}
\!\!\!\!\!\bfep(\bfx)\!\!&=&\!\!\bfep^{b(0)}(\bfx) +\!\bfGa^{(0)}*^L\!\bftau(\bfx), \\
\label{8.10}
\!\!\!\widehat\bfep(\bfk)\!\!&=&\!\!\widehat{\bfep^{b(0)}}(\bfk)+^L\!\widehat\bfGa^{(0)}(\bfk)
^L\!\widehat\bftau(\bfk), \ (\bfk\not={\bf 0}), 
\EEEQ
and $\widehat{\bfep}({\bf 0})=\widehat{\bfep^{b(0)}}({\bf 0})$. Under BFCS loading (\ref{2.4}), the field periodicity within the unit cell $\Omega_{00}$, which holds in Eq. (\ref{8.7}), is lost. As a result, 
$\Omega_{00}$ is replaced by a larger mesocell $w$ that fully contains the representative volume element (RVE), i.e., ${\rm RVE}\subset w$ (see Fig. 4). 
The size of the periodically distributed mesocells $w$ acts as a postprocessing learning parameter, chosen so that the strain field $\bfep(\bfy)$ vanishes in the boundary layer region $\bfy \in w \setminus \text{RVE}$. This outer region imposes vanishing periodic boundary conditions—either standard PBC (\ref{2.39}) and (\ref{2.40}) or their nonlocal VPBC counterparts (\ref{2.37}) and (\ref{2.38}).
Specifically, the vanishing of the strain field $\bfep(\bfy)$ in $\bfy\in w\setminus $RVE enables $\bfep(\bfx)$ ($\bfx\in w$) to be regarded as periodic within an extended medium where the mesocell $w$ serves as the new periodicity cell.
The resulting Eqs. (\ref{8.9}) and (\ref{8.10}) can be solved iteratively using the same Picard (fixed-point) method as used for (\ref{8.7}) and (\ref{8.8}). This approach yields estimates of the  effective dataset $\bfcD^{\rm d}_k$ (\ref{8.6}) for composite materials (CMs) with deterministic structure--a class for which FFT methods have not previously been applied.
For periodic structures, the same scheme is used to estimate the dataset $\bfcD^{\rm p}_k$ (\ref{5.12}). In this context, each specific translation $\bfchi\in {\cal V}_{\rm \bf x}$ (\ref{5.4}) is treated as a deterministic structure. Consequently, the  effective dataset $\bfcD^{\rm p}_k$ (\ref{5.12}) is obtained by averaging over these translations, using the statistical formulation given in (\ref{5.5}).

Just as the Picard iteration scheme used to solve Eqs. (\ref{8.9}) and (\ref{8.10}) under periodic boundary conditions (PBC) (\ref{2.39}) and (\ref{2.40}) served as the foundation for a broad class of FFT-based homogenization methods for periodic composites \cite{{Buryachenko2023j}, {Lucariniet2022}, {Schneider2021}, {Seguradoet2018}}, the corresponding iterative scheme applied to Eqs. (\ref{8.9}) and (\ref{8.10}) under BFCS loading (\ref{2.4}) may similarly serve as a conceptual and computational trigger for the development of a new generation of FFT approaches. 
Several pseudocode algorithms for both linear and nonlinear classical FFT-based methods were presented in \cite{Lucariniet2022}, focusing on periodic boundary conditions (PBC) as defined by Eqs. (\ref{2.28}) and (\ref{2.39}). In the corresponding new generation of FFT approaches, a formal substitution $\bfep^{w\Gamma} \to \bfep^{b(0)}$ is sufficient to implement the modified scheme.
These generalized FFT methods would extend existing frameworks to handle both deterministic and periodic structures within a unified formulation, accommodating non-periodic loading and field distributions
in $\Omega_{00}$ while retaining the computational efficiency of Fourier-based solvers for $w$'s periodic system.
In this extended setting, deterministic microstructures—lacking intrinsic statistical homogeneity or periodicity—can still be analyzed by embedding them within mesocells and applying BFCS-type loading (\ref{2.4}). This enables the use of FFT techniques even when classical
PBC-based assumptions on $\Omega_{00}$ are no longer valid. Moreover, for periodic structures, the method naturally incorporates statistical averaging over translations (as discussed with  effective dataset $\bfD^{\rm p}_k$ (\ref{5.12})), thereby linking deterministic and periodic analysis within the same numerical framework (see for details \cite{Buryachenko2025b}).
As with the original FFT scheme introduced by Moulinec and Suquet—which revolutionized computational micromechanics—the proposed extension using BFCS loading and its associated iterative solution scheme has the potential to similarly ignite progress in the modeling of advanced composites, especially those exhibiting high contrast, nonlinearity, or nonlocal behavior.
}

The fundamental framework and key results of this paper—embodied in Eqs. (\ref{3.23})-(\ref{3.27}), (\ref{4.7})-(\ref{4.22}), (\ref{5.1})-(\ref{5.12}), (\ref{6.2})-(\ref{6.12}), and Subsections 8.3-8.6—are formulated in an operator form that remains entirely independent of any specific constitutive law of elasticity. This universality ensures that these equations hold for any constitutive model, including, e.g., those detailed in Section 7.
In essence, the representation of PD using Eqs. (\ref{2.1})-(\ref{2.27}) and (\ref{3.1})-(\ref{3.7}) can be seamlessly substituted with the more detailed constitutive formulations presented in Section 7, without affecting the overall structure and validity of the results, see Sections 3-8. The choice of PD as the primary modeling approach in this work does not stem from an inherent superiority over other models but rather from the author’s long-standing dedication to peridynamic CMs—an expertise and intellectual commitment developed over more than a decade of focused research in this field \cite{{Buryachenko2014a}, {Buryachenko2015b}, {Buryachenko2017}, 
{Buryachenko2018}, {Buryachenko2018b}, {Buryachenko2018c}, {Buryachenko2019b}, {Buryachenko2020}, {Buryachenko2020b}, {Buryachenko2022a}, {Buryachenko2023}, {Buryachenko2023a}, {Buryachenko2023b}, {Buryachenko2023e}, {Buryachenko2023f}, 
{Buryachenko2023h}, {Buryachenko2023c}, {Buryachenko2023d}, {Buryachenko2023i}, {Buryachenko2023j}, {Buryachenko2023k}, 
{Buryachenko2024a}, {Buryachenko2024b}, {Buryachenko2025},  {Buryachenko2025a}, {Buryachenko2025b}, 
{Buryachenko2025c}, {Buryachenko2025d}, {Buryachenko2025e}} including proposal of background of PM
\cite{Buryachenko2014a}, \cite{Buryachenko2017}. 

\section {Estimation of a set of surrogate operators}
\setcounter{equation}{0}
\renewcommand{\theequation}{9.\arabic{equation}}
{\color{black} 
Sections 3–8 are devoted to the development of the AGIE-CAM approach, which is specifically formulated to establish the computational sequence
AGIE$\to$ new RVE$\to$ effective dataset $\bfcD^{\rm r}$ (\ref{4.22}) (or alternatively, $\bfcD^{\rm p}$ (\ref{5.12}) or $\bfcD^{\rm d}$ (\ref{8.6})).
These effective datasets are noteworthy in that they do not explicitly depend on either the microstructural characteristics (e.g., the inclusion volume fraction $c^{(1)}$ or their statistical descriptors (\ref{2.35})) or on the mechanical properties of the constituent phases (e.g., (\ref{2.2}) or (\ref{2.23})).
However, the direct use of these datasets as a final outcome of micromechanical modeling (for instance, as $^L!\bfL^*$ (\ref{8.6})) is of limited practical relevance.
As will be demonstrated in Section 9, it is only when the AGIE-CAM framework is integrated with ML\&NN   techniques that their combination yields fundamentally new possibilities for nonlocal surrogate operator modeling.}

Silling’s pioneering work \cite{{Silling2021},{Youet2020}}, later expanded in \cite{Youet2024}, introduced data-driven ML techniques for modeling composite materials (CMs) by constructing surrogate nonlocal operators from DNS. These studies focused on a finite 1D heterogeneous bar under wave loading at the boundary and oscillatory forcing.
In the current (see for details \cite{{Buryachenko2023},{Buryachenko2023a}}), Silling’s approach \cite{Silling2021} is adapted by replacing the DNS dataset $\bfcD^{\rm DNS}$ (\ref{8.4}) with $\bfcD^{\rm r}$ (\ref{4.22}), $\bfcD^{\rm p}$ (\ref{5.12}) or $\bfcD^{\rm d}$ (\ref{8.6}), corresponding to random, periodic, and deterministic structures, respectively. The inhomogeneous forcing term $\bfb(\bfx)$ with compact support (\ref{2.4}) serves both as a loading parameter and a tool for learning surrogate nonlocal constitutive laws for CMs.
The surrogate  effective  datasets $\widetilde{\bfcD}^{\rm r}$, $\widetilde\bfcD^{\rm p}$, $\widetilde\bfcD^{\rm d}$ approximate their respective  effective datasets while efficiently compressing micromechanical data. The field PM dataset $\widetilde{\bfcD}^{\rm r}$ is constructed via a surrogate model ${\cal S}$, mapping statistical micromechanical averages to macroscopic fields:
$\widetilde{\bfcD}^{\rm r}={\cal S}({\bfcD}^{\rm r})$.
This approach retains essential micromechanical features while simplifying data representation.
\vspace{-2.mm}
\BBEQ
\!\!\!\!\!\!\!\!\!\!\!\!\!\!\bfcL_{\rm \gamma}[\lle{\bfu}_k\rle](\bfx)\!\!\!&=&\!\!\!{\bfGa}(\bfx), \nonumber\\
\label{9.1}
\!\!\!\!\!\!\!\!\! \!\!\!\!\!\bfcL_{\rm \gamma}[\lle{\bfu}_k\rle](\bfx)\!\!\!&=&\!\!\!
\!\!\int \!\!\bfK_{\gamma}(|\bfx-\bfy|) (\lle{\bfu}_k\rle(\bfy)-\lle{\bfu}_k\rle(\bfx))~d\bfy,
\EEEQ
where $\gamma:=b,{\rm \sigma},{\rm u_i},{\rm \sigma}_i$ and ${ \bfGa}(\bfx):=-\bfb(\bfx), \lle\bfsi\rle(\bfx),$
$ \lle\bfu\rle_i(\bfz,\bfx),\lle\bfsi\rle_i(\bfz,\bfx)$, respectively, correspond to four surrogate operators $\bfcL_{\gamma}$.
Each of these fields corresponds to a surrogate operator $\bfcL_{\gamma}$ associated with the problem.
{\color{black}Thus, the external force--displacement function pairs
$\{\lle\bfu_k\rle(\bfx),\bfb_k(\bfx)\}$    of $\bfcD^{\rm r}$ (\ref{4.22}) within domain $\bfx\in w$ (see Fig.4),  
are employed to estimate $\bfK_b$ following \cite{{Silling2021},{Youet2020}}.
In an analogous manner, the pairs
$\{\lle\bfu_k\rle(\bfx),\lle\bfsi\rle_k(\bfx)\}$,
$\{\lle\bfu_k\rle(\bfx),\lle\bfu_{ik}\rle(\bfz,\bfx)\}$, and $\{\lle\bfu_k\rle(\bfx),\lle\bfsi_{ik}\rle(\bfz,\bfx)\}$
($\bfz\in v_i$) 
are utilized to identify the kernels $\bfK_{\sigma}$,  $\bfK_{u_i}$, and  $\bfK_{\sigma_i}$, respectively.  }
The objective is to construct an optimal surrogate model for the kernel functions
$\bfK_{\gamma}^*$, which define the surrogate datasett $\widetilde\bfcD^{\rm r}$.
This is formulated as four optimization problems, one for each $\gamma$
\BBEQ
\label{9.2}
\!\!\!\!\!\!\!\!\!\!\!\!\!\!\!\!\!\!\!\!\!\bfK_{\gamma}^*\!\!&=&\!\!{\rm arg}\!\min_{\!\!\!\!\!\!\!\!\!\! {{\bf K}_{\gamma}}}\!\sum_{k=1}^N\!|| \bfcL_{\rm {\gamma}}[\lle{\bfu_k}(\bfb_k)\rle](\bfx)- {\bf\Gamma}_k(\bfx)||^2_{l_2} \nonumber\\
\!\!&+&\!\!{\cal R}(\bfK_{\gamma}).
\EEEQ
The objective function quantifies the discrepancy between the surrogate model and the original effective dataset using an $l_2$-norm over $\bfx\in R^d$, with a regularization term ${\cal R}(\bfK_{\gamma})$ (e.g., Tikhonov regularization) to improve conditioning.
To optimize $\bfK_{\gamma}$, the Adam optimizer \cite{KingmaB2014} is employed in an iterative gradient descent scheme, where $\bfK_{\gamma}$ is expressed as a linear combination of Bernstein-based polynomials. Further details are available in \cite{Youet2020}, with additional insights in \cite{{Fanet2023},{Youet2021},{Youet2022},{Youet2024}}.

The approaches in \cite{Youet2020} and \cite{Youet2024} use uncompressed datasets (\ref{8.4}), unlike the compressed methodology in (\ref{4.22}) (or (\ref{5.12})). The DNS dataset ${D}^{\rm DNS}$ (\ref{8.4}) captures detailed microscale displacements
$\bfx\in w$ for each applied force $\bfb_k(\bfx)$, making it significantly larger than
${\bfcD}^{\rm I}$ ($I={\rm r,p,d}$) (\ref{4.22}) even for identical loadings.
Compression in ${\bfcD}^{\rm I}$ does not require full-field DNS computations but instead leverages micromechanics to estimate effective parameters more efficiently. On the other side, for a homogeneous linearized peridynamic medium (\ref{2.12}) and
(\ref{2.13}) subjected to remote homogeneous BC (\ref{2.28}) (e.g. $\bfeta(\bfxi)=\bfxi$), the corresponding local moduli can be obtained from the equation of constitutive law (see \cite{Sillinget2003}).
Seemingly, for surrogate homogeneous media subjected to remote homogeneous BC (\ref{2.28}), we obtain
\BBEQ
\label{9.3}
^L\!\bfL^*&=&\int\bfK_{\sigma}(|\bfx-\bfy|)(\bfy-\bfx)~d\bfy,
\EEEQ
\sffamily
{\noindent{\bf Comment 9.1}
The method in \cite{Xuet2021} (see also \cite{{XuF2020},{Youet2021}}) defines the averaged displacement field as
$\lle \widetilde {\bf u}\rle^{\Omega}_i(\bfx) =|\Omega_i|^{-1}$ $\int_{\Omega_i} \widetilde \bfu(\bfchi)~d\bfchi
$ where $\Omega_i$ is the averaging domain for a fixed grid point $\bfx$. This approach assigns the computed field to a fixed computational grid ${\bf \Lambda}$, producing a discrete kernel $\bfK^*_b$.
In contrast, translation-averaging methods (e.g., Eq. (\ref{5.6})) use moving averaging cells, yielding continuous kernels. For a detailed comparison of discrete vs. continuous formulations, see \cite{Buryachenko2022a}, p. 895.
\rmfamily

The original approach in \cite{{Silling2021},{Youet2020}} estimates the kernel $\bfK^*_b$ for 
${\bfcD}^{\rm DNS}$ by solving the optimization problem (\ref{9.2}). However, it does not establish that $\bfK^*_b$ can directly approximate statistical average stresses $\lle\bfsi\rle(\bfx)$, similar to how a micromodulus estimates the stress field $\bfsi(\bfx)$ for homogeneous medium (\ref{2.17}). Moreover, within ${\bfcD}^{\rm DNS}$, key displacement averages inside inclusions, $\lle\bfu\rle_i(\bfx)$ ($\bfx\in v_i$), are not retained during the online stage. Instead of relying on $\bfK^*_{b}$, kernels $\bfK^*_{\rm u_i}$ and $\bfK^*_{\rm \sigma_i}$ are more applicable for nonlinear problems like fracture mechanics and plasticity.
To address this, we use compactly supported forcing terms to construct surrogate operators, bypassing the need for effective moduli estimation via $ B^b\to \infty$.

The upscaled peridynamic model allows for a coarser spatial discretization (see \cite{{Youet2022},{Youet2023},{Sillinget2023}}) compared to DNS, greatly improving computational efficiency for large-scale simulations. However, estimating specific effective field parameters, such as $\lle\bfu\rle^{l(1)}(\bfx),\
\lle\bfsi\rle^{l(1)}(\bfx)$, ($\bfx\in R^d)$, is beyond the study's scope. This framework ensures scalability while preserving fidelity to micromechanical behavior within the RVE-based approach.

The surrogate operators in (\ref{9.1}) and (\ref{9.2}) are predefined and limited to linear response prediction. To address this, nonlocal neural operators have been introduced to learn mappings between function spaces \cite{{Lanthaleret2024},{Liet2003}}.
An ordinary artificial neural network (ANN) defines a nonlinear local operator. Consider an $L$-layer fully connected neural network (FCNN) $\Psi(\bfx)$: $\bfR^{\rm d_{\bf x}}\to \bfR^{\rm d_{\rm \bf u}}$, mapping input $\bfx$ to output $\bfu$ through multiple layers. Each layer processes the previous layer's output using a weight matrix ${\bf w}^l$ and bias vector $\bfb^l$ for $(1\leq l\leq L-1)$
\BBEQ
\label{9.4}
%\nonumber
\!\!\!\!\!\!\!\!\!\!\!\!\!\!\!\!\!\bfz^l(\bfx)=\bfcA(\bfw^l \bfz^{l-1}(\bfx)+\bfb^l), \ \bfu(\bfx)=\bfw^L \bfz^{L-1}(\bfx)+\bfb^L,
\EEEQ
where $\bfcA$ is a nonlinear activation function (e.g., tanh). The tractable parameters $\bfthe:= \{\bfw^l,\bfb^l\}_{1\leq l\leq L}$ are optimized via a loss function. This function $\Psi(\bfx)$ is local since $\Psi(\bfx)$ depends only on $\bfz(\bfx)$ at the same point $\bfx$. In contrast, nonlocal neural operators incorporate integral operators to capture long-range dependencies:
\BB
\label{9.5}
\bfz^l(\bfx)= \bfcA(\bfw^l \bfz^{l-1}(\bfx)+\bfb^l+(\bfcK^l(\bfz^{l-1})(\bfx)),
\EE
where $\bfcK^l$ integrates against a matrix-valued kernel $\bfK^l$.
Various architectures-DeepONet, PCA-Net, graph neural operators, Fourier neural operator (FNO), and Laplace neural operator (LNO)—have been developed, with detailed comparisons available in \cite{{Gosmaniet2022},{HuZet2024},{KumaraY2023},{Lanthaleret2024}}.

The Peridynamic Neural Operator (PNO) \cite{Jafarzadehet2024} introduces a surrogate operator $\bfcG$ ($\bfcG(\lle\bfu\rle)(\bfx)\approx -\bfb(\bfx)$) for predicting the behavior of highly nonlinear, anisotropic, and heterogeneous materials, offering greater accuracy and efficiency than traditional models based on predefined constitutive laws (e.g., Eq. (\ref{9.1})).
An extension, Heterogeneous PNO (HeteroPNO) \cite{Jafarzadehet2024b}, enables data-driven constitutive modeling of fiber orientation fields in anisotropic materials. Two loading cases were analyzed for a finite square sample: (1) volumetric Dirichlet boundary conditions without body forces, and (2) body forces $\bfb(\bfx)$ generated via FFT from a Gaussian white noise field rather than a compactly supported function (\ref{2.4}). However, extending the surrogate operator to an infinite medium while eliminating sample size and boundary layer effects without relying on the RVE concept (see Definition 8.3 and Fig. 12 in \cite{Sillinget2024}) remains a challenge.

Physics-Informed Neural Networks (PINNs) \cite{{Cuomoet2022},{Haghighataet2021},{Harandiet2024},{Huet2024},{Karniadakiset2021},{KimL2024},{Raissiet2019},{RenL2024}} enforce physical equations (e.g., (\ref{2.5}), (\ref{2.6})) as constraints within neural networks, ensuring physically consistent training. Residuals of these equations are incorporated into the loss function.
By integrating neural operators with PINNs \cite{{Faroughiet2024},{Gosmaniet2022},{WangY2024}}, models can efficiently capture complex nonlinearities, heterogeneity, and nonlocal effects with high generalization. However, these methods are typically applied to finite-size samples without direct extension to infinite media.
In particular, peridynamic (PD) differential operators are incorporated into PINNs for problems with sharp gradients \cite{Haghighatet2021}, though the constraint equations are based on solid mechanics PDEs. PINNs with PD governing equations have been used to analyze displacement fields in homogeneous and heterogeneous elastic plates \cite{Ninget2023} and to predict quasi-static damage and crack propagation in brittle materials \cite{EghbalpoorS2024}.
PINN approach establishes a relationship between the material parameters of a mesoscale
model and the material parameters with constraints based on known physical relationships \cite{Linet2025}.
The PD-PINN framework effectively captures complex displacement patterns influenced by geometric parameters like pre-crack position and length. The total loss function $\cL_{\rm tot}$ (see \cite{{Gosmaniet2022},{Karniadakiset2021},{Raissiet2019}}) combines governing equation loss $\cL_{\rm gov}$, boundary condition losses $\cL^u_{\rm BC}$, a local balance between the internal and external forces $\cL_{\rm BC}^f$, and data loss $\cL_{\rm data}$, each weighted appropriately. Model parameters
$\bfthe^*$ are optimized by minimizing $\cL_{\rm tot}$ until a set accuracy or iteration limit is met \cite{{KingmaB2014},{Paszkeet2019}}.

Instead, nonlocal energy-informed neural networks (EINN) \cite{{YuZ2024},{YuZ2024b},{ZhouY2024}} define the total loss function based on the system’s total potential energy, combining internal strain energy loss and external work loss while enforcing boundary conditions.
To solve the inverse problem of computing the peridynamic kernel, \cite{Difonzoet2024} proposed using radial basis functions (RBFs) as activation functions in PINNs. It was demonstrated that selecting an appropriate RBF is crucial for obtaining physically meaningful solutions consistent with the data.

The nonlinear micromechanical model of CAM (see Sections 4 and 5) can be integrated into PNO and PINN by replacing the full-field dataset ${\bfcD}^{\rm DNS}$ with the compressed  effective dataset ${\bfcD}^{\rm r}$ (\ref{4.22}) (or
${\bfcD}^{\rm p}$, ${\bfcD}^{\rm d}$), forming AGIE-CAMNN (CAM Neural Network approach). This approach eliminates boundary layer and size effects,
and the known difficulties for generalizability to different domain shapes for neural operators
(see for references \cite{Jafarzadehet2024}),
a known issue in local micromechanics \cite{Buryachenko2007}. The proposed approach ensures generalizability since the domain of interest is the entire space $R^d$.
Unlike previous methods (\cite{{EghbalpoorS2024},{Ninget2023},{YuZ2024},{YuZ2024b},{ZhouY2024}}), it removes boundary condition residual losses. Boundary layer and edge effects (appearing in \cite{{Jafarzadehet2024},{Jafarzadehet2024b},{Youet2022b}}), previously addressed through inefficient minus-sampling \cite{Buryachenko2022a}, are inherently avoided. {\color{black} So, a naive way proposed in the minus-sampling method is to consider Minkowski subtraction $w^*=w\ominus b({\bf 0}, r)$ (see p. 169 in \cite{Buryachenko2022a}) within domain $w$ and allow measurements from an object in $w^*$ to an object in $w$. Accordingly, the radius distribution function \cite{Buryachenko2022a}, along with the kernel of the surrogate operators \cite{Youet2020}, \cite{Youet2024}, remains invariant with respect to the distance from the boundary.
}
Using
${\cal D}^{\rm r}$ instead of ${\bfcD}^{\rm DNS}$ results in four surrogate operators $\bfcG_{\gamma}(\lle\bfu\rle)(\bfx)$ ($\gamma=b,\sigma,u_i,\sigma_i$) instead of a single operator $\bfcG(\lle\bfu\rle)(\bfx)$ \cite{Jafarzadehet2024}, akin to mixed PINN formulations \cite{{Haghighataet2021},{Harandiet2024}}. Crucially, this enables a nonlocal counterpart to the effective concentration factor (\ref{8.2}$_2$), making it applicable to nonlinear problems like fracture and plasticity.

The key advantage of substituting $\bfcD^{\rm DNS}\to \bfcD^{\rm r}$ (\ref{4.22}) is that $\bfcD^{\rm r}$ is derived for a compact-support forcing term (\ref{2.4}) and incorporates the crucial RVE concept (Subsection 8.3). This ensures the elimination of size-scale and boundary layer effects at the dataset preparation stage. Without RVE, these effects inevitably arise, compromising the accuracy of the four surrogate operators $\bfcG_{\gamma}(\lle\bfu\rle)(\bfx)$ $(\gamma = b, \sigma, u_i, \sigma_i$, Eq. (\ref{9.1})), and cannot be reliably corrected later. Thus, RVE is essential for ensuring accurate, scale-independent ML and NN applications in computational micromechanics.
The process for generating surrogate models is illustrated in Fig. 8. Block 1 DNS $\bfcD^{\rm DNS}$ (\ref{8.4}) has been 

\vspace{-1.mm} \noindent \hspace{-8.mm} 
\parbox{10.2cm}{
\centering \epsfig{figure=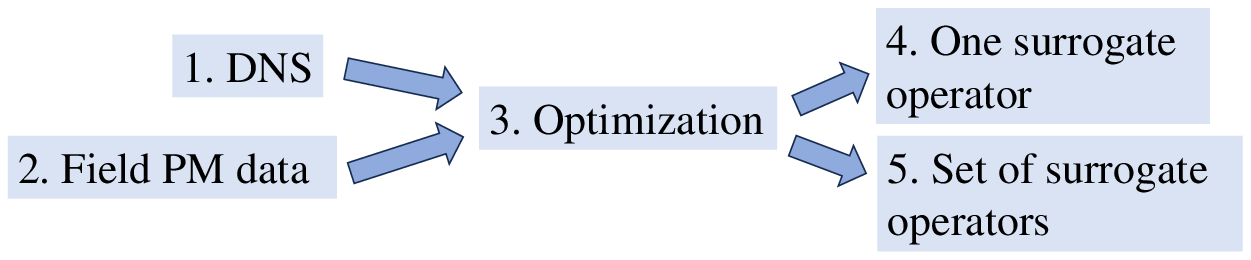, width=8.8cm}\\ \vspace{-119.mm}
\vspace{119.mm}
\vspace{-89.mm} \tenrm \baselineskip=8pt
{{\sc Fig. 8:} The scheme of obtaining of surrogate model set}}
\vspace{4.mm}

\noindent utilized in studies such as \cite{{Fanet2023},{Jafarzadehet2024},{Youet2020},{Youet2021},{Youet2022}}.
In the proposed AGIE-CAMNN approach, we analyze either $\bfcD^{\rm r}$ (\ref{4.22}), $\bfcD^{\rm p}$ (\ref{5.12}) or $\bfcD^{\rm d}$ (\ref{8.6}), employing AGIE (\ref{3.23}) in AMic and CMic tools, respectively. These datasets are compressed in Block 2 (Field PM Data) and used as input in Block 3 (Optimization), which applies established ML and NN techniques (Section 9).
By replacing the large dataset $\bfcD^{\rm DNS}$ (\ref{8.4}) with the more compact $\bfcD^{\rm r}$, $\bfcD^{\rm p}$,
or $\bfcD^{\rm d}$, computational efficiency is significantly improved. Only the alignment between Blocks 2 and 3 is required, with no changes to Block 3 itself. Solving the optimization problem in Block 3 results in either a single surrogate operator (Block 4 corresponding to Block 1) or a set of surrogate models (Block 5 corresponding to Block 2).

\sffamily
\noindent{\bf Comment 9.2} To the best of the author's knowledge, the analysis of composite materials (CMs) subjected to a forcing term with compact support was first explored in \cite{{Buryachenko2023},{Buryachenko2023a}}, where this issue was framed as a highly specific example of the interaction between nonlocal effects, driven by the distinct material and field scale $a, l_{\delta}, B^b$. However, it has now become clear that the systematic application of a forcing term with compact support, as articulated in (\ref{2.4}) \cite{{Buryachenko2023},{Buryachenko2023a}} (and further explored in \cite{{Buryachenko2024b},{Buryachenko2025},{Jafarzadehet2024},{Sillinget2024},{Yuet2024}}), far exceeds its initial role as a mere specialized loading scenario for composite materials. This methodology has evolved into a formidable and highly versatile framework for probing nonlocal effects in CMs, unlocking new dimensions in the study of heterogeneous media.
When integrated with the RVE, the forcing term (\ref{2.4}) emerges as an indispensable element in advancing the application of LM and NN techniques, facilitating the construction of surrogate operators within the powerful context of the PM. This marks a transformative leap, pushing the boundaries of traditional micromechanics to a more sophisticated and adaptable paradigm capable of addressing complex material behaviors with unprecedented precision and scalability.
\rmfamily

\sffamily
\noindent{\bf Comment 9.3.} PNO was utilized in \cite{Jafarzadehet2024} to develop a continuum constitutive model for single-layer graphene at zero temperature using synthetic data from molecular dynamics (MD) simulations. Graphene, a two-dimensional carbon allotrope with a hexagonal lattice structure (with an interatomic distance of 1.46\,\,$^{^o}\!\!\!\!\!A$), possesses remarkable mechanical properties and various applications \cite{Akinwandeet2017}. The MD code calculates the equilibrium atomic displacements and interatomic forces under external force fields and boundary conditions, with data generation details in \cite{{Sillinget2023},{Youet2022}}. Integrating AGIE0-CAMNN into the model \cite{Jafarzadehet2024} is straightforward. At each $\bfb_k(\bfx)$ ($k=1,\ldots,N)$ (\ref{2.4}), we conducted an MD simulation $\bfcD^{\rm DNS}_k(\bfLa^k_j)$ as described in \cite{Jafarzadehet2024}, for each specific grid $\bfLa_j$ ($j=1,\ldots,M)$.

For each $\bfb_k(\bfx)$, we estimate $\bfcD^{\rm DNS}_k(\bfLa^k_j)$ for the sets of the grids $\bfLa^k_j$ ($j=1,\ldots,M)$ and compute $\bfcD^{\rm p}_k$ by applying a translation average to $\bfcD^{\rm DNS}_k(\bfLa^k_j)$ (with the step, e.g., $d|\bfchi|=0.1\,\,^{^o}\!\!\!\!\!A$ in Eq. (\ref{5.6})).
This leads to the calculation of $\bfcD^{\rm p}$ (\ref{5.12}), which then transitions from Block 2 Field PM data to Block 3 Optimization in Fig. 8, followed by the construction of Block 5 Set of surrogate operators. To fully align with the original PNO approach \cite{Jafarzadehet2024}, $\bfcD^{\rm DNS}_k(\bfLa^k_j)$ can be estimated using a coarse-grained method \cite{Jafarzadehet2024}.
The principal advantage of $\bfcD^{\rm p}$ (\ref{5.12}) lies in its ability to eliminate both size and boundary effects, while remaining independent of any particular computational grid $\bfLa^k_j$. {\color{black} Conversely, the surrogate operators proposed in \cite{Jafarzadehet2024, Jafarzadehet2024b} (see Comment 9.1) are constructed with respect to a predefined grid $\bfLa^k_j$.} 
Extending AGIE-CAMNN to CMs with random or periodic distributions of inclusions from different materials, to a two-dimensional lattice with random properties \cite{Sillinget2024}, or to the modeling of defects in 2D materials (\cite{{Jiet2025},{Zaeemet2024}}) is straightforward and provides the same benefits as those seen with the uniform grid $\bfLa$. 
The effectiveness and broad applicability of AGIE-CAMNN to periodic systems stem from the generality of the operator form of solutions (\ref{5.1}) and (\ref{5.2}), which can be derived through methods such as, e.g., MD simulation or continuum-based molecular mechanics (see for references \cite{{Jiet2025},{Zaeemet2024}}).

\rmfamily

\sffamily
\noindent{\bf Comment 9.4.} In the LM, the ML and ANNs are being increasingly utilized to model composite systems, demonstrating their power as predictive tools for data-driven multi-physical modeling. These techniques provide insights into system properties that go beyond the capabilities of traditional computational and experimental analyses. Comprehensive reviews on these advancements can be found in the works \cite{{Agarwalet2024},{Guet2018},{LiuTet2020},{Liuet2024},{Sharmaet2022}}.
For example, in the work \cite{Guet2018} (see also \cite{Bhaduri et2022}), a Convolutional NN (CNN) was applied to predict key mechanical properties such as toughness and strength of materials.

Additionally, in a study \cite{Yinet2024}, a combined approach of CNN and Principal Component Analysis (PCA) was proposed to efficiently predict stress-strain curves, incorporating three critical material features: tensile strength, modulus, and toughness for fiber-reinforced composites.
Lefik {\it at al.} \cite{Lefiket2009} applied ANN models to predict the nonlinear elastic-plastic behavior (e.g., the strain-stress curve) of a two-phase composite material, while Le {\it et al.} \cite{Leet2015} used RVE analyses with periodic boundary conditions to generate training data for constructing a constitutive model for nonlinear elastic behavior.
Finally, the PINN homogenization theory, proposed by Chen {\it et al.} \cite{Chenet2024}, is a method for identifying homogenized moduli and local electromechanical fields in periodic piezoelectric composites.
In the study of micromechanics of periodic CMs, Buryachenko \cite{Buryachenko2023g} demonstrated that Dvorak’s \cite{Dvorak1992} transformation field analysis provides the foundation for the self-consistent clustering analysis (SCA), also known as clustering discretization methods (CDM), developed by Liu {\it et al.} \cite{Liuet2016} (see also
\cite{{Liet2019},{Tanget2018},{Zhanget2019}}). CDM has been applied to a broad range of nonlinear problems, such as nonlinear elasticity \cite{Liet2019}, various nonlinear interface properties \cite{Chenet2024}, elastoplasticity \cite{Huanget2023}, elastic-viscoplasticity \cite{Yuet2019}, and damage analysis \cite{{ChaouchY2024},{Heet2020}}.
A further step in the generalization of data-driven ML approaches in local micromechanics (LM) \cite{{Buryachenko2022a},{Hanet2020},{Hanet2025},{Heet2020},{Kafka2018},{Liet2019},{Liuet2018b}} can be easily extended to their peridynamic equivalents \cite{Buryachenko2023d}. This includes modeling different physical phenomena, such as state-based models, diffusion, viscosity, thermoelastoplasticity, damage accumulation, debonding, plastic localization, and wave propagation, for composite materials with various periodic structures (e.g., polycrystals, fiber networks, hybrid structures, foam materials).
Recently, Liu {\it et al.} \cite{Liuet2020} (see also \cite{Nguyenet2022})
developed a deep material network for
for process modeling, material homogenization, machine learning, and multiscale simulation.

The main limitation of these ML and ANN techniques (referred in Comment 9.4) in CMs modeling lies in the implicit use of Definition 8.1, which implies either a finite sample size \cite{{Bhaduri et2022},{ChaouchY2024},{Hanet2025}} or a periodic system \cite{{Hanet2020},{Heet2020},{Kafka2018},{Liet2019},{Liuet2018b},{Yanget2019},{Yuet2019}} with corresponding boundary conditions (BC). 
The requirement of periodicity is essential for methods that couple the FFT approach (see Subsection 8.6) with image-learning techniques. In such frameworks, periodicity is enforced through a specialized preprocessing step that involves locally reassigning the material properties of voxels situated along the edges or faces of the voxelized domain (see \cite{{Leclercet2025},{Liaoet2024}}).
As a result, factors like sample size, boundary layers, and edge effects are typically not considered in the analysis. 
Extending the mentioned methods based on Definition 8.1 to cases corresponding to Definition 8.3, and then constructing surrogate nonlocal operators, is a challenging and uncertain task. However,
all these methods can be easily recast in the framework of proposed Comments 9.1 and 9.2 (and Fig. 8) using RVE Definition 8.3. 
For periodic structures in CMs, PBC, as shown in equations (\ref{2.39}) and (\ref{2.40}), are replaced by the BFCS (\ref{2.4}). 
In this context, the conventional periodic unit cell $\Omega$ with PBCs (\ref{2.39}) and (\ref{2.40}) is replaced by a periodically repeated mesocell $w$ that fully contains the RVE and is subjected to BFCS loading (\ref{2.4}). The size of these mesocells $w$ serves as a postprocessing learning parameter, selected to ensure that the strain field $\bfep(\bfy)$ vanishes in the boundary layer $w \setminus \text{RVE}$. This layer enforces vanishing boundary conditions—either the classical PBCs (\ref{2.39}), (\ref{2.40}) or their nonlocal variants VPBC (\ref{2.37}), (\ref{2.38}).
The classical CDM framework developed for periodic unit cells $\Omega_{00}$ can be directly reformulated for RVEs embedded in mesocells $w$. Notably, the same computational scheme extends naturally to deterministic structure composites with RVEs contained in $w$--a generalization not addressed in traditional CDM formulations.
\rmfamily
%$v_i(\bfx), \bfbC(\bfx), \Omega^{\rm sq}$
%$X^{\partial v}, \Omega^{\rm sq}$ \ \ \ $\bfu(\bfx), \bfsi(\bfx), \bfL^*$

\section{Conclusion}
%\vspace{-2.mm}
The term ``micromechanics" traditionally brings to mind concepts such as effective moduli, linear local elasticity, RVE by Hill \cite{Hill1963}, and remote homogeneous boundary conditions. These foundational ideas have long defined the field, providing a structured and rigorous framework for analyzing material behavior at the microscale. However, we have moved beyond these conventional constraints—breaking free from this rigid, {\it Procrustean bed} of assumptions. By challenging and redefining the fundamental principles of micromechanics, we have expanded its scope, introducing new methodologies and perspectives that transcend the limitations of traditional approaches. This shift allows for a more flexible and comprehensive understanding of microscale interactions, paving the way for innovative applications and deeper insights into material behavior.

In particular, a systematic analysis of composite materials (CMs) subjected to a BFCS (\ref{2.4})—rather than remote homogeneous boundary conditions—immediately removes all the aforementioned restrictions. More specifically, the proposed universal framework, referred to as AGIE-CAM, is highly adaptable and built upon physically intuitive hypotheses. These hypotheses can be modified or refined as needed, even to the extent of being discarded, not merely for the sake of theoretical complexity but in response to practical challenges and advancements. Within a unified analytical scheme, AGIE-CAM enables the study of a broad spectrum of micromechanical problems.
This approach encompasses CMs with random (statistically homogeneous and inhomogeneous), periodic, and deterministic (neither random nor periodic) structures. It also addresses materials exhibiting linear and nonlinear behavior, coupled and uncoupled locally elastic responses, as well as weakly nonlocal (strain gradient and stress gradient) and strongly nonlocal (strain-type and displacement-type, peridynamics) phase properties.
Although the PM of both the random and periodic structure CMs was historically developed as a generalization of the corresponding methods of the LM,
it does not mean that PM's society manifests itself as a user of the LM methods;
we also demonstrated that some methods developed in PM initiate the new methods of LM unknown before (i.e., the methods of both the LM and PM mutually enrich one another).

To present LM and PM within a unified theoretical framework, we introduce—for the first time—a structured, modular approach. Both LM and PM are formulated as block-based (or modular) schemes, allowing specialists in one block to contribute without requiring expertise in the underlying details of other blocks.
This modular structure consists of distinct block teams, including the image technique team (see Figs. 2 and 4), the Block 3 Perturbator and Block 4 Micromechanics teams (see Fig. 6), and the ML and NN technique team (see Figs 6 and 7). Each team (of Block 3 and Block 4) operates independently within its designated block while ensuring seamless integration within the larger framework.
This structured approach facilitates effective collaboration while maintaining flexibility and efficiency in the overall system.
The development of this modular structure in PM is enabled by a critical generalization of CAM, which builds on the new AGIE (\ref{3.23}). AGIE-CAM is an exceptionally flexible, robust, and physics-based framework that integrates data-driven, multi-scale, and multi-physics modeling, accelerating both fundamental and applied research in random, periodic, and deterministic heterogeneous media.
However, fully leveraging AGIE-CAM’s potential requires collaborative efforts across multiple disciplines, including image processing, computational mathematics, micromechanics, material science, physics, and data science. The proposed modular 

\vspace{-0.mm} \noindent \hspace{-3mm} \parbox{18.2cm}{
\centering \epsfig{figure=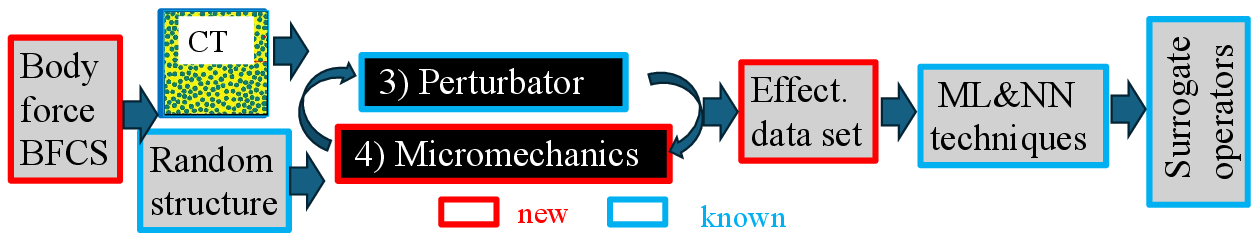, width=16.2cm}\\ \vspace{-179.mm}
\vspace{119.mm}
\vspace{-115.mm} \tenrm \baselineskip=8pt
{{\sc Fig. 9:} The total scheme of AGIE-CAM}}
\vspace{2.mm}

\noindent PM structure provides an ideal foundation for fostering effective interdisciplinary collaboration in this complex and multidisciplinary field.
The blocked (or modular) structures, as depicted in Figures 2, 4, 6, 7, and 8, are meticulously designed to  allow experts working on one block to 
%\noindent   
focus on their specialized area without needing to be well-versed in the intricacies of other blocks. This modularity fosters independence among teams, enabling them to work concurrently with minimal overlap of expertise. Furthermore, when developing joint software, each block team (e.g., such as Teams 3 or 4, as shown in Fig. 6) has the autonomy to make 
modifications or improvements to their block at any stage of its development, without the need to inform or disrupt the partner team (Teams 4 or 3, respectively).
The only necessary coordination arises between Blocks 3 and 4, where effective collaboration requires alignment on specific adjustments between the two blocks.

Broadly speaking, this review focuses on the development of Block 4 Micromechanics and its integration with Block 3 Perturbator (see Fig. 6), along with its alignment with other blocks. The specific methods for solving Block 3 Perturbator are not examined in detail; they are only briefly mentioned in the Introduction and Subsection 3.2 and are assumed to be known.
The review does not include comparisons between different peridynamic models, numerical results, or discussions on the significance, applications, and limitations of peridynamic theory—these topics are beyond their scope (refer to the Introduction for relevant references). 
Additionally, while the blocks depicted in Figs. 7 and 8 are only briefly mentioned, their interaction with Blocks 3 and 4 in Fig. 6a is analyzed in greater detail. The overall architecture of the AGIE-CAM framework is illustrated in Fig. 9, where the newly introduced 
$\bfcD^{\rm p}_k$, and $\bfcD^{\rm d}_k$ are established through a revolutionary RVE concept that extends Hill’s classical framework \cite{Hill1963}. Unlike conventional RVE definitions, which are constrained by the constitutive laws of material phases or the functional forms of surrogate operators, this new approach is based on intrinsic field concentration factors within the phases of both random and periodic CMs. This makes it a universal and highly adaptable tool applicable across a broad range of material systems.
By integrating this generalized RVE concept into  effective dataset generation, the framework becomes inherently suited for ML and NN techniques in predicting nonlocal surrogate operators. 
Owing to the presence of the domain $\overline{\rm RVE}$, characterized by a vanishing strain field, this innovation effectively overcomes major challenges such as sample-size dependence, boundary-layer effects, and edge-induced inaccuracies.

As a result, the approach ensures highly accurate and reliable

\vspace{40.mm}

\noindent 
 models for any structure CMs mentioned, regardless of whether their phases exhibit local or nonlocal, linear or nonlinear elastic properties. 
         The surrogate operators derived from these  effective  datasets
 demonstrate exceptional robustness and reliability,   maintaining consistent performance even in complex micromechanical systems. This breakthrough significantly expands the capabilities of ML and NN models, enhancing their accuracy and generalization potential in applications requiring precise micromechanical analysis of intricate material systems.

{\color{black}It is remarkable that whenever one hears the term “micromechanics,” the following notions are almost invariably the first to come to mind: 1) Effective moduli; 2) Homogeneous boundary conditions; 3) Hill's RVE; 4) Scale separation hypothesis;
5) Effective field hypothesis.  These notions constitute the conceptual foundation of micromechanics, forming its structural framework. At the same time, they act as constraints inherent 
%\vspace{39.mm}
\noindent to both 
Amic (\ref{3.9}) and Cmic (\ref{3.10}), thereby functioning as a kind of {\it Procrustean bed}. Overcoming this {\it  Procrustean bed} (marked with red crosses in Fig. 10) is not the primary aim of the present study but rather emerges as a significant byproduct of the innovative AGIE-CAM formulation.
Although replacing one class of loadings (\ref{2.30}) or (\ref{2.31}) (or (\ref{2.28}) or (\ref{2.29})) with another type (\ref{2.4}) may seem a purely technical modification, it in fact represents a fundamental shift in the conceptual philosophy of micromechanical research.

Broadly speaking, conventional micromechanical approaches based on the GIE—whether expressed through Eqs. (\ref{3.13}), (\ref{3.14}), or (\ref{3.24})—have historically been devoted to addressing classical problems, such as the determination of effective material properties, a research direction dating back to 1824–1879 (see Subsection 3.2).
Although these traditional frameworks have evolved through advanced formulations such as the Amic and Cmic methods (Eqs. (\ref{3.9}) and (\ref{3.10})), they remain anchored to long-established objectives and constrained by the conceptual limitations of the {\it Procrustean bed}.
In contrast, the emerging paradigm of micromechanics, founded on the AGIE-CAM framework (Eqs. (\ref{3.25}) or (\ref{3.23})), introduces a fundamentally new problem governed by the BFCS loading (\ref{2.4}).
Rather than simply extending traditional methodologies, it redefines and repurposes them—particularly those dealing with a finite set of interacting inclusions within an RVE (see Fig. 7)—to tackle new, nonclassical challenges that extend far beyond the traditional boundaries of micromechanics.
The restrictive concepts constituting the {\it Procrustean bed} are no longer required within the AGIE-CAM framework and are marked with red crosses in Fig. 10.
Moreover, the removal of the EFH results in the complete collapse of the Amic formulation (\ref{3.9}), which is likewise
marked with red crosses
in Figs. 7 and 10.
Furthermore, the application of BFCS loading (\ref{2.4}) to periodic systems leads to the loss of solution periodicity, implying that the methods of Blocks 1 and 2 (\ref{3.10})—also marked by red crosses in Figs. 7 and 10 are inapplicable under such loading conditions. Ultimately, the AGIE-CAM approach reduces the solution of both Amic and Cmic problems to the analysis of a finite inclusion field within a new RVE, represented by the modified Block 3 in Figs. 7 and 10, entirely free from the {\it Procrustean bed} constraints of conventional micromechanics (see Fig. 10).
The solution of this modified Block 3 provides the basis for the complete AGIE-CAM computational chain:
AGIE $\to$ RVE $\to$ Effective dataset $\to$ ML\&NN technique (see Fig.9).}

{\color {black} In addition to the theoretical distinctions between AGIE and GIE, it is particularly compelling to evaluate their practical relevance. The classical EFH and GIE (as well as GIE-CAM) frameworks rely on solving the governing GIE under BC (\ref{2.28}) to estimate effective moduli and field concentration factors. The utility of these methods has been firmly established for nearly two centuries since 1824.
By contrast, the practical impact of solving AGIE under BFCS loading (\ref{2.28}) is negligible unless accompanied by ML\&NN methods. In fact, without the incorporation of the ML\&NN block (if it is explicitly omitted from Fig. 9), the AGIE solution has little to no real-world significance. This is precisely why AGIEs, despite the simplicity of the foundational form (\ref{3.25})—which is even more straightforward than the first GIE (\ref{3.13})—have historically been overlooked.
It is only through the integrated AGIE-CAMNN approach, combining AGIE with ML\&NN tools, that one achieves meaningful outcomes. This synthesis enables the prediction of a wide range of predefined surrogate operators—both for overall effective properties and for local concentration fields—while eliminating the influence of sample size, boundary layers, and edge effects. This marks a substantial leap forward and underscores the transformative potential of AGIE when empowered by modern computational intelligence.} 

\vspace{1.mm} \noindent \hspace{-8mm} 
\parbox{10.2cm}{
\centering \epsfig{figure=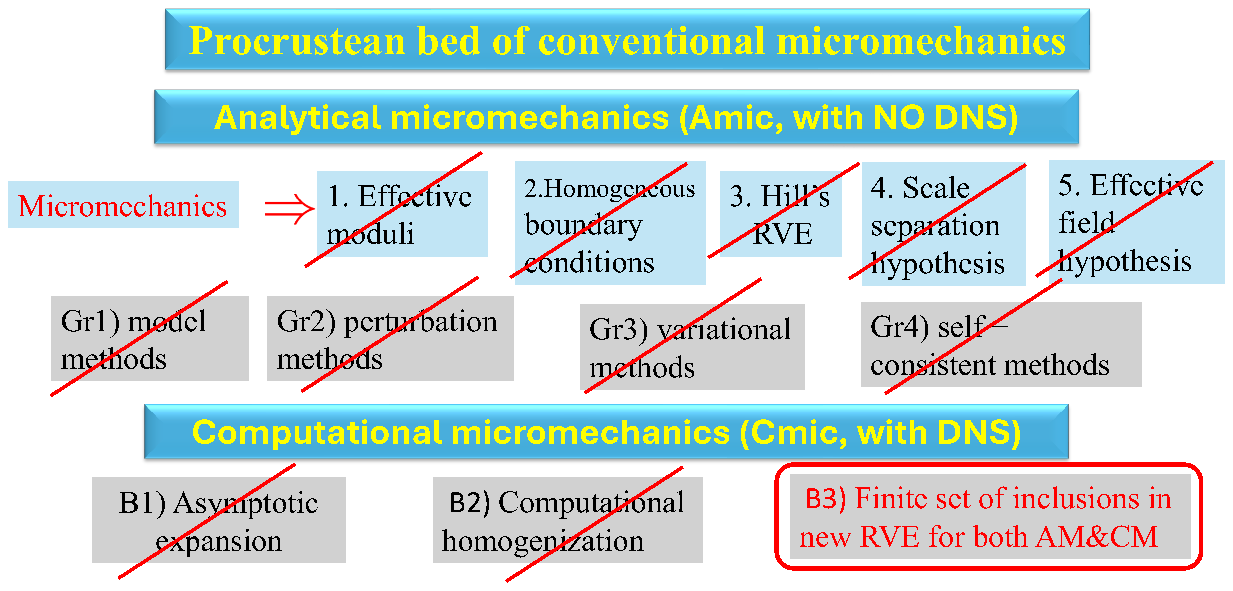, width=9.4cm}\\ \vspace{-122.mm}
\vspace{122.mm}
\vspace{-78.mm} \tenrm \baselineskip=8pt
{{\sc Fig. 10:} Procrustean bed of micromechanics}}
\vspace{3.mm}

While preserving the generality of classical GIE-CAM based on GIE (\ref{3.24}), the specific nature of the loading condition (\ref{2.4}) allows AGIE (\ref{3.23}) to be formulated with the same precision as GIE (\ref{3.24}). This, however, is not merely a restatement; it serves as a pivotal step toward a more comprehensive generalization, enabling GIE (\ref{3.24}) to handle arbitrary loading scenarios. These include (i) general body forces, which may or may not have compact support (\ref{2.4}), and (ii) remote boundary conditions (VBC), which are not necessarily homogeneous (\ref{2.28}).
What distinguishes AGIE (\ref{3.23}) is its remarkable increase in flexibility and the broader spectrum of surrogate operators it accommodates, extending far beyond the capabilities of GIE (\ref{3.24}). This leap forward leads to the development of a truly universal AGIE-CAM framework for studying composites (CMs) with arbitrary microstructures and phase properties, as outlined in the paper. More significantly, this next-generation AGIE-CAM serves as a unifying framework, harmonizing various analytical approaches, as illustrated in Figures 2, 4, 6, 7, and 8. It ensures that these methodologies can operate both in synergy and independently, thus establishing a more integrated yet modular analytical paradigm.
This development represents the establishment of a {\it Unified Micromechanics Theory} for heterogeneous media, signifying a pioneering breakthrough in the field and introducing, in essence, a {\it new philosophy of micromechanics}.

{\color{black}
\smallskip
{\bf Acknowledgments:}
The author acknowledges Dr. Stewart A. Silling for the fruitful personal discussions,
encouragements, helpful comments, and suggestions. The author also acknowledges the reviewers for the
encouraging comments that initiated a significant correction of the manuscript
Permissions to reproduce Fig. 5  from Silling {\it et al.} \cite{Sillinget2024} and Fig. 3 from \cite{Youet2022b} were granted
by Springer Nature (License Number 6044880317237) and ASME (License Number 1631311), respectively.
}

\end{document}